\documentclass[12pt]{article}

\AtBeginDocument{%
   \setlength\abovedisplayskip{0.05in}
   \setlength\belowdisplayskip{0.08in}}

\usepackage[onehalfspacing]{setspace}

\makeatletter
\renewcommand\@footnotetext[1]{%
  \insert\footins{%
    \reset@font\footnotesize
    \interlinepenalty\interfootnotelinepenalty
    \splittopskip\footnotesep
    \splitmaxdepth \dp\strutbox \floatingpenalty \@MM
    \hsize\columnwidth \@parboxrestore
    {\setstretch{1.0}\protect\@makefntext{%
      \rule{\z@}{\footnotesep}\ignorespaces#1}}}}
\makeatother

\usepackage[top=50mm, bottom=-500mm, left=50mm, right=50mm]{geometry}
\usepackage{lineno}
\usepackage{amssymb}
\usepackage{amsthm}
\usepackage{amsmath}
\usepackage{cases}
\usepackage{caption}
\usepackage{epsfig}
\usepackage{graphicx}
\usepackage{graphics}
\usepackage{float}
\usepackage{subfigure}
\usepackage{multirow}
\usepackage{xcolor}
\usepackage{lineno}
\usepackage{fullpage}
\usepackage[normalem]{ulem} 
\usepackage{makeidx}
\usepackage{xspace}
\usepackage{wrapfig}
\usepackage{amsmath}
\usepackage{lscape}
\usepackage{mathrsfs}
\usepackage{adjustbox}
\usepackage{multirow}
\usepackage{varwidth}
\usepackage{rotating}
\usepackage{lscape}
\usepackage[bottom]{footmisc}
\usepackage[colorlinks,citecolor=darkblue,urlcolor=darkblue,linkcolor=darkblue]{hyperref} 
\usepackage{cleveref}
\makeindex
\newtheorem{theorem}{Theorem}

\newtheorem{des}{Desideratum}
\newtheorem{corollary}{Corollary}

\newtheorem{lemma}{Lemma}
\newtheorem{ex}{Example}

\newtheorem{proposition}{Proposition}
\newtheorem{definition}{Definition}

\newtheorem{axiom}{Axiom}

\newtheorem{obs}{Observation}

\definecolor{purple}{rgb}{0.6, 0.4, 0.8}
\definecolor{darkred}{rgb}{1, 0.1, 0.3}
\definecolor{darkblue}{rgb}{0.0, 0.0, 0.55}
\definecolor{darkgreen}{rgb}{0,0.6,0.5}
\definecolor{forestgreen}{rgb}{0.0, 0.46, 0.37}
\definecolor{bittersweet}{rgb}{1.0, 0.44, 0.37}
\definecolor{navy}{rgb}{0.0, 0.0, 0.55}
\definecolor{brown}{rgb}{0.53, 0.18, 0.09}
\definecolor{Green}{rgb}{0.0, 0.47, 0.44}

\newcommand {\mm}[1] {\ifmmode{#1}\else{\mbox{\(#1\)}}\fi}

\newcommand\E{\mathbb{E}}

\newcommand\R{\mathbb{R}}

\usepackage{dsfont}

\newcommand{\mb}[1]{\textbf{#1}}

\usepackage{sectsty}
\sectionfont{\centering\Large}
\subsectionfont{\large}
\usepackage{fancyhdr}
\usepackage{natbib}

\newtheorem{remark}{Remark}

\usepackage{pst-node, pst-plot, pst-poly}
\usepackage{rotating}

\usepackage{tikz}
\usepackage{venndiagram}
\usepackage{pstricks}
\usepackage{tikz-cd}
\usepackage{siunitx}
\usepackage{pgfplots}
\usepgfplotslibrary{fillbetween}
\usepackage{algorithm}
\usepackage{algorithmic}
\usepackage{bbm}
\usepackage{csquotes}

\usepackage{scalerel}
\usepackage[T1]{fontenc}
\usepackage{titling} 
\usetikzlibrary{shapes,arrows,trees,calc}
\pgfplotsset{compat=1.12}

\allowdisplaybreaks 
\begin{document}

\setlength{\droptitle}{-0.95in}   
\title{\small\bf \mbox{\MakeUppercase{Robust Aggregation of Preferences}}\thanks{I sincerely thank Drew Fudenberg and Stephen Morris for their guidance and support during this project. I also want to thank Hengjie Ai, Ian Ball, Roberto Corrao, Simone Cerreia-Vioglio, Bach Dong-Xuan, Glenn Ellison, Itzhak Gilboa, Nima Haghpanah, Matthew Jackson, Giacomo Lanzani, Ricky Li, Charles Manski, Massimo Marinacci, Esfandiar Maasoumi, Anna Merotto, Marcus Pivato, Lorenzo Stanca, Tomasz Strzalecki, Alex Wolitzky, Muhamet Yildiz, and seminar participants at the MIT Theory Lunch and the Online SCW Seminar for insightful comments and discussions. \textit{JEL codes}: D71, D81.
}
\vspace{-0.2in}
}

\author{
\textsc{Florian Mudekereza}\thanks{Department of Economics, MIT, \href{mailto:florianm@mit.edu}{\texttt{\footnotesize florianm@mit.edu}}.}
}

\date{}
\maketitle
\thispagestyle{empty}
\setcounter{page}{0}
\vspace{-0.88in}
\begin{abstract}
This paper analyzes a society composed of individuals who have diverse sets of beliefs (or models) and diverse tastes (or utility functions). It characterizes the model selection process of a social planner who wishes to aggregate individuals' beliefs and tastes but is concerned that their beliefs are \textit{misspecified} (or incorrect). A novel \textit{impossibility} result emerges under several desiderata: a utilitarian social planner who prioritizes \textit{robustness} to misspecification never aggregates individuals' beliefs but instead behaves as a \textit{dictator} by adopting one individual's belief as the social belief. This tension between robustness and aggregation exists because aggregation yields policy-contingent beliefs, which are very sensitive to policy outcomes. The impossibility can be resolved, but it would require assuming individuals have heterogeneous tastes and some common beliefs. 
Applications in treatment choice and dynamic macroeconomics are explored.
\newline\textit{Keywords}: welfare aggregation, utilitarianism, robustness, misspecification, ambiguity 
\end{abstract}
{\footnotesize  \textit{For every bias identified for individuals, there is an accompanying bias in the public sphere.}
   \par\hfill-- \citet[][p. 102]{cass14}}
\vspace{-0.24in}

\section{Introduction}
  Policymakers are entrusted with choosing policies that maximize the welfare of all members of society. This is a challenging task, however, because such policies often yield \textit{uncertain} outcomes, and the members' preferences may not align \citep{manski23}. That is, their beliefs about the contingencies may differ, and their tastes may also differ. Consider surveying experts about the future impact of artificial intelligence (AI) on humanity. The fact that AI operates as a ``black box'' yields conflicting predictions about its impact. Some experts believe that AI will be beneficial due to technological advancement, e.g., in medicine \citep{med22}, whereas others fear that AI will eliminate jobs due to excessive automation, lead to privacy violations \citep{daron21}, and even cause human extinction \citep{chad24}.  
Given these disparate opinions, how should a policymaker regulate AI? We operationalize normative principles to  guide the choice of welfare-improving policies in such environments.
\par The welfare-aggregation literature recommends forming a social preference by aggregating individuals' beliefs and tastes. 
 \citeauthor{robust16}'s (\citeyear{robust16}) \textit{revealed common-taste unambiguous Pareto dominance} yields the most general utilitarian aggregation of preferences: it restricts the social utility function to be a linear combination of individuals' utility functions and the set of social beliefs to be any subset of the convex hull of all individuals' beliefs. This convex-hull restriction is interpreted as a ``utilitarian aggregation of individuals' beliefs.''\footnote{For example, \citet[][p. 1183]{util16} note: ``Society’s belief consists of a set of prior probabilities that are weighted averages of individual priors [...] Thus [...] each social prior is a `utilitarian' aggregation of individual priors.'' Similarly, \citet[][p. 112]{billot21} note: ``The utilitarian aggregation rule requires social utility and beliefs to be a convex combination of individual utilities and beliefs, respectively.''} However, since social beliefs are allowed to be \textit{any} subset of the convex hull, the social preference is not uniquely \textit{identified}. This lack of identification poses a major challenge for implementation because, as \citet[][p. 1754]{simsek14} ask, ``which belief should the planner use?'' We contribute to this literature by showing that an \textit{impossibility} result emerges when we attempt to identify the beliefs of a social planner who is concerned that individuals' beliefs are \textit{misspecified}: she always behaves like a ``probability dictator'' by choosing one individual's belief as the social belief. We show this separately using three social planning \textit{desiderata}: 
\begin{itemize}
    \item \textit{Desideratum} \ref{des:welfare}: \textit{Welfare Dominance}. Let the welfare criterion be convex in individuals' beliefs and consider the following model selection: The social planner chooses a subset of the convex hull that yields the highest welfare uniformly across  every policy that all individuals agree on. Proposition \ref{thm:welfare} demonstrates that this subset must be a  singleton consisting of one unique individual’s belief. Being a singleton suggests that a misspecification-averse social planner aims to minimize ambiguity at the social level.
        \item \textit{Desideratum} \ref{des:max}: \textit{Maximum Ambiguity}. Instead of allowing the social planner to choose any subset of the convex hull, suppose we follow \citet{simsek14} and \citet{util16} by insisting that she chooses the entire convex hull. Proposition \ref{thm:max} shows that the social planner will still choose one individual’s belief as the social belief.
    \item \textit{Desideratum} \ref{des:hybrid}: \textit{Robust-Control Axiomatization}. Suppose the welfare criterion is strictly convex in individuals' beliefs and differs from that of a Bayesian planner. These properties correspond to behavioral axioms that are satisfied by many robust criteria such as \citeauthor{hansen01}'s (\citeyear{hansen01}) \textit{multiplier} criterion. Proposition \ref{thm:hybrid} shows that a social planner who satisfies these axioms must choose one individual's belief as the social belief. This happens because aggregating beliefs leads to violations of Pareto dominance. 
\end{itemize}
 
\par Propositions \ref{thm:welfare}--\ref{thm:hybrid} constitute our impossibility results. Let's illustrate them in the context of AI. Consider the U.S. agency tasked with regulating AI, the Center for AI Standards and Innovation (CAISI). Since CAISI's main goal is AI safety, suppose its director requests each of her advisors to report the probability that AI will cause ``extremely bad'' outcomes (e.g., human extinction) in the foreseeable future.\footnote{\citet[][Section II.A]{chad24} uses a continuous-time ``utilitarian social welfare'' to answer this question.} Assuming these advisors are AI experts, \citet{ai24} find, in a survey of  2,778 AI experts, that 57.8\% of them reported at least a 5\% probability of extremely bad outcomes, and 1 in 10 experts reported at least a 25\% probability. Assuming the director can also report her own belief, suppose she is very optimistic about AI and reports that the probability is zero. Revealed common-taste unambiguous Pareto dominance requires her to choose any probability in the convex hull of everyone's (including her own) reported probabilities. Our impossibility results show that if she is concerned that the advisors' probabilities are incorrect and wishes to hedge against this,  
then revealed common-taste unambiguous Pareto dominance would lead her to use her own belief and disregard all her advisors' reports. This dictatorial behavior is undesirable, however, because it conflicts with CAISI's mission: ``CAISI will serve as industry's primary point of contact within the U.S. government to facilitate testing and collaborative research.''\footnote{This quote is part of the first sentence on CAISI's webpage: \url{https://www.nist.gov/caisi}.} 

\par Our impossibility results exploit the fact that allowing the social planner to choose any probability in the convex hull is too permissive. Resolving these impossibility results, if possible, will therefore require working with a less permissive Pareto principle. The rest of this paper shows that this is possible using \citeauthor{robust16}'s (\citeyear{robust16}) \textit{revealed unambiguous Pareto dominance}, which prescribes that if all individuals unambiguously prefer a policy over another, then so should the social planner. Section \ref{sec:rep2} shows that satisfying revealed unambiguous Pareto dominance is equivalent to restricting the social utility function to be a linear combination of individuals' utility functions and the social beliefs to be any subset of the \textit{intersection} of individuals' sets of beliefs. \citet{manski95} interprets such an intersection as a \textit{domain of consensus}---it consists of those probabilities that all AI advisors agree on in the AI example, which is desirable  because it rules out the director's zero-probability belief. 
\par However, this new aggregation requires imposing restrictive conditions on individuals' tastes and beliefs: (1) individuals must have diverse tastes and (2) they must have some beliefs in common. Notice also that the identification problem reemerges here because social beliefs are allowed to be \textit{any} subset of the intersection. To resolve this, assume also (3) individuals have \textit{regular} beliefs---their sets of beliefs are \textit{Bregman} balls (e.g., entropy balls), which are popular in economics and information geometry \citep{chern13}. The center of a Bregman ball is interpreted as an individual's ``reference'' model. In the AI example, (3) requires a standardized reporting format wherein each advisor reports only two things: a probability they think is most plausible and a ``radius'' capturing their own \textit{confidence} in it. Under (1)--(3), Corollary \ref{thm:pos1} establishes the following utilitarian aggregation of beliefs: The unique social belief is a convex combination of individuals' reference models.  The intuition for this result is that the intersection of Bregman balls can be shrunk to a singleton, which would force the social planner---regardless of her ambiguity attitude---to adopt a utilitarian subjective expected utility (SEU) preference using the unique belief in the intersection. Unlike all prior studies that assume every individual is SEU to obtain utilitarian aggregation of beliefs, our possibility result reveals a new insight: A \textit{tradeoff} between robustness and belief aggregation is possible because what matters is the structure of individuals'  beliefs, rather than their attitudes toward uncertainty. In the AI example, this tradeoff will be reflected in the weights assigned to the advisors' reference probabilities---the weight on each advisor must be \textit{inversely} related to the size of their reported range of probabilities (Proposition \ref{thm:mono}). Thus, despite the director's concerns for misspecification, belief aggregation is possible. 
\par Section \ref{sec:tension} aims to explain the mechanism driving the \textit{tension} between robustness and belief aggregation. To obtain clear insights, we focus on settings where individuals have entropy balls, and in the spirit of Desideratum \ref{des:max}, the social planner is forced to use the entire intersection of individuals' beliefs. Theorem \ref{thm:rep} shows that such a social planner will use a social belief that is policy-contingent---it is a convex combination of individuals' reference models whose weights depend on policies. The policy dependence reflects the fact that the social planner's concerns for misspecification are \textit{contextual}---some policies may involve higher stakes and hence require more caution than others. In the AI example, policies pushing for rapid AI deployment and experimentation may warrant a thorough assessment. Thus,  the chosen social belief is very sensitive to policy outcomes.  Corollary \ref{thm:constant} then reveals that this sensitivity vanishes precisely when the set of social beliefs is a singleton.   
 \par Our framework allows the social planner to have a wide range of attitudes toward misspecification. Following the \textit{robust-control} literature, our welfare criterion is based on \citeauthor{hansenmiss22}'s (\citeyear{hansenmiss22}) novel variational representation, which allows a misspecification-averse decision maker to entertain both \textit{structured} and \textit{unstructured} models by penalizing the latter based on their statistical ``distance'' from the former. We often employ their most tractable criterion whose penalty is the relative entropy. This criterion has four key special cases: (i) When the set of structured models is a singleton, our criterion becomes the multiplier criterion of \citet{hansen01}, which is very popular in the macro-finance literature. (ii) When this singleton is a convex combination of several models,  our criterion resembles \citeauthor{giac22}'s (\citeyear{giac22}) criterion for single-agent problems. (iii) A simpler version of our criterion resembles the \textit{smooth ambiguity} criterion of \citet{smooth05} and coincides with it in special cases. (iv) When the social planner has no concern for misspecification, our criterion becomes the \textit{maxmin} expected utility (MEU) criterion of \citet{gilboa89}. Now, regarding the individuals, we allow them to have ambiguity-sensitive preferences, but they are not necessarily concerned about misspecification. 
Their ambiguity attitudes are described by \citeauthor{cerre11}'s (\citeyear{cerre11}) ``Monotonic Bernoullian Archimedean'' preferences, which is a broad class that includes all popular ambiguity preferences. 
 \par Our multiplier welfare criterion is very tractable, which is relevant because \citet[][p. 63]{strz11} notes the challenge of finding decision models that are ``easy to incorporate into economic models of aggregate behavior.''  
Section \ref{sec:appli} illustrates this tractability in two applications. 
First, we explore treatment choice, where a public authority has to decide the fraction of a society that should receive a treatment. We show that our criterion provides a justification for treatment diversification and reveals some new comparative statics. Second, we revisit \citeauthor{ball18}'s (\citeyear{ball18}) dynamic macro model consisting of a non-SEU representative-agent economy where macro announcements generate a premium by resolving uncertainty about the future. However, in a critique of representative-agent macro, \citet[][p. 118]{kirman92} notes: ``First, whatever the objective of the modeler, there is no plausible formal justification for the assumption that the aggregate of individuals, even maximizers, acts itself like an individual maximizer [...]  There is simply no direct relation between individual and collective behavior.'' We address these concerns by showing that our misspecification-averse social planner can behave identically to \citeauthor{ball18}'s (\citeyear{ball18}) representative agent. 
\par\noindent--- \textit{Related Work}: \textit{Utilitarianism} is perhaps the simplest and most influential welfare aggregation principle in social sciences. It is generally applied when social planning is linked to the individuals' preferences via the (standard) Pareto principle. \citet{hars55} proposes the first equivalence between the Pareto principle and utilitarianism when individuals are EU maximizers with diverse tastes but identical beliefs. This turns out to be impossible for SEU individuals \citep[][]{impo79,mongin95}. 
Then, \citet{gil04} ingeniously restored possibility of aggregation by  relaxing the Pareto principle while arguing that a unanimous choice is not always compelling because it may arise due to contradictory beliefs and tastes---a phenomenon famously known as ``spurious unanimity.'' 
\par \citet{simsek14} have pioneered the study of social planning when individuals have misspecified beliefs and tackled the identification problem of the social preference noted earlier. They propose a ``belief-neutral'' welfare criterion based on the idea that a social planner worries that SEU individuals  suffer from behavioral biases that have heterogeneously distorted their beliefs. However, their welfare criterion provides an \textit{incomplete} ranking of policies because it requires the social planner to prefer a policy over another across the entire convex hull of individuals' beliefs. From a normative and axiomatic perspective, \citeauthor{simsek14}'s (\citeyear{simsek14}) framework is a special case of \citeauthor{robust16}'s (\citeyear{robust16}) framework wherein the social preference is the least complete unambiguous preference that satisfies \textit{common-taste unambiguous Pareto dominance} with respect to all individuals' preferences.   
\par This paper can be viewed as an extension of \citet{simsek14} in three main ways. First, we combine \citeauthor{robust16}'s (\citeyear{robust16}) axiomatic framework and \citeauthor{hansenmiss22}'s (\citeyear{hansenmiss22}) decision-theoretic framework to formalize the idea of misspecification concerns in social planning. Second, our welfare criterion provides a \textit{complete} ranking of policies and reveals new (im)possibility results. Third, it also reveals that a \textit{sequential} (rather than a \textit{simultaneous}) aggregation of beliefs and tastes is required under misspecification.  
The presence of misspecified beliefs prompts major welfare concerns because, as \citet[][p. 1760]{simsek14} note, individuals may be unaware of their misspecifications and take actions that hurt their own and others’ welfare. A social planner who relies
on misspecified beliefs may inadvertently choose irreversible
policies that are detrimental to society. \citet[][]{banuri19} find experimental evidence that policy advisors tend to have biased beliefs. 
\par There are other papers that are related to our framework. For example, \citet{util16} assume SEU individuals, an MEU social planner, and propose modifications of the Pareto principle that are equivalent to utilitarianism.
 To obtain similar aggregations, \citet{qu17} assumes individuals and social planner are MEU, and restricts the Pareto principle. \citet{agg21} proposes an ambiguity aversion axiom that leads to a smooth ambiguity criterion, and assumes SEU individuals. \citet{billot21} assume  SEU individuals and social planner, and propose a ``belief-proof'' Pareto principle to address spurious unanimity. 
\par\noindent--- \textit{Organization}: Section \ref{sec:model} describes our framework. The first aggregation and impossibility results are presented in Section \ref{sec:rep1}. The second aggregation and (im)possibility results are in Section \ref{sec:rep2}. Section \ref{sec:tension} analyzes the tension between robustness and aggregation, followed by some properties and comparative statics in Section \ref{sec:prop}. Section \ref{sec:appli} explores some applications, and Section \ref{sec:conc}  is a discussion. \hyperref[app:online]{Online Appendix} provides more applications and extensions.

\section{Framework}\label{sec:model}

\subsection{Preliminaries}
We consider a society consisting of $n\geq1$ individuals. Let $s$ be a \textit{state of the world}, $S$ be a finite set of all such states,\footnote{Almost all our results continue to hold when $S$ is not finite. This is discussed in Appendix \hyperref[app:compact]{A}.} and $X$ be a set of outcomes. A social planner has to choose an act $f$, i.e., a function $f: S \rightarrow X$, and the set of all such acts is $F$. An outcome $x\in X$ is identified with the constant act yielding outcome $x$ no matter which state occurs, so $X\subsetneq F$. \par An element of $X$ specifies an outcome for all individuals in  society. Let $X$ be a convex subset of a Euclidean space. For example, $X$ can be the set of lotteries over a finite set of prizes \citep{ansc63} or the set $\R^{kn}$ of allocations of a finite number $k$ of commodities. 
Let $\Delta:=\Delta(S)$ denote the set of all probability distributions over $S$.  
 \subsection{Preferences}\label{sec:pref}
 A preference over acts is described by a binary relation $\succsim$ defined on $F$. We write $f \succsim g$ when act $f$ is weakly preferred to act $g$. Strict preference and indifference are $\succ$ and $\sim$, respectively. We consider the broadest class of ambiguity-sensitive preferences called ``Monotonic Bernoullian Archimedean'' (MBA) preferences, which satisfy \citet[][Axioms 1--4]{cerre11} in Online Appendix \hyperref[app:mba]{D}. 
 This class includes most of the popular ambiguity models such as MEU \citep{gilboa89}, Choquet expected utility \citep{choq89}, smooth ambiguity \citep{smooth05}, and variational \citep{mmr06}.
\par \citet[][Proposition 2]{cerre11} shows that any MBA preference induces a ``revealed'' \textit{unambiguous} preference as in \citet{bew02}, which captures the component of the preference ranking that is unaffected by the ambiguity that the individual perceives. 
\begin{definition}\label{def:unamb} \normalfont A binary relation $\succsim^*$ on $F$ is an \textit{unambiguous} preference relation if there exists a pair $(u,P)$, where $u: X \rightarrow \R$ is a nonconstant, affine utility function,  and $P\subseteq\Delta$ is a closed, convex set of probability distributions on $S$, such that, for every acts $f,g \in F$, 
$$f\succsim^* g  \quad\text{ if and only if }\quad \E_p[u(f)] \geq \E_p[u(g)]\quad \forall p\in P,$$ where $\E_q[u(f)]$ denotes the expected value of $u(f(s))$ with respect to some belief $q\in \Delta$.  
\end{definition}
The set $P$ captures a decision maker's confidence about the unknown states, 
so it may not be shrunk to a singleton without compromising the notion of ``confidence'' \citep{gil11}. 
It is unique for $\succsim^*$ whereas the utility function $u$ is unique up to a positive affine transformation. When $P$ is a singleton, $\succsim^*$ is SEU, whereas when it contains multiple beliefs, $\succsim^*$ is an unambiguous preference, which satisfies all SEU properties except \textit{completeness}. 
\par We can now relate an MBA preference $\succsim$ to its unambiguous component $\succsim^*$ according to \citet[][Definition 1, Proposition 5.(i)-(ii)]{cerre11}.  This is formalized below.
\begin{definition}\label{def:rev}\normalfont
The unambiguous preference $\succsim^*$ represented by $(u,P)$ in Definition \ref{def:unamb} is called a \textit{revealed} unambiguous preference with respect to an MBA preference $\succsim$ if there exists a $\alpha: F\rightarrow [0, 1]$ such that, for every $f,g\in F$,
$f\succsim g \text{ if and only if }V(f|P)\geq V(g|P),$
where \begin{align}\label{eq:mba}
    V(f|P)=\alpha(f)\hspace{0.04in}\underset{p\in P}{\text{min}}\hspace{0.04in}\E_{p}[u(f)]+(1-\alpha(f))\hspace{0.04in}\underset{p\in P}{\text{max}}\hspace{0.04in}\E_{p}[u(f)].
\end{align}  Then, for every $f,g\in F$ and $P\subseteq\Delta$, $f$ is said to be \textit{unambiguously preferred} to $g$, denoted $f \succsim^* g$, if and only if $V\big(\zeta f + (1-\zeta)h\big|P\big)\geq V\big(\zeta g +(1-\zeta)h\big|P\big),$ $\forall h \in F,\zeta\in(0,1]$.
\end{definition}
This definition shows that unambiguous preferences are the cornerstones of all ambiguity-sensitive preferences and have been used in the literature \citep{simsek14,robust16,pivato24}. Here, $\alpha:F\rightarrow[0,1]$ captures the degree of caution with which the act $f$ is evaluated, and it is unique if the minimal and maximal expected utilities of $f$ do not coincide. The pair $(u,P)$ is independent of $\alpha$. The most cautious rule is $\alpha(f)=1$ for all $f\in F$ (i.e., MEU), whereas the least cautious rule is $\alpha(f)=0$ for all $f\in F$. 

\subsubsection{Individuals}\label{sec:ind}
 
We follow \citet[][Section IV.B]{robust16} by assuming every individual $i$ has an MBA preference $\succsim_i$ represented by $(u_i,P_i,\alpha_i)$ and $V_i(.|P_i)$  in Definition \ref{def:rev}.  
The next example illustrates a specific structure of the set $P_i$ that will play a key role in our ensuing analysis.
\begin{ex}[\textit{Regular} beliefs]\normalfont
    Suppose each individual $i$ is endowed with  a probability distribution $p_i\in\Delta$.
Here, $p_i$ is interpreted as individual $i$'s ``reference model.'' Now, each $i$ is allowed to entertain other models constrained within a \textit{Bregman} ball centered around $p_i$:\footnote{\citet{topo18} refer to eq. (\ref{eq:breg}) as a \textit{primal} ball and provide graphical illustrations.}
\begin{align}\label{eq:breg}
    B^G_{\eta_i}(p_i)=\big\{q\in\Delta:D_G(p_i\lVert q)\leq\eta_i\big\},
\end{align}
where $D_G(p_i\lVert q)=G(p_i)-G(q)-\langle\nabla G(q),p_i-q\rangle$ is the Bregman divergence, for any function $G$ of the ``Legendre type'' such that $B^G_{\eta_i}(p_i)$ is convex.\footnote{Here, $\langle.,.\rangle$ is the inner product. $G$ is defined on any open convex subset $\Omega$ of a Euclidean space, and being of the Legendre type means that $G$ is strictly convex and differentiable, and the length of its gradient $\nabla G$ must go to infinity when approaching the boundary of $\Omega$ \citep[][Section 2]{topo18}.} The radius $\eta_i\geq0$ quantifies individual $i$'s confidence in $p_i$. 
The most popular Bregman ball is the \textit{entropy ball} defined as
\begin{align}\label{eq:const}
    \varGamma_{\eta_i}(p_i)=\big\{{q}\in \Delta: {R}({p}_i\lVert{q})\leq \eta_i \big\},
\end{align}
 where ${R}(.\lVert .)$ is the relative entropy:  
 ${R}(p\lVert{q})=\E_{p}\big[\text{log}\frac{p(s)}{q(s)}\big]$, i.e., when $G(z)=\sum_{j}\big(z_j\text{log}\hspace{0.02in}z_j-z_j\big)$ in eq. (\ref{eq:breg}).  Bregman balls originate from information/computational geometry \citep[][]{topo18}. Entropy balls are central in economics \citep{hansen01,hansen08,ball18}, econometrics \citep{tim23}, and statistics \citep[][]{watson16}. Online Appendix \hyperref[sec:extballs]{C.I} explores other families of sets.   
\end{ex}
\begin{definition}\label{def:reg}    \normalfont Individual $i$ is said to have \textit{regular} beliefs if $P_i=B^G_{\eta_i}(p_i)$, for $p_i\in\Delta$, $\eta_i>0$.\end{definition}

 \par \noindent--- \textit{Interpretation}: In the AI example, while advisor $i$ believes her reference probability $p_i$ is the best approximation of the correct model, she still considers nearby models in $B^G_{\eta_i}(p_i)$ because such models may capture features of the correct model that were missed by $p_i$. An MEU $i$ with beliefs $\varGamma_{\eta_i}(p_i)$ has a \textit{constraint preference} \`a la   \citet{hansen01}.

\subsubsection{Social Planner}\label{sec:social}
The social planner, $i=0$, is not assumed to be more informed than the individuals, but unlike them, she is concerned about \textit{misspecification}. Such concerns are captured  using the variational preference $\succsim_{0,c,Q}$ introduced in \citet[][eq. (1)]{hansenmiss22} 
with criterion
\begin{align}\label{eq:rep2}
V^{c}_0(f|Q)=\underset{p\in\Delta}{\text{min}}\Big\{\E_{p}\big[u_0(f)\big]+\underset{q\in Q}{\text{min}}\hspace{0.03in} c(p,q)\Big\}\quad \forall f\in F,
\end{align}
where $c$ is nonnegative, lower semicontinuous, convex in $p$, and $c(p, q)=0$ if and only if $p=q$. This preference embeds a hierarchy among beliefs: $Q$ contains ``structured models,'' i.e., beliefs that the social planner finds most plausible,\footnote{As \citet[][p. 2]{hansenmiss22} remark, ``These structured models are ones that are explicitly motivated or featured, such as ones with substantive motivation or scientific underpinnings.''} 
and every $p\notin Q$ is ``unstructured.'' $\succsim_{0,c,Q}$ also has an MBA representation $(u_0,P_0,\alpha^c_0)$, where $Q\subsetneq P_0$ (Online Appendix \hyperref[app:mbamiss]{D.I}). 
\par\noindent--- \textit{Interpretation}: After CAISI's director chooses an AI policy without knowing the correct model, an adversary---``Nature''---chooses a model $p\in\Delta$ so as to minimize expected social welfare. Nature incurs a ``penalty'' for choosing models $p$ that are farther from the set $Q$ because such models may be more harmful to society. A director represented by eq. (\ref{eq:rep2}) foresees this, leading her to be pessimistic about the outcome of her policy, so in compliance with the Precautionary Principle, she exercises caution in choosing the course of action.\footnote{For example, \citet{daron24} advocate for ``precautionary motives'' when regulating AI.} 
\par To highlight that $Q$ is special, set $c=\lambda D_{\phi}$,\footnote{Here, $\phi:[0,\infty)\rightarrow[0,\infty)$ is a continuous and strictly convex function, and denote by $\phi^*$ its convex Fenchel conjugate. $\phi$-divergences include popular divergences such as the relative entropy, $\chi^2$ divergence, and Hellinger divergence. More formally, $\phi$ satisfies $\phi(1)=0$ and $\text{lim}_{t\rightarrow\infty}\phi(t)/t=+\infty$. Also, the $\phi$-divergence is defined as $D_{\phi}(p\lVert q)=\E_q\big[\phi\big(\frac{p(s)}{q(s)}\big)\big]$ if $p$ is absolutely continuous with respect to $q$ and $+\infty$ otherwise. 
} where $D_{\phi}$ is a ``$\phi$-divergence''  and the parameter $\lambda\in(0,\infty]$ quantifies the social planner's concern for misspecification---larger values imply lower concern for misspecification.  Observation \ref{thm:phi} \citep[][Proposition 1]{hansenmiss22} demonstrates that $Q$ is the only set that drives the decision-making process.

\begin{obs}\label{thm:phi}
    For every $f\in F$, $Q\subseteq\Delta,$ $\lambda\in(0,\infty]$, and $\phi$,
    \begin{align*}
       V^{\lambda D_{\phi}}_{0}(f|Q)= \underset{p\in\Delta}{\text{\normalfont min}}\Big\{\E_{p}\big[u_0(f)\big]+\lambda \hspace{0.02in}\underset{q\in Q}{\text{\normalfont min}}\hspace{0.03in} D_{\phi}(p\lVert q)\Big\}=\lambda\hspace{0.02in}\underset{q\in Q}{\text{\normalfont min}}\hspace{0.05in}\underset{\psi\in \R}{\text{\normalfont sup}}\Big\{\psi-\E_{q}\Big[\phi^*\big(\psi-u_0(f)/\lambda\big)\Big]\Big\}.
    \end{align*}
\end{obs}
When $\phi(t)=(t-1)^2/2$, $D_{\phi}=\chi^2$ yields the \textit{Gini} criterion---a monotone version of the maxmin \textit{mean-variance} criterion \citep[][eq. (5)]{mmr06}. When $\phi(t)=t\hspace{0.02in}\text{log}\hspace{0.02in}t-t+1$, $D_{\phi}=R$ yields the \textit{entropic} preference $\succsim_{0,\lambda R,Q}$, which has the most tractable criterion
\begin{align}\label{eq:rep}
V^{\lambda R}_{0}(f|Q)=\underset{p\in\Delta}{\text{min}}\Big\{\E_{p}\big[u_0(f)\big]+\lambda\hspace{0.02in}\underset{q\in Q}{\text{min}}\hspace{0.02in} R(p\lVert q)\Big\}\quad \forall f\in F.
\end{align}
 There are two notable special cases of the entropic criterion: (1) When $Q$ is a singleton, the criterion in eq. (\ref{eq:rep}) reduces to \citeauthor{hansen01}'s (\citeyear{hansen01}) multiplier criterion  \citep[][eq. (2)]{strz11}. (2) When her concerns for misspecification vanish (i.e., $\lambda=\infty$), eq. (\ref{eq:rep}) becomes $V^{\infty R}_{0}(.|Q)= \text{min}_{q\in Q}\E_q[u_0(.)]$, which is the MEU criterion (see, Section \ref{sec:discussion}).
\section{Part I: Aggregation and Impossibility}\label{sec:rep1}
The first Pareto principle is introduced in Section \ref{sec:par1}, then an aggregation result is presented in Section \ref{sec:agg1}, followed by all our impossibility results in Section \ref{sec:impo}.

\subsection{Pareto Principle}\label{sec:par1}
A weak notion of agreement among individuals' tastes is required for the first aggregation scheme---individuals must agree on the strict ranking of a pair of outcomes, as defined below.  
\begin{definition}[c-minimal agreement]\normalfont\label{def:min}
  The profile $(\succsim^*_i)_{i=1}^n$ is said to satisfy c-\textit{minimal agreement} if there exist two constant acts $x,y\in X$ such that $x \succ^*_iy$, for all $i=1,\dots,n$.  
\end{definition}
 Let $\text{co}(.)$ denote the convex hull. The first Pareto principle requires the notion of  \textit{common-taste} acts \citep[][Section II.B]{robust16}: Two acts $f$ and $g$ are common-taste acts if $x \succsim^*_i y$ is equivalent to $x \succsim^*_j y$, for all $x, y \in \text{co}\big(f(S) \cup g(S)\big)$ and $i,j = 1,\dots,n$. 
That is, $f$ and $g$ are common-taste acts if all individuals' utility functions are identical up to positive affine transformation when restricted to the set of all outcomes of these two acts. Common-taste acts are \textit{consensual} in the sense that all individuals agree state by state on the ranking of their respective outcomes. Let $F_{\star}$ denote the set of all common-taste acts. 
\begin{definition}\label{def:compareto}\normalfont
 The social unambiguous preference $\succsim^*_0$ satisfies \textit{common-taste unambiguous Pareto dominance} with respect to the profile $(\succsim^*_i)_{i=1}^n$ of individual unambiguous preferences if for all $f,g\in F_{\star}$, $f \succsim^*_0 g$ whenever $f \succsim^*_i g$ for all $i = 1,\dots,n$.    
\end{definition}
 
Since we will allow the social planner and individuals to have arbitrary ambiguity-sensitive preferences,
we focus on the following extension \citep[][Section IV.B]{robust16}. 
\begin{definition}\label{def:revcompareto}\normalfont The social MBA preference  $\succsim_0$ satisfies \textit{revealed common-taste unambiguous Pareto dominance}  with respect to the profile $(\succsim_i)_{i=1}^n$ of individual MBA preferences if the social revealed unambiguous preference  $\succsim^*_0$ satisfies common-taste unambiguous Pareto dominance with respect to the profile $(\succsim^*_i)_{i=1}^n$ of individual revealed unambiguous preferences.  
\end{definition}

\subsection{Aggregation}\label{sec:agg1}
We now present the first utilitarian aggregation result \citep[][Corollary 2]{robust16}.
\begin{obs}\label{thm:agg2}
    Let $(\succsim_i)_{i=1}^n$ be a profile of arbitrary MBA preference relations represented by $\big\{(u_i,P_i,\alpha_i)\big\}_{i=1}^n$ and satisfy {\normalfont c}-minimal agreement. Then, an MBA preference $\succsim_0$ represented by $(u_0,P_0,\alpha_0)$ satisfies revealed common-taste unambiguous Pareto dominance with respect to $(\succsim_i)_{i=1}^n$ if and only if there exists a nonzero $\beta\in\R^n_+$ and a constant $\gamma\in\R$ such that\begin{align}\label{eq:socialrep2}
        u_0=\sum_{i=1}^n\beta_iu_i+\gamma \quad \text{ and }\quad P_0\subseteq\text{\normalfont co}\Bigg(\bigcup_{i=1}^nP_i\Bigg).
    \end{align}
\end{obs}
This result is the most general welfare-aggregation result in the literature because it allows everyone to have arbitrary MBA preferences---the $\alpha_i$'s and $\alpha_0$ are free parameters. 

\subsubsection{Identification of Social Preference}
A key practical limitation of Observation \ref{thm:agg2} is the following \textit{identification} problem: Observation \ref{thm:agg2} does \textit{not} uniquely identify a social preference because the social planner is allowed to choose $P_0$ as any subset of $\text{co}(\bigcup_{i=1}^nP_i)$.\footnote{\citet[][p. 1184]{util16} note: ``The social planner is granted discretion in choosing the set of weights over individual beliefs, thus bounding their relative weights or even ignoring some of them altogether.''} In fact, below are two papers that attempt to resolve this identification problem by assuming the social planner chooses $P_0=\text{co}(\bigcup_{i=1}^nP_i)$.
\begin{ex}\label{ex:neutral}\normalfont
 \citet{simsek14} consider SEU individuals $\{(u_i,p_i)\}_{i=1}^n$, an unambiguous social preference $(u_0,P_0)$ satisfying (\ref{eq:socialrep2}), and assume $P_0=\text{co}\big(\{p_1,\dots,p_n\}\big)$---they name this assumption a ``belief-neutral'' approach. They resolved the identification problem but at the cost of having an \textit{incomplete} welfare criterion. Alternatively, \citet{util16} consider SEU individuals, an MEU social preference $(u_0,P_0,1)$ satisfying (\ref{eq:socialrep2}), and assume an axiom called ``social ambiguity avoidance,'' which uniquely identifies $P_0=\text{co}\big(\{p_1,\dots,p_n\}\big)$. 
     
\end{ex}

\subsection{Impossibility Results}\label{sec:impo}

Observation \ref{thm:agg2} and, more generally, the welfare-aggregation literature do not provide  guidance on how a social planner should choose the set of social beliefs $P_0$, except that it must be a nonempty subset of $\text{\normalfont co}(\bigcup_{i=1}^nP_i)$. This lack of guidance poses a challenge for implementation. 
\par Recall that for the variational preference $\succsim_{0,c,Q}$ in eq. (\ref{eq:rep2}), structured models $Q$ are the social beliefs of interest, and Observation \ref{thm:agg2}'s (\ref{eq:socialrep2}) implies $Q\subseteq \text{\normalfont co}(\bigcup_{i=1}^nP_i)$. Our main question becomes: \textit{Which structured models} $Q$ \textit{should the social planner use?} The subsequent sections propose three different social planning desiderata that identify a specific choice of $Q$.

\subsubsection{Impossibility I:  Welfare Dominance}\label{sec:impo1}

Our main impossibility result is based on a model selection process that prioritizes maximizing social welfare across all consensual policies. We call this process Desideratum \ref{des:welfare}. 
 \begin{des}\label{des:welfare}
 \textit{Choose} $Q\subseteq\text{\normalfont co}(\bigcup_{i=1}^nP_i)$ \textit{if, for every} $Q'\subseteq\text{\normalfont co}(\bigcup_{i=1}^nP_i)$, 
\begin{align}\label{eq:modelselect3}
 V^{c}_0(f|Q')\leq V^{c}_0(f|Q) \quad \forall f\in F_{\star}.
\end{align}

    \end{des} 

   This desideratum describes a social planner who applies a \textit{no-regret} model selection rule to all relevant decision problems. Specifically, this social planner prioritizes beliefs that place the highest value on  policies that all individuals agree are better in every state.  
 \begin{proposition}[Impossibility I]\label{thm:welfare}
        Suppose $\succsim_{0,c,Q}$ satisfies Desideratum \ref{des:welfare}, $q\mapsto c(.,q)$ is convex,  and the MBA profile $(\succsim_i)_{i=1}^n$ satisfies {\normalfont c}-minimal agreement with beliefs $(P_i)_{i=1}^n$. If $\succsim_{0,c,Q}$ satisfies revealed common-taste unambiguous Pareto dominance with respect to  $(\succsim_i)_{i=1}^n$, then $Q$ in (\ref{eq:modelselect3}) must be $Q=\{p_j\}$, where $p_j\in P_j$ is unique for some individual $j\in\{1,\dots,n\}$.
    \end{proposition}

 Proposition \ref{thm:welfare} reveals a sharp tension between robustness and belief aggregation: The ``welfare-dominant'' strategy of a utilitarian social planner who prioritizes robustness to misspecification is to act as a \textit{dictator} in the following sense. (1) The welfare-dominant set $Q$ in (\ref{eq:modelselect3}) is a singleton $\{p_j\}$ and (2) $p_j$ is uniquely chosen from some individual $j$'s set of beliefs $P_j$,\footnote{Online Appendix \hyperref[app:dom]{E} illustrates such a singleton and indicates that it avoids ``double counting'' caution.} who then acts as a ``probability dictator''  in  society \citep{mongin98}. We indicate in Section \ref{sec:impo} that convexity of the penalty function $q\mapsto c(.,q)$ captures a very pervasive behavioral property of robustness called ``model hybridization aversion.'' Notice also that replacing $F_{\star}$ with $F$ in Desideratum \ref{des:welfare} would only strengthen the impossibility result. 
 
 \par \noindent--- \textit{Interpretation}: Recall that in Observation \ref{thm:phi}, only structured models mattered for the social preference $\succsim_{0,\lambda D_{\phi},Q}$. Thus, the welfare-dominant $Q$ being a singleton in Proposition \ref{thm:welfare} can be interpreted as the social planner preferring to be \textit{ambiguity neutral}.\footnote{\citet[][p. 28]{hansenmiss22} note: ``Smaller sets of structured models are, thus, more valuable. Indeed, in decision problems that feature a larger set of structured models--so, a more discordant information--the decision maker exhibits, ceteris paribus, a higher uncertainty aversion due to a larger model ambiguity.''} This interpretation can be formalized axiomatically for the entropic preference $\succsim_{0,\lambda R,Q}$. For any $x,y\in X$, let $\mathfrak{m}(x,y)$ denote their ``preference average'' \citep[][]{ghir03}. Given any $f,g\in F$, define $\mathfrak{m}(f,g)$ as any act $h\in F$ such that $h(s)$ is indifferent to $\mathfrak{m}(f(s),g(s))$ for all $s\in S$.
\begin{axiom}[Social Ambiguity Neutrality]\normalfont The entropic preference $\succsim_{0,\lambda R,Q}$  is said to satisfy \textit{social ambiguity neutrality} if, for all $f,g\in F$, $f\succsim_{0,\lambda R,Q}g\implies f\succsim_{0,\lambda R,Q}\mathfrak{m}(f,g)\succsim_{0,\lambda R,Q}g$. 
\end{axiom}
\begin{obs}\label{thm:neutral} $Q$ is a singleton if and only if $\succsim_{0,\lambda R,Q}$ satisfies social ambiguity neutrality.
\end{obs}

 \subsubsection{Impossibility II: Maximum Ambiguity}

  The first impossibility result exploited the fact that the social planner could choose any beliefs in the convex hull of individuals' beliefs. Desideratum \ref{des:max} contrasts this by considering the extreme case where the social planner must use the entire convex hull as social beliefs. 
    \begin{des}\label{des:max}
        \normalfont \textit{Choose} $Q=\text{\normalfont co}(\bigcup_{i=1}^nP_i)$.
    \end{des}
Desideratum \ref{des:max} captures a \textit{control} over social ambiguity by forcing/requiring the social planner to face more ambiguity than all individuals combined. Notably, this desideratum is consistent with \citeauthor{simsek14}'s (\citeyear{simsek14}) belief-neutral approach and \citeauthor{util16}'s (\citeyear{util16}) social ambiguity avoidance axiom (see, Example \ref{ex:neutral}).  Proposition \ref{thm:max} demonstrates that forcing the social planner to face maximum ambiguity will \textit{not} resolve the impossibility. 
    
    \begin{proposition}[Impossibility II]\label{thm:max}
        Suppose $\succsim_{0,\lambda R,Q}$  satisfies Desideratum \ref{des:max}. Then,
        \begin{align*}
  V^{\lambda R}_{0}\Bigg(f\Bigg|\text{\normalfont co}\Big(\bigcup_{i=1}^nP_i\Big)\Bigg)=  V^{\lambda R}_{0}\Bigg(f\Bigg|\bigcup_{i=1}^nP_i\Bigg) \quad \forall f\in F,\lambda>0.
        \end{align*}
    \end{proposition}
 Proposition \ref{thm:max}  shows that the entropic welfare criterion in eq. (\ref{eq:rep}) is \textit{invariant} to  convex combinations of individuals' beliefs. That is, belief aggregation has no welfare value to a social planner who seeks robustness to misspecification. Notice that the inner minimizer $q_0$ of $V^{\lambda R}_{0}\big(.\big|\bigcup_{i=1}^nP_i\big)$ in eq. (\ref{eq:rep}) will satisfy $q_0\in\bigcup_{i=1}^nP_i$, so it will necessarily be one individual's belief, and hence this individual will once again play the role of a probability dictator.

 \subsubsection{Impossibility III: Robust-Control Axiomatization}
We now abstract from model selections by proposing two behavioral axioms that are consistent with robust control but incompatible with aggregation. This section therefore complements the previous two by supplying a microfoundation for our impossibility results. Let $\succsim^{\scriptscriptstyle\wedge}_{0,c,Q}$ be the \textit{dominance} relation of $\succsim_{0,c,Q}$ \citep[][eq. (15)]{hansenmiss22},\footnote{$\succsim^{\scriptscriptstyle\wedge}_{0,c,Q}$ is a dominance relation of $\succsim_{0,c,Q}$ because, for all $f,g\in F$, $ f\succsim^{\scriptscriptstyle\wedge}_{0,c,Q}g\implies  f\succsim_{0,c,Q}g$.} where 
\begin{align*}
 f\succsim^{\scriptscriptstyle\wedge}_{0,c,Q}g \iff   V^c_0(f|\{q\})\geq V^c_0(g|\{q\}) \quad\forall q\in Q.
\end{align*}
Let $x_{f,q}$ be the consequence indifferent to act $f$ under the preference $\succsim^{\scriptscriptstyle\wedge}_{0,c,q}$, i.e., $x_{f,q}\sim^{\scriptscriptstyle\wedge}_{0,c,q} f$.
\begin{axiom}[Model hybridization aversion]\label{ax:hybrid}
 \normalfont   For all $q,q'\in\Delta$, $\zeta\in(0,1)$, and $f\in F_{\star}$,
 \begin{align*}
     \zeta x_{f,q}+(1-\zeta)x_{f,q'}\succsim^{\scriptscriptstyle\wedge}_{0,c,Q}x_{f,\zeta q+(1-\zeta)q'}.
 \end{align*}
\end{axiom}
Model hybridization aversion is the main behavioral axiom that drives our impossibility results. It captures the fact that a misspecification-averse social planner dislikes convex combinations of beliefs.  Notice that this axiom requires dominance only across consensual acts $F_{\star}$, so it is weaker than \citet[][Axiom A.9]{hansenmiss22}, which requires the same dominance to hold across all acts $F$.  This is a very pervasive property because it is satisfied when the penalty $q\mapsto c(.,q)$ is convex, which holds for all $\phi$-divergences (Observation \ref{thm:phi}). 
\par We start by strengthening Axiom \ref{ax:hybrid} with a \textit{strict} preference over mixtures of beliefs.\footnote{To build intuition for Axioms \ref{ax:hybrid}--\ref{ax:shybrid}, suppose the correct model is a Markov process \citep[e.g., as in][]{hansens22}. Then, mixing models is problematic because a convex combination of Markov processes is not a Markov process. Similarly, if the correct model is a rational expectations model, then mixing is also problematic because as \citet[][footnote 15]{hansens22} remark: ``mixtures of rational expectations models are not rational expectations models.'' Thus, a misspecification-averse social planner is justified to be averse to mixing models because the process of mixing models often destroys the probabilistic structure. 
}
\begin{axiom}\label{ax:shybrid}
 \normalfont   \textit{For all} $q\neq q'\in \Delta$, $\zeta\in(0,1)$, \textit{and nonconstant} $f\in F_{\star}$,
 \begin{align*}
     \zeta x_{f,q}+(1-\zeta)x_{f,q'}\succ^{\scriptscriptstyle\wedge}_{0,c,Q}x_{f,\zeta q+(1-\zeta)q'}.
 \end{align*}
\end{axiom}
 For every consensual state-dependent policy, $f\in F_{\star}$, Axiom \ref{ax:shybrid} requires $q\mapsto V^c_0(f|\{q\})$ to be strictly convex. This holds for many criteria such as the entropic/multiplier criteria. 
\par The next axiom is new and relies on robust control to describe the behavior of the social planner.
For $p\in\Delta$ and $f\in F$, define $x^p_f\in X$ such that $u_0(x^p_f)=\E_p[u_0(f)]$. 
\citet[][p. 15, eq. (14)]{hansenmiss22} interpret $x^p_f$ as ``a consequence that would be indifferent, so equivalent, to act $f$ if $p$ were the correct model.'' We use this to define a Bayesian planner.\footnote{Our Bayesian planner with full-support prior is consistent with the Harsanyi representation in \citet[][Proposition 2]{meyer95} and the ``Bayesian planner'' characterized in \citet[][Lemma 9]{agg21}.}
\begin{definition}\label{def:bayes}
    \normalfont A \textit{Bayesian planner with SEU welfare criterion} $(u_0,p)$ evaluates policies \textit{as if} $p:=\sum_{i=1}^n\pi_i\hspace{0.02in}p_i$ were the correct model, for $p_i\in P_i$ $\forall i$ and a full-support prior distribution $\pi=(\pi_i)_{i=1}^n$. For each policy $f\in F$, her certainty equivalent $x^p_f$ satisfies  $u_0(x^p_f)=\E_p[u_0(f)]$.
\end{definition}
In robust control, \citet[][p. 1101]{hansen10} note that behaving as a Bayesian planner would undermine the social planner's concerns for misspecification:\footnote{\citet[][p. 44]{kasa01} also adds: ``Any other objective implicitly commits you to a probability weighting across potential misspecifications, which is counter to the whole spirit of robust control.''} ``By averaging over the array of candidate models using a prior (subjective) distribution, say $\pi$, we can form a `hyper model' that we regard as correctly specified [...] In this way, specifying the family of potential models and assigning a subjective probability distribution to them removes model misspecification.'' Thus, a social planner who insists on maintaining her concerns for misspecification would not behave as a Bayesian planner. Axiom \ref{ax:acc} captures this behavior.
\begin{axiom}\label{ax:acc}
    For every Bayesian planner with SEU welfare criterion $(u_0,p)$, there exists an outcome $x\in X$ and a consensual policy $f_{\star}\in F_{\star}$ such that:
    \hspace{0.08in}$$x\succ^{\scriptscriptstyle\wedge}_{0,c,Q}x^p_{f_{\star}} \quad\text{and}\quad f_{\star}\succsim_{0,c,Q}x.$$
\end{axiom}
\par  Axiom \ref{ax:acc} indicates that there exists a decision problem $(x,f_{\star})$ such that a misspecification-averse planner would behave differently from a Bayesian planner.  To see this, recall that $x^p_f$ is the sure value of policy $f$ to a Bayesian planner who \textit{na{\"i}vely} treats $p$ as the correct model. Axiom \ref{ax:acc} reflects an interaction between two planners: Given $(x,f_{\star})$, suppose a Bayesian planner recommends ranking $x$ strictly over $x^p_{f_{\star}}$. Axiom \ref{ax:acc} states that, in some decision problems, a misspecification-averse planner would reject this recommendation. Specifically, she would rather choose a policy $f_{\star}$ that everyone agrees on because she is concerned that the recommendation $x$ may have been driven partly by the Bayesian planner's na{\"i}ve trust in $p$. Given any $Q$,  Axiom \ref{ax:acc} is satisfied by all criteria in eq. (\ref{eq:rep2}) because structured models $Q$ are exogenous,\footnote{\citet[][p. 16]{hansenmiss22} remark: ``In the traditional purely subjective axiomatizations, there is no way (actually, no language) to embed the probabilistic information that $Q$ represents in the decision maker preference.''  Axiom \ref{ax:acc} is closely related to \citet[][Axiom A.4]{hansenmiss22} (see, Axiom \ref{ax:subjective}).} and this axiom  disciplines only $Q$, without restricting the representation. 
\begin{des}\label{des:hybrid}
For a penalty $c$ and structured models $Q$, $\succsim_{0,c,Q}$ satisfies Axioms \ref{ax:shybrid} and \ref{ax:acc}.
\end{des}

\begin{proposition}[Impossibility III]\label{thm:hybrid}
 Suppose $\succsim_{0,c,Q}$ satisfies Desideratum \ref{des:hybrid}, and the MBA profile $(\succsim_i)_{i=1}^n$ satisfies {\normalfont c}-minimal agreement with beliefs $(P_i)_{i=1}^n$. If $\succsim_{0,c,Q}$ satisfies revealed common-taste unambiguous Pareto dominance with respect to  $(\succsim_i)_{i=1}^n$, then  $Q\subseteq\bigcup_{i=1}^nP_i$.  
\end{proposition}
For the multiplier criterion ($Q=\{\hat{q}\}$ in eq. (\ref{eq:rep})), this result shows that $\hat{q}\in P_j$ for some individual $j\geq1$, who is the probability dictator. Thus, Proposition \ref{thm:hybrid} shows that Desideratum \ref{des:hybrid} and aggregation are incompatible in a new way: A misspecification-averse planner who satisfies our ``robust-control'' axioms (Axioms \ref{ax:shybrid}--\ref{ax:acc}) must abandon belief aggregation. 

 \subsubsection{Summary of Impossibility Results}
  
 \par Our main impossibility result, Proposition \ref{thm:welfare}, establishes the incompatibility of robustness and aggregation in general settings---MBA individuals and robust criteria with convex penalties. Proposition \ref{thm:max} shows that it persists under a canonical ``maximum ambiguity'' restriction. Proposition \ref{thm:hybrid} derives it independently from some axioms. This raises a question: \textit{Is it ever possible to ensure that a misspecification-averse social planner aggregates individuals' beliefs?} The rest of this paper provides a positive answer using a different aggregation scheme.

\section{Part II: Aggregation, Impossibility, and Possibility}\label{sec:rep2}
This section refines the previous analysis to demonstrate that it is possible to aggregate beliefs under misspecification. Section \ref{sec:pareto2} presents a different Pareto principle, followed by an aggregation and impossibility in Section \ref{sec:util}, and  a possibility result in Section \ref{sec:pos}.
\subsection{Pareto Principle}\label{sec:pareto2}
The second aggregation scheme will require individuals' tastes to be sufficiently \textit{diverse}. Specifically, for each individual, there must exist two constant acts between which an individual is the only one to have a strict preference whereas all other individuals are indifferent. 
\begin{definition}[c-diversity]\label{def:div}\normalfont The profile $(\succsim^*_i)_{i=1}^n$ is said to satisfy c-\textit{diversity} if for all $i = 1 ,\dots, n$, there exists $x,y\in X$ such that $x\succ_i^* y$ whereas $x \sim^*_j y$ for all $j \neq i$.     
\end{definition}
C-diversity, also called ``independent prospects,'' is equivalent  to the individuals' utility functions being linearly independent (when $X$ is at least $n$-dimensional) \citep{weymark93}. Notice that this property is  restrictive because it implies c-minimal agreement (Definition \ref{def:min}), and it also does not allow individuals to have identical tastes. It is also popular in the literature \citep[e.g.,][]{weymark91,mongin98,robust16,zuber16}. 
\par Below is the analogue of revealed common-taste unambiguous Pareto dominance, which extends \citeauthor{robust16}'s (\citeyear[][Definition 2]{robust16})   \textit{unambiguous Pareto dominance}---it prescribes that if all individuals unambiguously prefer act $f$ to $g$, then so should the social planner. 

\begin{definition}\label{def:revpareto}\normalfont The social MBA preference $\succsim_0$ satisfies \textit{revealed unambiguous Pareto dominance}  with respect to the profile $(\succsim_i)_{i=1}^n$ of individual MBA preferences if the social revealed unambiguous preference $\succsim^*_0$ satisfies unambiguous Pareto dominance with respect to the profile $(\succsim^*_i)_{i=1}^n$ of individual revealed unambiguous preferences.  
\end{definition}

\subsection{Aggregation and Impossibility Result}\label{sec:util}
\par\noindent --- \textit{Aggregation}: The next result shows that combining c-diversity and revealed unambiguous Pareto dominance yields a sharp aggregation of individuals' beliefs and tastes. 
\begin{obs}\label{thm:agg}
    Let $(\succsim_i)_{i=1}^n$ be a profile of arbitrary MBA preference relations with representation $\big\{(u_i,P_i,\alpha_i)\big\}_{i=1}^n$ that satisfy {\normalfont c}-diversity. Then, an MBA preference relation $\succsim_0$ with representation $(u_0,P_0,\alpha_0)$ satisfies revealed unambiguous Pareto dominance with respect to $(\succsim_i)_{i=1}^n$ if and only if there exists a nonzero $\beta\in\R^n_+$ and a constant $\gamma\in\R$ such that
    \begin{align}\label{eq:socialrep}
        u_0=\sum_{i=1}^n\beta_iu_i+\gamma \quad \text{ and }\quad P_0\subseteq\underset{\beta_i>0}{\bigcap_{i=1,}^n}P_i.
    \end{align}
\end{obs}
Observation \ref{thm:agg} is a corollary of \citet[][Theorem 1]{robust16}. Just like Observation \ref{thm:agg2}, it does not impose any restrictions on the $\alpha_i$'s and $\alpha_0$. However, unlike the aggregation in Observation \ref{thm:agg2}, which is \textit{simultaneous}, the aggregation in Observation \ref{thm:agg} is \textit{sequential} in the sense that the social planner first aggregates individuals' utility functions, and then aggregates only the beliefs of those individuals who received nonzero utility weights.
\begin{remark}\normalfont\label{rem:agg}
  There exist several aggregation procedures in the literature. On one extreme, there are \textit{simultaneous} aggregations, which require a social planner to aggregate both beliefs and tastes at the same time \citep[e.g.,][]{gil04,util16}. On the other extreme, there are \textit{separate} aggregations, which allow a social planner to either aggregate only beliefs or only tastes \citep[e.g.,][]{pivato24}. In contrast, the aggregation in Observation \ref{thm:agg} is sequential, and hence it falls somewhere between these two extremes.  
   
\end{remark}
Since Observation \ref{thm:agg} imposes a more restrictive aggregation of individuals' beliefs than Observation \ref{thm:agg2}, the social planner will now face less ambiguity than every individual.

\begin{remark}\normalfont
     Restricting social beliefs to the intersection of individuals' beliefs is natural.  \citet{manski95}, \citet{niel18}, and \citet{hill19,hill23} refer to such an intersection as the \textit{domain of consensus}, \textit{common core of agreement}, \textit{accepted credal statements}, and \textit{corpus-level of consensus}. 
     This is perhaps a minimal requirement for a ``democratic'' decision rule.
      
\end{remark}

\par\noindent--- \textit{Impossibility result}: Similarly to the aggregation in Observation \ref{thm:agg2}, the aggregation in Observation \ref{thm:agg} is also prone to an impossibility result, which arises immediately when the individuals' sets of beliefs are pairwise disjoint. In such cases, the only social preference that satisfies revealed unambiguous Pareto dominance is dictatorial. This impossibility arises, for example, when all individuals have SEU preferences $\{(u_i,p_i)\}_{i=1}^n$ and the $p_i$'s are distinct.

\subsection{Possibility Result}\label{sec:pos}
 
Theorem \ref{thm:pos} is our main \textit{possibility} result---it shows that a utilitarian aggregation of beliefs is possible when individuals' sets of beliefs are arbitrary Bregman balls $\big\{B^G_{\eta_i}(p_i)\big\}_{i=1}^n$ (eq. (\ref{eq:breg})). Assume $G$ has a positive definite Hessian---a technical assumption needed for identification. 

\begin{theorem}\label{thm:pos}
    Let $\big\{(u_i,B^G_{\eta_i}(p_i),\alpha_i)\big\}_{i=1}^n$ be MBA preferences satisfying {\normalfont c}-diversity. If an MBA preference $(u_0,P_0,\alpha_0)$ satisfies revealed unambiguous Pareto dominance with respect to $\big\{\big(u_i,B^G_{\eta_i}(p_i),\alpha_i\big)\big\}_{i=1}^n$, then $(u_0,P_0)$ is in (\ref{eq:socialrep}) and $\bigcap_{i=1,\beta_i>0}^nB^G_{\eta_i}(p_i)\cap\text{\normalfont co}\big(\{p_1,\dots,p_n\}\big)\neq\varnothing$.
\end{theorem}
This result shows that if every individual has an arbitrary Bregman ball, then the intersection of their balls always contains at least one convex combination of their reference models. Corollary \ref{thm:pos1} provides a sharper prediction when individuals share the same radius.
   \begin{corollary}\label{thm:pos1}
In Theorem \ref{thm:pos}, there exists a unique $r^*\geq0$ such that, if $\eta_i=r^*\geq0$ for every $i$, then $P_0=\{q^{*}_0\}$ and  $q^{*}_0\in\text{\normalfont co}\big(\{p_1,\dots,p_n\}\big)$, so $(u_0,P_0,\alpha_0)$ becomes SEU $(u_0,q^{*}_0)$ for all $\alpha_0$.
    \end{corollary}
       We see that if individuals share the radius $r^*$, then $\bigcap_{i=1,\beta_i>0}^nB^G_{r^*}(p_i)=\{q^{*}_0\}$, where  $q^{*}_0=\sum_{i=1}^n\mu^{*}_i\hspace{0.01in}p_i$ for some convex weights $\{\mu^{*}_i\}_{i=1}^n$.  Thus, regardless of her ambiguity attitude, the social planner must adopt the SEU preference $(u_0,q^{*}_0)$. In information geometry, $q^{*}_0$---the unique point of intersection of Bregman balls---is known as the \textit{Chernoff point} of the reference models $\{p_1,\dots,p_n\}$ \citep[][]{chern13,small18}.  Corollary \ref{thm:pos1} complements the aggregation results  in \citet{gil04}, \citet{util16}, and \citet{billot21}, which obtain belief aggregation by assuming all individuals are SEU. In contrast, Corollary \ref{thm:pos} shows that belief aggregation is possible even when individuals have arbitrary MBA preferences as long as their sets of beliefs are Bregman balls with common radius $r^*$. 
\par\noindent \textit{--- Interpretation}: $r^*$ has two interpretations. 
(1) $r^*$ is the radius of the \textit{smallest enclosing} Bregman ball containing all the reference models \citep[][Section 4]{small18}; (2) Corollary \ref{thm:r} shows that $r^*$ is also the unique radius that maximizes the $c$-\textit{robust social value function}, defined as $V^c_{0}(.):=\text{sup}_{f\in F}\hspace{0.02in}V^c_0(f|.)$, assuming it is well-defined and nonconstant. 
\begin{obs}\label{thm:r}
  For any $\{p_1,\dots,p_n\}$,  nonzero $\beta\in\R^{n}_+$, $G$, and penalty function $c$, $$r^*=\underset{r\geq 0}{\text{\normalfont arg max}}\hspace{0.08in}V^c_{0}\Bigg(\underset{\beta_i>0}{\bigcap_{i=1,}^n}B^G_{r}(p_i)\Bigg).$$
\end{obs}

\section{Part III: Robustness vs. Aggregation}\label{sec:tension}

Thus far, it is not clear why there is a tension between robustness and aggregation. This section explores the source of this tension by carefully analyzing the model selection process. 

\subsection{A New Pareto Principle}
The aggregation schemes in Observations \ref{thm:agg2} and \ref{thm:agg} apply to every MBA preference $(u_0,P_0,\alpha_0)$, so their belief aggregations restrict the entire set of beliefs $P_0$. However, for the variational preference $\succsim_{0,c,Q}$ in eq. (\ref{eq:rep2}), we are interested only in the structured models $Q\subsetneq P_0$. This section introduces a new Pareto principle that will allow us to restrict $Q$ directly. To this end, observe that $\succsim_{0,c,Q}$ has a \textit{misspecification-neutral} component, denoted $\succsim^{\star}_{0,c,Q}$, which arises whenever the social planner has no concern for misspecification. Specifically, $\succsim^{\star}_{0,c,Q}$ is represented by the MEU criterion $\text{min}_{q\in Q}\E_q[u_0(.)]$ \citep[][Proposition 4]{hansenmiss22}, whose set of beliefs is $Q$, so we can apply a Pareto principle through this component.
\begin{definition}\label{def:revparetomiss}\normalfont The preference $\succsim_{0,c,Q}$ satisfies \textit{neutral unambiguous Pareto dominance}  with respect to the profile $(\succsim_i)_{i=1}^n$ of individual MBA preferences if the misspecification-neutral preference $\succsim^{\star}_{0,c,Q}$ satisfies revealed unambiguous Pareto dominance with respect to $(\succsim_i)_{i=1}^n$.  
\end{definition}
Since $\succsim^{\star}_{0,c,Q}$ is MEU, it is an MBA preference, so a version of Observation \ref{thm:agg} follows immediately, where $P_0$ in (\ref{eq:socialrep}) is replaced by $Q$. The next sections leverage this insight.

\subsection{Aggregation and Model Selection}\label{sec:analysis}
Let $\theta:=(\beta,\eta)\in\R^{2n}_+$ and, as in \citet[][eq. (6)]{strz11}, define the function $\phi_{\lambda}$ as follows:
      \begin{align*}
     \mathcal{Q}_{\theta}:=\underset{\beta_i>0}{\bigcap_{i=1,}^n}\varGamma_{\eta_i}(p_i) \quad\quad \text{and} \quad\quad \phi_{\lambda}(u):=\begin{cases}
         -\text{exp}(-u/\lambda) &\text{ for }\lambda\in(0,\infty)\\
         u &\text{ for }\lambda=\infty,
    \end{cases}
  \end{align*}
where $\varGamma_{\eta_i}(p_i)$ is the entropy ball in eq. (\ref{eq:const}). To ease notation, let's also define $\theta_{\lambda}:=(\theta,\lambda)\in\R^{2n+1}_+$. Paralleling Desideratum \ref{des:max} and Proposition \ref{thm:max}, Theorem \ref{thm:rep} leverages the entropic preference $\succsim_{0,\lambda R,Q}$ to analyze the tension between robustness and aggregation by focusing on the model selection process when $Q=\mathcal{Q}_{\theta}$---the entire intersection in (\ref{eq:socialrep}). Section \ref{sec:tensiondisc} discusses how this setting allows us to determine the mechanism driving the tension.
    \begin{theorem}\label{thm:rep}
   Suppose the MBA profile $(\succsim_i)_{i=1}^n$ with representation $\{(u_i,\varGamma_{\eta_i}(p_i),\alpha_i)\}_{i=1}^n$ satisfies {\normalfont c}-diversity. If $\succsim_{0,\lambda R,Q}$ satisfies neutral unambiguous Pareto dominance with respect to $(\succsim_i)_{i=1}^n$ and $Q=\mathcal{Q}_{\theta}$, then, for each $i\geq1$, there exists a unique $\mu^{\theta_{\lambda}}_i:X \rightarrow\R_+$ such that
     \begin{align}\label{eq:modelf}
q^{f,\theta_{\lambda}}_{0}(s)=\sum_{i=1}^n\mu^{\theta_{\lambda}}_i(f(s))\hspace{0.015in}p_i(s)\quad \forall s\in S
 \end{align}
is the unique solution to the inner minimization of  $V^{\lambda R}_{0}(f|\mathcal{Q}_{\theta})$, for all $f\in F$. For each $i\geq1$ and $f\in F$, $\mu^{\theta_{\lambda}}_i(f)=w_i(f,\theta_{\lambda})\hspace{0.02in}\mathds{1}_{\beta_i>0}$, where the $w_i$'s are nonnegative functions that ensure $q^{f,\theta_{\lambda}}_{0}\in \mathcal{Q}_{\theta}$. Moreover, $u_0=\sum_{i=1}^n\beta_i u_i+\gamma$, where $\beta\in\R^n_+$, $\beta\neq0$, $\gamma\in \R$, and 
\begin{align*}
V^{\lambda R}_{0}(f|\mathcal{Q}_{\theta})=\phi^{-1}_{\lambda}\Bigg(\sum_{i=1}^n\E_{p_i}\Big[\mu^{\theta_{\lambda}}_i(f)\hspace{0.02in}\phi_{\lambda}\big(u_0(f)\big)\Big]\Bigg)\quad \forall f\in F.
\end{align*}
    \end{theorem}
  We sketch Theorem \ref{thm:rep}'s proof in Appendix \hyperref[sec:sketch]{B}, where the functional form of $\mu^{\theta_{\lambda}}_i$ appears.

\subsection{Discussion of Mechanisms}\label{sec:tensiondisc}
Theorem \ref{thm:rep} highlights two key mechanisms that describe the model selection process of a social planner with misspecification concerns. (1) When restricted to use the entire intersection of entropy balls $\mathcal{Q}_{\theta}$, the social planner chooses a social belief that is a weighted average of reference models whose weights depend on policies. This means that, in her search for robustness to misspecification, she chooses a policy-indexed family of beliefs $\big\{q^{f,\theta_{\lambda}}_{0}\big\}_{f\in F}$. (2) The belief $q^{f,\theta_{\lambda}}_{0}$ also depends on $\lambda$---her concern for misspecification. Corollary \ref{thm:constant} will show shortly that the dependence on $f$ and $\lambda$ ceases to exist when $\mathcal{Q}_{\theta}$ is a singleton. 
\par These two mechanisms contrast the welfare-aggregation literature, which proposes social beliefs as convex combinations of individuals' beliefs whose weights do not depend on policies \citep[][]{gil04,simsek14,util16,robust16,agg21,billot21}. Theorem \ref{thm:rep} shows that this is no longer possible when the social planner has concerns for misspecification. Specifically, her concerns are captured by the weights $\big\{\mu^{\theta_{\lambda}}_i(f)\big\}_{i=1}^n$ in eq. (\ref{eq:modelf}), 
which measure the degree of caution with which she weighs every reference model $p_i$ depending on each policy $f\in F$. That is, each $p_i$ is weighed depending on the ``context'' in the sense that there may be some policies that involve high stakes, which may require more caution than other policies as in \citet{hill13}. In fact, this behavior is consistent with the statistics literature, where \citet[][Principle 1a]{watson16} argue that the impact of misspecification on decisions should be contextual.
\subsection{A Key Special Case}

When $\mathcal{Q}_{\theta}$ is a singleton ($\succsim_{0,\lambda R,\mathcal{Q}_{\theta}}$ is a multiplier preference), Theorem \ref{thm:rep} takes a sharper form. 
\begin{corollary}\label{thm:constant}
 In Theorem  \ref{thm:rep}, 
 if $\mathcal{Q}_{\theta}$ is a singleton, then the social belief in eq. (\ref{eq:modelf}) becomes
    \begin{align}\label{eq:model}
q^{\theta}_0(s)=\sum_{i=1}^n\mu^{\theta}_i\hspace{0.015in}p_i(s) \quad \forall s\in S,
    \end{align}
 where $\sum_{i=1}^n\mu^{\theta}_i=1$ and, for every $i=1,\dots,n$, $\mu^{\theta}_i\geq0$ is a unique constant that does not depend on either $f$ or $\lambda$. The criterion in Theorem \ref{thm:rep} becomes a multiplier welfare criterion
      \begin{align}\label{eq:repclose}
V^{\lambda R}_{0}(f|\mathcal{Q}_{\theta})=\phi^{-1}_{\lambda}\Bigg(\sum_{i=1}^n\mu^{\theta}_i\hspace{0.02in}\E_{p_i}\big[\phi_{\lambda}(u_0(f))\big]\Bigg)\quad \forall f\in F.
\end{align}
\end{corollary}

Recall that $\mathcal{Q}_{\theta}$ being a singleton captures social ambiguity neutrality (Observation \ref{thm:neutral}). Corollary \ref{thm:constant} is useful because it yields a tractable aggregation of preferences---the multiplier welfare criterion has the closed-form expression in eq. (\ref{eq:repclose}).\footnote{In single-agent decision problems, \citeauthor{giac22}'s (\citeyear{giac22}) \textit{structured average robust control} criterion is $\E_{\mu}\phi^{-1}_{\lambda}\big(\hspace{0.02in}\E_{q}\big[\phi_{\lambda}(u_0(f))\big]\big)$, for some $\mu\in\Delta(\{p_1,\dots,p_n\})$. Notice that \citeauthor{giac22}'s (\citeyear{giac22}) static criterion resembles closely our multiplier criterion in eq. (\ref{eq:repclose}) because they differ only in the order of $\phi^{-1}_{\lambda}$ and $\E_{\mu}$.} We refer hereafter to any convex combination $q^{\theta}_0\in \mathcal{Q}_{\theta}$ of individuals' reference models as a ``utilitarian social belief.'' For the rest of this paper, we will leverage the multiplier welfare criterion in Corollary \ref{thm:constant}. 

\section{Properties, Comparative Statics, and Optimal Policies}\label{sec:prop}
We analyze the utilitarian social belief and multiplier welfare criterion. Sections \ref{sec:propbel}, \ref{sec:comp}, and \ref{sec:optim} study, respectively, some properties, comparative statics, and optimal policies.
\subsection{Properties of Utilitarian Social Belief}\label{sec:propbel}
 
A utilitarian social belief $q^{\theta}_0\in \mathcal{Q}_{\theta}$ has many desirable properties. 
Proposition \ref{thm:accurate} shows that, under misspecification, $q^{\theta}_0$ is the closest model (in $\mathcal{Q}_{\theta}$) to the correct model, denoted $p^*$. 
\begin{proposition}\label{thm:accurate}
    Let $p^*\in\Delta$ be absolutely continuous with respect to $\mathcal{Q}_{\theta}$. Then, for all $s\in S$, $$\sigma p^*(s)+(1-\sigma)q^{\theta}_\sigma(s)=\underset{q\in \mathcal{Q}_{\theta}}{\text{\normalfont arg min}}\hspace{0.03in}R(p^*\lVert q),$$  where $\sigma\in[0,1]$ is a unique constant, and $q^{\theta}_{\sigma}:=\sum_{i=1}^n\mu^{\sigma,\theta}_ip_i\in\mathcal{Q}_{\theta}$ for some unique convex weights $\big\{\mu^{\sigma,\theta}_i\big\}_{i=1}^n$. When $\sigma=0$, $q^{\theta}_{\sigma}=q^{\theta}_0\in \mathcal{Q}_{\theta}$ is a utilitarian social belief.
\end{proposition}
\par Proposition \ref{thm:accurate} \textit{projects} the truth $p^*$ on the intersection of individuals' entropy balls $\mathcal{Q}_{\theta}$, i.e., it solves the inner minimization in eq. (\ref{eq:rep}) when $p=p^*$ and $Q=\mathcal{Q}_{\theta}$. 
The projection of $p^*$ in $\mathcal{Q}_{\theta}$ is the convex combination $\sigma p^*+(1-\sigma)q^{\theta}_\sigma$, where  $\sigma\in[0,1]$ is the probability that $p^*$ is an element of $\mathcal{Q}_{\theta}$. Hence, with probability $1-\sigma$, $p^*$ is not in $\mathcal{Q}_{\theta}$, in which case Proposition \ref{thm:accurate} shows that its \textit{best} approximation within $\mathcal{Q}_{\theta}$ is a convex combination $q^{\theta}_0=\sum_{i=1}^n\mu^{\theta}_ip_i$. This result formalizes \citeauthor[][]{hansen14}'s (\citeyear{hansen14}) taxonomy of uncertainty in our context: 
On one extreme, \textit{risk} trivially implies $\sigma=1$; \textit{ambiguity} implies $\sigma\in(0,1]$, i.e., positive probability that $p^*\in\mathcal{Q}_{\theta}$; on the other extreme, \textit{misspecification} implies $\sigma=0$, i.e., $p^*\notin\mathcal{Q}_{\theta}$. Thus, a social planner who is concerned about misspecification believes $\sigma=0$, so $q^{\theta}_0\in \mathcal{Q}_{\theta}$ is the unique structured model that hedges against such concerns uniformly across all policies.
\begin{corollary}\label{thm:bound}

 When $\sigma=0$ in Proposition \ref{thm:accurate}, there exists a constant $\kappa^*_q\geq0$ such that $R(p^*\lVert q)\geq R(p^*\lVert q^{\theta}_{0})+\kappa^*_q$, for all $q\in\mathcal{Q}_{\theta}$, with equality if and only if $q=q^{\theta}_0$.
\end{corollary}
Although our social planner believes $\sigma=0$, i.e., the correct model $p^*$ is not contained in $\mathcal{Q}_{\theta}$, she is confident that $p^*\in W_R\big(q^{\theta}_0,\mathcal{Q}_{\theta}\big)=\Big\{p\in\Delta:R(p\lVert q^{\theta}_0)=\underset{q\in \mathcal{Q}_{\theta}}{\text{\normalfont min}}\hspace{0.02in}R(p\lVert q)\Big\}$. Here, $W_R\big(q^{\theta}_0,\mathcal{Q}_{\theta}\big)$ is the partial identification set capturing her belief that $q^{\theta}_0$ best approximates $p^*$, which satisfies $W_R(q^{\theta}_0,\mathcal{Q}_{\theta})\cap \mathcal{Q}_{\theta}=\{q^{\theta}_0\}$ \citep[][Lemma 2.(ii)]{hansenmiss22}. 
\par When $\sigma=0$, it may be useful to have a way to assess the \textit{goodness of fit} of $q^{\theta}_0$, i.e., when the truth is not in $\mathcal{Q}_{\theta}$, is there a way to tell whether $q^{\theta}_0$ is a good fit? We answer this in two steps; we first show that ${q}^{\theta}_0$ is the ``closest'' belief to all the individuals' reference models. 
  \begin{proposition}\label{thm:mindist}
     The unique solution to $\underset{q\in\Delta}{\text{\normalfont min}}\sum_{i=1}^n\mu^{\theta}_iR(p_i\lVert q)$ is  $q^{\theta}_0=\sum_{i=1}^n\mu^{\theta}_ip_i$ in eq. (\ref{eq:model}).
 \end{proposition}
Analogues of the above objective function are used to measure goodness of fit  in \citet{asset21} and \citet[][eq. (11)]{hansenmiss22}. When the social planner trusts all $n$ individuals equally, i.e., $\mu^{\theta}_i=1/n$, then ${q}^{\theta}_0=\frac{1}{n}\sum_{i=1}^n{p}_i$ is in the ``center'' of $\mathcal{Q}_{\theta}$, and hence \citet{stone61} refers to this belief as a democratic ``opinion pool.'' Now, to assess goodness of fit, let $H(q)$ denote the Shannon entropy of any model $q\in\Delta$.
   \begin{corollary}\label{thm:mindist2}
    $\sum_{i=1}^n\mu^{\theta}_iR(p_i\lVert q^{\theta}_0)=0$ in Proposition \ref{thm:mindist} if and only if $H(q^{\theta}_0)=\sum_{i=1}^n\mu^{\theta}_iH(p_i)$.
 \end{corollary}
Since Shannon entropy is concave, $H(q^{\theta}_0)\geq\sum_{i=1}^n\mu^{\theta}_iH(p_i)$ by Jensen's inequality, so their absolute difference is a goodness-of-fit measure, where lower values indicate better fit. This raises another question: When is $q^{\theta}_0$ most ``informative'' of the truth? This happens when the $p_i$'s are linearly independent, which  means that the $p_i$'s would ``span'' a reasonable range of models in $\Delta$. This idea has appeared in the literature \citep[][]{mongin98,agg21}.

\subsection{Comparative Statics}\label{sec:comp}
\par\noindent--- \textit{Comparative statics for $\eta$}: The next result studies how the weights $\{\mu^{\theta}_i\}_{i=1}^n$ in the utilitarian belief $q^{\theta}_0=\sum_{i=1}^n\mu^{\theta}_ip_i$ change when an individual's radius $\eta_i$ in eq. (\ref{eq:const}) increases. 
\begin{proposition}\label{thm:mono}
 Suppose each $\mu_i\in\R$ depends on $\bar{\eta}=(\bar{\eta}_1,\dots,\bar{\eta}_n)\in \R^n_+$ in such a way that there exists an arbitrary function $\bar{q}=\sum_{i=1}^n\mu_i\hspace{0.015in}p_i$ that satisfies all individuals' equality constraints in eq. (\ref{eq:const}), where $\sum_{i=1}^n\mu_i=1$. Moreover, let $|\eta_i-\bar{\eta}_i|<\bar{\delta}_i$ for any constants $(\bar{\delta}_1,\dots,\bar{\delta}_n)\in\R^n_+$ and $\eta=(\eta_1,\dots,\eta_n)\in \R^n_+$. Then, it must be that $\bar{q}\in\mathcal{Q}_{\theta}$, where, for all i,
 \begin{align*}
     \frac{\partial \mu_i}{\partial \eta_i}&\leq0\quad \text{and}\quad
     \sum_{j\neq i}\frac{\partial}{\partial \eta_i}\mu_j\geq0.
 \end{align*}
\end{proposition}
Proposition \ref{thm:mono} finds a key relationship between each weight $\mu^{\theta}_i$ and individual $i$'s radius/confidence $\eta_i$: $\mu^{\theta}_i$ decreases in $\eta_i$. Intuitively,  a looser constraint in eq. (\ref{eq:const}) implies that $i$ has low confidence in her own reference model $p_i$, which signals that $i$ is less knowledgeable of the correct model, so the social planner responds by lowering her trust in $p_i$. Notably, this relationship captures the \textit{tradeoff} between robustness and aggregation in social planning. 
 
\par\noindent--- \textit{Comparative statics for $\beta$ and $\lambda$}: Each weight $\mu^{\theta}_i$ in the utilitarian social belief $q^{\theta}_0$ in eq. (\ref{eq:model}) depends on the utility weight $\beta_i$ in a straightforward way. As Theorem \ref{thm:rep} shows, $\mu^{\theta}_i=0$ whenever $\beta_i=0$, which is consistent with the aggregation scheme in (\ref{eq:socialrep}).
\par Let's now explore how the criterion $V^{\lambda R}_{0}(.|\mathcal{Q}_{\theta})$ in eq. (\ref{eq:repclose}) changes when $\lambda$ changes. Notice that, for all $f\in F$ and  $(\theta,\lambda)\in\R^{2n+1}_+$, $V^{\lambda R}_{0}(f|\mathcal{Q}_{\theta})=\phi^{-1}_{\lambda}\big(\sum_{i=1}^n\mu^{\theta}_i\E_{p_i}[\phi_{\lambda}(u_0(f))]\big)$ is ordinally equivalent to   $\sum_{i=1}^n\mu^{\theta}_i\E_{p_i}[\phi_{\lambda}(u_0(f))]$. Thus, more concerns for misspecification (i.e., smaller $\lambda$) correspond to more aversion to social risk (i.e., more concave $\phi_{\lambda}$).
\begin{ex}\label{ex:dis}\normalfont
         Set $p_i(s)=\mathds{1}_{s\geq s_i}$ in eq. (\ref{eq:model}) with $s_i\in S$, for all $i$ and $s$. Then, eq. (\ref{eq:repclose}) is $\phi^{-1}_{\lambda}\big(\sum_{i=1}^n\mu^{\theta}_i\hspace{0.01in}\phi_{\lambda}(u_0(f_i))\big)$, which resembles the smooth ambiguity criterion in \citet[][Section 4.2]{agg21},\footnote{In \citeauthor{smooth05}'s (\citeyear{smooth05}) terminology, $(\mu^{\theta}_1,\dots,\mu^{\theta}_n)\in\Delta\big(\{p_1,\dots,p_n\}\big)$ in Corollary \ref{thm:constant} can be interpreted as the social planner's ``second order'' probability over the ``first order'' probabilities $\{p_1,\dots,p_n\}$.} 
         where $f_i:=f(s_i)$. 
         Here, $\phi_{\lambda}$ captures the social planner's attitude toward individuals' disagreements about the $s_i$'s, so  $\lambda$ measures her degree of ``disagreement aversion.'' 
          
     \end{ex}

\subsection{Optimal Policies}\label{sec:optim}
The social planner ultimately wishes to choose an act from a set ${F}_0\subseteq F$ of \textit{optimal} acts---those acts that yield the highest value of the multiplier criterion in eq. (\ref{eq:repclose}). Formally,
\begin{align}\label{eq:opt}
    {F}_0:=\underset{f\in F}{\text{arg sup}}\hspace{0.07in}\phi^{-1}_{\lambda}\Bigg(\sum_{i=1}^n\mu^{\theta}_i\hspace{0.02in}\E_{p_i}\big[\phi_{\lambda}(u_0(f))\big]\Bigg).
\end{align} 
We say $f$ \textit{strongly dominates} $g$, denoted $f\vartriangleright^{\scriptscriptstyle\wedge}_{0,c,Q} g$, if for all acts $h,w\in F$, $(1-\zeta)f+\zeta h\succ^{\scriptscriptstyle\wedge}_{0,c,Q}(1-\zeta)g+\zeta w$, for some $\zeta\in[0,1]$. Hence, $f\vartriangleright^{\scriptscriptstyle\wedge}_{0,c,Q}g$ implies $f\succ^{\scriptscriptstyle\wedge}_{0,c,Q}g$. Strong dominance strengthens strict dominance in the sense that the social planner can convince others ``beyond reasonable doubt.'' We then say an act $f\in F$ is (weakly) \textit{admissible} if there is no act $g\in F$ that (strongly) strictly dominates $f$. The next result provides a description of ${F}_0$ in eq. (\ref{eq:opt}).

\begin{proposition}[Optimal Policy]\label{thm:optim}
 Let $F$ be a compact and convex subset of a reflexive Banach space and $u_i$ be strictly concave and continuous for all $i=1,\dots,n$. Then, ${F}_0=\{f_0\}$ and the unique optimal act $f_0$ is admissible. If, in addition, $u_i$ is differentiable for any $i$ with $\beta_i>0$, and for all such $i$, $p_i(s)>0$ $\forall s\in S$, then $f_0$ solves the equation $\sum_{i=1}^n\beta_i\nabla u_i(f_0)=0$.
\end{proposition}

Proposition \ref{thm:optim} characterizes the uniqueness of optimal policies under natural conditions such as individual-level risk aversion. The conditions on  $F$ allow for a broad class of functions, e.g., simple (or finite-valued) functions in $L^p$ space, for $p\in(1,\infty)$, which are used in Anscombe-Aumann settings. 
If ${F}_0$ is not a singleton, then choices are restricted to  weakly admissible acts $\big\{f\in F:\nexists g\in F, g\vartriangleright^{\scriptscriptstyle\wedge}_{0,c,Q}f\big\}$ 
\citep[][Proposition 8.(i)]{hansenmiss22}. 
\section{Applications}\label{sec:appli}
 Sections \ref{sec:div} and \ref{sec:macro} apply our multiplier welfare criterion, respectively, in treatment choice problems and dynamic macroeconomic models. Online Appendix \hyperref[sec:app2]{B} considers other applications in asset pricing, Ellsberg experiments, and empirical measurements of parameters.

\subsection{Treatment Choice}\label{sec:div}

This application shows that our multiplier welfare criterion provides a justification for treatment diversification, and its tractability reveals new comparative statics. The motivation is that public authorities often have to decide which type of treatment to administer to heterogeneous members of a population. \citet{manski09} points out the technical challenges of treatment choice under ambiguity because a SEU or MEU social planner typically chooses to assign all the population to only one treatment. Hence, to justify diversification, \citet{manski09} considers Savage's minimax regret criterion as the welfare criterion. However, \citet{agg21} notes that it is difficult to obtain closed-form solutions when using this criterion. This application also shows that our welfare criterion yields simple closed-form solutions. 

\par Suppose there are only two states of treatment responses $S=\{s_1,s_2\}$ in a population. Let $n=2$ so that each treatment response corresponds to an expert’s opinion about the effectiveness of the treatment. Thus, for all $s\in S$, $p_1(s)=\mathds{1}_{s\geq s_1}$ and $p_2(s)=\mathds{1}_{s\geq s_2}$, i.e., each expert $i$'s reference model is that the state of the world is $s_i\in S$ (Example \ref{ex:dis}). Treatment $a$'s response is known, whereas $b$ is a newly proposed treatment with uncertain effectiveness. The social welfare of each treatment in each state $s_i\in S$ are given in Table \ref{tab:joint}.
\begin{table}[hbt!]
\centering
\caption{Social Welfare}
\begin{tabular}{ll|lll}
\hline\hline
                            &                        & \multicolumn{3}{c}{states} \\
                            &                        & $s_1$     &     & $s_2$    \\\hline
\multirow{2}{*}{treatments} & \multicolumn{1}{c|}{$a$} &       2    &     &   2       \\
                            & \multicolumn{1}{c|}{$b$} &   1        &     & 4 \\\hline\hline 
\end{tabular}\label{tab:joint}
\end{table}
This table indicates that the experts disagree on the effectiveness of treatment $b$; expert 1 is pessimistic whereas expert 2 is optimistic. The social planner is tasked to aggregate these conflicting opinions and decide whether or not to diversify the treatment in the population.

Let $u_0=\beta u_1+(1-\beta)u_2$, where $\beta:=\beta_1\in[0,1]$ and set $\gamma=0$ in (\ref{eq:socialrep}). In this application, we write $u_1(s_1)$ and $u_2(s_1)$ to denote the social welfare generated by implementing, respectively, treatment $b$ and $a$ in state $s_1$. 
Hence, $u^1_0(\beta):=u_0(s_1)=\beta +2(1-\beta)=2-\beta$, since $u_1(s_1)=1$ and $u_2(s_1)=2$ in Table \ref{tab:joint}, and similarly $u^2_0(\beta):=u_0(s_2)=4\beta +2(1-\beta)=2+2\beta$. Thus, the key parameter in this analysis is $\beta$, which measures the treatment allocation, so we may write $u^i_0(\beta)$ to denote the expected social welfare for any treatment allocation $\beta\in[0,1]$ in state $i=1,2$. Our multiplier welfare criterion (eq. (\ref{eq:repclose})) can be written in closed-form as 
\begin{align*}
V^{\lambda R}_{0}(\beta|\mathcal{Q}_{\theta})=\phi^{-1}_{\lambda}\Bigg(\sum_{i=1}^{2}\mu^{\theta}_i\hspace{0.015in}\phi_{\lambda}\big(u^i_0(\beta)\big)\Bigg)=\phi^{-1}_{\lambda}\Big(\mu\phi_{\lambda}(2-\beta)+(1-\mu)\phi_{\lambda}(2+2\beta)\Big),
\end{align*}
where we define $\mu:=\mu^{\theta}_1\in(0,1)$.  
Hence, $\mu$ denotes the social planner's belief that the true treatment response is $s_1$, i.e., it measures how much she trusts expert 1's opinion. 
\par When $\lambda=\infty$, then $\phi_{\lambda}(u_0)=u_0$ (by definition of $\phi_{\lambda}$), so the criterion above reduces to that of a MEU social planner (Section \ref{sec:discussion}), which is linear in $\beta$ and hence the optimal would be to set $\beta\in\{0,1\}$. When instead $\lambda<\infty$,  our criterion is much more nuanced and may favor diversification. To see this, the first-order condition of $V^{\lambda R}_{0}(\beta)$ with respect to $\beta$ is
\begin{align*}
    -\mu e^{-(2-\beta)/\lambda}+2(1-\mu)e^{-(2+2\beta)/\lambda}=0,
\end{align*}
whose unique solution is $\hat{\beta}(\lambda,\mu)=\lambda\frac{1}{3}\text{log}\frac{1-\mu}{2\mu}$, where we require $0<\lambda< 3(\text{log}\frac{1-\mu}{2\mu})^{-1}$ and $\mu<1/3$ so that $\hat{\beta}(\lambda,\mu)\in(0,1)$. In Example \ref{ex:dis}, we show that lower values of $\lambda$ in such situations capture the social planner's aversion to the experts' disagreement on the state of the world. These comparative statics therefore show that when she is more averse to disagreements (i.e., lower $\lambda$), she is less willing to diversify the treatment (because $\hat{\beta}(\lambda,\mu)$ is increasing in $\lambda$). In contrast, $\hat{\beta}(\lambda,\mu)$ is decreasing in $\mu$, which is perhaps intuitive. As $\mu$ increases, the social planner believes it is more likely that the state of the world is $s_1$ in which case treatment $b$ is ineffective, so she is less willing to diversify treatment then.
\par The comparative statics in $\mu$ above are new, whereas those in $\lambda$ are consistent with \citet[][]{agg21}, who studies a social planner represented by the smooth ambiguity criterion 
    $V_0^{\phi}(f)=\sum_{i=1}^n\mu_i\hspace{0.015in}\phi\big(\E_{p_i}[u_0(f)]\big),$
for a strictly increasing and concave function $\phi$ \citep{smooth05}.  
As we illustrated earlier in Example \ref{ex:dis}, this criterion is ordinally equivalent to our welfare criterion in eq. (\ref{eq:repclose}) when $\phi=\phi_\lambda$, $\mu_i=\mu^{\theta}_i$, and $p_i(s)=\mathds{1}_{s\geq s_i}$, for all $s\in S$.\footnote{Specifically, in  \citet[][Section 4.1]{agg21}, $\phi(x)=x^{1-a}/(1-a)$, for $a\in(0,1)$, $n=2$, and $\mu=1/2$ to obtain the first-order condition $-(2-\beta)^{-a}/2+(2+2\beta)^{-a}=0$, whose unique solution is $\beta^*(a)=2\big(\frac{2^{1/a}-1}{2+2^{1/a}}\big)$. Here, $a$ plays the role of $1/\lambda$, so our comparative statics agree because $\beta^*(a)$ decreases in $a$.}

\subsection{Dynamic Macroeconomic Model}\label{sec:macro}
We revisit  \citeauthor{ball18}'s (\citeyear{ball18}) two-period macro model to show that our misspecification-averse social planner behaves identically to their \textit{representative} agent. They argue that  macro announcements, e.g., the release of the employment report and the Federal Open Market Committee statements, \textit{resolve} uncertainty about the future course of the macro economy. 

\par In \citet[][Section 3.2]{ball18}, there is a representative-agent economy with two periods, 0 and 1. Period 0 has no uncertainty and the aggregate consumption, $C_0$, is a known constant. The aggregate consumption in period 1, $C_1$, is a random variable that depends on a state $s\in S$, with realization $C_1(s)$, and let $Y(s)$ denote the realization of asset payoff. Each state occurs with positive probability, and the announcements fully reveal the states.
\par  The timeline is as follows. Period 0 is divided into two subperiods. In period $0^-$, before any information about $C_1$ is revealed, the \textit{pre}-announcement market opens and asset prices at this point are called pre-announcement prices and are denoted $P^-$, so $P^-$ cannot depend on the realization of $C_1$, which is still unknown at this point. In period $0^+$, the agent receives an announcement $s$ that carries information about $C_1$. Subsequently after this announcement, the \textit{post}-announcement asset market opens. The post-announcement asset prices depend on $s$ and are denoted $P^+$.  The announcement return of an asset, $R_A(s)$, can now be defined as the return of a strategy that buys the asset before the pre-scheduled announcement and sells immediately afterwards (assuming zero dividend at $0^+$), i.e., $R_A(s)=\frac{P^+(s)}{P^-}$. Then, an asset is said to require a positive announcement \textit{premium} whenever $\E_q[R_A(s)]>1$, where $q\in\Delta$ is the agent's reference model  under which the equity premium is evaluated.
\par  Since there is no uncertainty after time $0^+$, \citet[][Section 3.2]{ball18} assume the agent ranks consumption streams according to a time-separable and differentiable utility function $u$, so the continuation utility conditional upon announcement of $s$ is $u(C_0)+\psi u(C_1(s))$, where $\psi$ is the discount rate. The agent has a constraint preference: 
\begin{align}\label{eq:repconst}
    \underset{p:R(p\lVert q)\leq r}{\text{min}}\E_p\Big[u(C_0)+\psi u(C_1(s))\Big],
\end{align}
\citep[][eq. (5)]{ball18}, where our notation differs from theirs since we make explicit the reference model $q$. The pre-announcement price of an asset with payoff $Y(s)$ becomes
\begin{align*}
    P^-=\E_q\Bigg[\frac{\dot{p}^{C_1,\ell}(s)}{q(s)}\frac{\psi u'(C_1(s))}{u'(C_0)}Y(s)\Bigg],
\end{align*}
\citep[][eq. (6)]{ball18}, $\frac{\psi u'(C_1(s))}{u'(C_0)}$ is a ratio of marginal utilities and, for all $s\in S$,   
\begin{align}\label{eq:premium}
    \dot{p}^{C_1,\ell}(s):=\frac{e^{-u(C_1(s))/\ell}}{\E_q\big[e^{-u(C_1(s))/\ell}\big]}\hspace{0.02in}q(s)
\end{align}
is the unique minimizer of the criterion in eq. (\ref{eq:repconst}), and the constant $\ell$ is uniquely identified by the entropy constraint $R\big(\dot{p}^{C_1,\ell}\big\lVert q\big)=r$ \citep[][eq. (7)]{ball18}.
\par The agent's preference in period $0^+$ is $u(C_0)+\psi u(C_1)$. The post-announcement price is $P^+(s)=\frac{\psi u'(C_1(s))}{u'(C_0)}Y(s)$, whereas the pre-announcement price is $P^-=\E_q\big[\frac{\dot{p}^{C_1,\ell}(s)}{q(s)}P^+(s)\big]$. Thus,  $\dot{p}^{\ell}$ is referred to as an adjusted stochastic discount factor (A-SDF). Importantly, $\dot{p}^{C_1,\ell}$ is decreasing in $u(C_1)$, and under a ``co-monotonicity'' assumption, \citet[][Claim 1]{ball18} shows that the announcement premium is nonnegative, i.e., $P^-\leq \E_q[P^+(s)]$.\footnote{Under SEU, pre- and post-announcement prices are, respectively, $P^-=\E_q\big[\frac{\psi u'(C_1(s))}{u'(C_0)}Y(s)\big]$ and $P^+=\frac{\psi u'(C_1(s))}{u'(C_0)}Y(s)$ \citep[][eqs. (2)-(3)]{ball18}, so no announcement premium: $\E_q[R_A(s)]=\frac{\E_q[P^+]}{P^-}=1$.} 
\par As noted before, there are several limitations of representative-agent macro models \citep{kirman92}. We address the issue of whether a representative agent can act consistently with respect to individuals in society. To this end, let's extend the above environment to an economy consisting of $n>1$ heterogeneous agents along with a social planner. Each agent $i$ has an MBA preference with utility function $u_i(C_0)+\psi u_i(C_1(s))$ and beliefs $P_i=\varGamma_{\eta_i}(p_i)$. 
\par The social planner has to aggregate agents' preferences over consumption plans $C_0$ and $C_1$, and she uses the entropic welfare criterion in eq. (\ref{eq:rep})   \citep[][Supplementary Material, Section S.2]{ball18}. Assume c-diversity, the $\varGamma_{\eta_i}(p_i)$'s have a unique intersection $\mathcal{Q}_{\theta}=\{q^{\theta}_0\}$, and $\succsim_{0,\lambda R}$ satisfies neutral unambiguous Pareto dominance with respect to agents' preferences $(\succsim_i)_{i=1}^n$. Then,  Theorem \ref{thm:rep} shows that the social utility function is $u_0(C_0)+\psi u_0(C_1(s))$, where $u_0=\sum_{i=1}^n\beta_iu_i$ (setting $\gamma=0$), and $q^{\theta}_0=\sum_{i=1}^n\mu^{\theta}_ip_i$.  The A-SDF becomes
\begin{align*}
   \dot{p}^{C_1,\theta_{\lambda}}_0(s):=\frac{e^{-\psi u_0(C_1(s))/\lambda}}{\E_{ q^{\theta}_0}\big[e^{-\psi u_0(C_1(s))/\lambda}\big]}\hspace{0.02in} q^{\theta}_0(s)= \underset{p\in \Delta}{\text{arg min}}\hspace{0.02in}\Big\{\E_p\big[u_0(C_0)+\psi u_0(C_1(s))\big]+\lambda R\big(p\big\lVert q^{\theta}_0\big)\Big\},
\end{align*}
for all $s\in S$. This indicates that our social planner behaves just like the representative agent because their A-SDFs, $\dot{p}^{C_1,\theta_{\lambda}}_0$ and $\dot{p}^{C_1,\ell}$ (eq. (\ref{eq:premium})), are identical after change of variables: $q=q^{\theta}_0$ and $\ell=\lambda/\psi$. Notably,  $\dot{p}^{C_1,\theta_{\lambda}}_0$ is decreasing in period 1's aggregate utility $u_0(C_1)$, so \citet[][Claim 1]{ball18} holds in our framework---the announcement premium is negative. Thus, our utilitarian social planner can play the role of the representative agent in \citet[][]{ball18}. This application has therefore addressed \citeauthor{kirman92}'s (\citeyear{kirman92}) concern by showing how our framework can enrich dynamic macro models of aggregate behavior. 

\section{Discussion and Conclusion}\label{sec:conc}
\subsection{Connections to the Literature}\label{sec:discussion}
Our framework is closely related to the existing literature on social choice under uncertainty. When ambiguity-sensitive preferences are considered, this literature typically assumes an MEU social planner \citep{gil11,util16,qu17}.
However, MEU implies that a social planner ignores (or is not aware of) misspecification. This follows from 
\begin{align}\label{eq:MEU}
\underset{\lambda\uparrow\infty}{\text{ lim }}V^{\lambda R}_{0}(f|Q)=\underset{\lambda\uparrow\infty}{\text{ lim }}\underset{p\in\Delta}{\text{ min }}\Big\{\E_p\big[u_0(f)]+\lambda\hspace{0.02in}\underset{q\in Q}{\text{min}}\hspace{0.02in} R(p\lVert q)\Big\}=\underset{q\in Q}{\text{min}}\hspace{0.02in}\E_q\big[u_0(f)] \quad \forall f\in F,
\end{align}
which is the MEU criterion \citep[see,][Proposition 2]{hansenmiss22}. 
\par In what follows, we discuss three existing frameworks that are special cases of our framework.\footnote{Online Appendix \hyperref[sec:axiom]{A} relates our framework to \citet{miss24} to microfound our welfare criteria.} \citet{util16} consider two Pareto principles: (1) \textit{Lottery Pareto} states that for two acts that involve events whose probabilities are agreed upon by all, consensus on the ranking of these acts will have to be respected by the social planner. (2) \textit{Likelihood Pareto} compels the social planner to accept any unanimous preference concerning acts that are contingent upon the same pair of identically ranked outcomes. Their Theorem 1 shows that satisfying these two principles is equivalent to utilitarianism. 
Given Observation \ref{thm:agg} and Corollary \ref{thm:constant}, both Lottery and Likelihood Pareto can be viewed as the limit of revealed unambiguous Pareto dominance (Definition \ref{def:revpareto}) when (1) the social planner's fear of misspecification vanishes ($\lambda\rightarrow\infty$, eq. (\ref{eq:MEU})) and (2) each individual's set of beliefs shrinks to a singleton, i.e., $P_i=\varGamma_{\eta_i}(p_i)\rightarrow\{p_i\}$ ($\eta_i\rightarrow0$, eq. (\ref{eq:const})), for $i=1,\dots,n$. Similarly, \citet[][Theorem 2]{util16} is a special case of Proposition \ref{thm:welfare} when $\alpha_0(f)=1$ for all $f\in F$ (i.e., MEU social planner), $P_i=\{p_i\}$ for all $i$ (i.e., SEU individuals), and $P_0=\text{co}\big(\{p_1,\dots,p_n\}\big)$. 
\par \citet{qu17} proposes an aggregation of preferences with an MEU social planner. He considers two intuitive Pareto principles. (1) Constant Pareto Condition: if $x \succsim_i y$ for all $i=1,\dots,n$, then
$x \succsim_0 y$, for all $x, y \in X$. (2) Restricted Pareto Condition: if $f\succsim_ig$ for all $i=1,\dots,n$ and
$f \succ_j g$ for some $j$, then $f \succsim_0 g$, for all $f, g \in F$. Consider SEU individuals satisfying c-minimal agreement, and their beliefs are singletons $P_i=\{p_i\}$. 
Observation \ref{thm:agg2} extends \citeauthor{qu17}'s (\citeyear{qu17}) main result (Theorem 3) from MEU to arbitrary MBA preferences.
\par \citet{billot21} consider a Pareto principle to address spurious unanimity when the social planner and all individuals have SEU preferences $\{(u_i,p_i)\}_{i=0}^n$. They propose a ``belief-proof'' Pareto condition, which states that: for all $f,g\in F$, if, for every  $i, j=1,\dots,n$,  $\E_{p_j}[u_i(f)]\geq\E_{p_j}[u_i(g)]$, then $\E_{p_0}[u_0(f)]\geq\E_{p_0}[u_0(g)]$. \citet[][Theorem 1]{billot21} shows that satisfying belief-proof Pareto condition is equivalent to the social planner preference satisfying $u_0\in\text{co}\big(\{u_1,\dots,u_n\}\big)$ and $p_0\in\text{co}\big(\{p_1,\dots,p_n\}\big)$. Notice that this representation is a special case of Observation \ref{thm:agg2} when each individual's set of beliefs is a singleton $P_i=\{p_i\}$, $\sum_{i=1}^n\beta_i=1$, $\gamma=0$, and $P_0=\{p_0\}$ in (\ref{eq:socialrep2}), and $\succsim_0$ also satisfies c-minimal agreement.

\subsection{Conclusion}\label{sec:conc2}
This paper has revealed novel insights regarding welfare aggregation when a social planner is concerned about misspecification. We find an impossibility result under several desiderata: a misspecification-averse social planner behaves like a dictator by choosing one individual's belief as the social belief. This impossibility can be resolved, but there is no ``free lunch'' in the sense that it requires imposing restrictive conditions on individuals' beliefs and tastes such as existence of common ``regular'' beliefs and diverse tastes. It also requires a ``sequential'' rather than a ``simultaneous'' aggregation scheme. The tension between robustness and aggregation exists because the latter yields social beliefs that are sensitive to policy outcomes. Thus, we have shown that there exists a tradeoff between robustness and aggregation. We have also illustrated the tractability of our welfare criteria in treatment choice and dynamic macro.  
  \par Some extensions are in our appendices. Online Appendix \hyperref[sec:axiom]{A} provides a microfoundation for our welfare criteria using behavioral axioms. 
  Online Appendix \hyperref[sec:app2]{B} explores three applications: asset pricing, Ellsberg experiments, and empirical measurements of parameters in our welfare criteria from choice data. Online Appendix \hyperref[sec:extballs]{C.I} extends our analysis to settings where sets of beliefs are different from Bregman balls, which yields popular non-utilitarian aggregations from econometrics and finance. Online Appendix \hyperref[sec:extpar]{C.II} allows the social planner to have her own subjective beliefs, which delivers popular aggregations from statistics.

\phantomsection\label{app:compact}
\section*{Appendix A: Arbitrary State Space}
We have assumed throughout that the state space $S$ is finite. This assumption is adopted because \citet{robust16} require finiteness to establish their aggregation results (Observations \ref{thm:agg2} and \ref{thm:agg}). Since most of our results are about model selection, we mostly deal with optimization (rather than aggregation) problems. Our optimization proofs hold even when $S$ is an arbitrary subset of a Euclidean space. Thus, we will prove most of our results without restrictions on $S$. 
To this end, let $\Delta:=\Delta(S)$ be the space of probability measures over the Borel subsets of $S$. Also, let $\Delta_d$ be the corresponding set of probability density functions.   
\par  As Appendix \hyperref[app:reflex]{C} highlights, working with density functions will facilitate the use of calculus of variation in many of our proofs. Let $\nu$ be a sigma-finite measure that dominates all probability measures in $\Delta$. Then, for each $\pi_q\in\Delta$, define the density $q\in\Delta_d$ as the Radon-Nikodym derivative of $\pi_q$ with respect to $\nu$. Our proofs will require $\Delta_d$ to be a subset of a \textit{reflexive} Banach space,\footnote{A Banach space  is said to be reflexive if and only if from any bounded sequence it is possible to extract a weakly convergent subsequence  \citep[see,][Theorem 2.28.(i)]{shap00}.} e.g., any Hilbert spaces or any Lebesgue spaces $L^p(S,\nu)$ with $p\in(1,\infty)$. In the remark that follows \citet[][Theorem 1.1]{dak89}, an example illustrates that the restriction to reflexive Banach spaces is \textit{necessary} and hence cannot be dropped in general. Lastly, when $S$ is not finite, we assume all acts are measurable and simple functions.

\phantomsection\label{sec:sketch}
\section*{Appendix B: Sketch of Theorem \ref{thm:rep}}

There are two steps to prove Theorem \ref{thm:rep}. (1) Appendix \hyperref[sec:first]{B.I} simplifies the representation eq. (\ref{eq:rep}) and (2) Appendix \hyperref[sec:second]{B.II} obtains a closed-form solution for the unique social belief. 

\subsection*{B.I: First step}\label{sec:first}
Our goal is to identify the unique social belief as an element of $\mathcal{Q}_{\theta}={\bigcap_{i=1,\beta_i>0}^n}\varGamma_{\eta_i}(p_i)$. To this end, we start by  exchanging the two minimum operators in eq. (\ref{eq:rep}), which yields 
 \begin{align*}
V^{\lambda R}_{0}(f|\mathcal{Q}_{\theta})=\underset{p\in\Delta_d}{\text{min}}\hspace{0.03in}\underset{q\in \mathcal{Q}_{\theta}}{\text{min}}\Bigg\{\int_Su_0(f)p\hspace{0.01in}d\nu+\lambda\hspace{0.02in} R(p\lVert q)\Bigg\}=\underset{q\in \mathcal{Q}_{\theta}}{\text{min}}\Bigg\{\underset{p\in\Delta_d}{\text{min}}\Bigg\{\int_Su_0(f)p\hspace{0.01in}d\nu+\lambda\hspace{0.02in} R(p\lVert q)\Bigg\}\Bigg\}.
 \end{align*}
 Thus, applying \citet[][Proposition 1.4.2]{dupuis97}, we get
\begin{align}\label{eq:newrep}
V^{\lambda R}_{0}(f|\mathcal{Q}_{\theta})=\underset{q\in \mathcal{Q}_{\theta}}{\text{min}}\hspace{0.07in}\phi^{-1}_{\lambda}\Big(\int_S\phi_{\lambda}\big(u_0(f)\big)\hspace{0.03in}q\hspace{0.03in}d\nu\Big). 
 \end{align} 
 The proof of \citet[][Proposition 1.4.2]{dupuis97} applies to probability measures, so we give an alternative proof that applies directly to densities using calculus of variations.

\subsection*{B.II: Second step}\label{sec:second}
This step establishes, for every act $f\in F$ and parameters $\theta_{\lambda}=(\beta,\eta,\lambda)\in\R^{2n+1}_+$, the existence and uniqueness of a solution to eq. (\ref{eq:newrep}). Any solution to eq. (\ref{eq:newrep}) is an element of $\mathcal{Q}_{\theta}\neq\varnothing$, and hence must satisfy each individual constraint preference in eq. (\ref{eq:const}) for all $i=1,\dots,n$ provided $\beta_i>0$ in $u_0=\sum_{i=1}^n\beta_iu_i+\gamma$  and also has to be a valid density. Thus, the Lagrangian of the minimization in eq. (\ref{eq:newrep}) becomes
\begin{align}\label{eq:lag1}
    \mathscr{L}({q})=\E_q\big[\phi_{\lambda}(u_0(f))\big]+\sum_{i=1}^n\ell_i\mathds{1}_{\beta_i>0}\big(R(p_i\lVert q)-\eta_i\big)+\ell_0\Big(\int_Sq\hspace{0.03in}d\nu-1\Big),
\end{align}
where each $\ell_i$ denotes the Lagrange multiplier of individual $i$'s constraint in eq. (\ref{eq:const}), the indicator $\mathds{1}_{\beta_i>0}$ ensures $i$ has nonzero utility weight in (\ref{eq:socialrep}) for each $i=1,\dots,n$, and $\ell_0$ denotes the multiplier of the normalizing constraint $\int_Sqd\nu=1$. We then apply standard techniques from calculus of variations, which state that a minimizer of the Lagrangian in eq. (\ref{eq:lag1}) must also be a solution to an \textit{Euler-Lagrange} equation \citep[][]{var00}.
\begin{lemma}\label{thm:sol}
   Fix any $f$ and $\theta_{\lambda}$. Then, there exists a unique solution $q^{f,\theta_{\lambda}}_{0}\in \mathcal{Q}_{\theta}$ to eq. (\ref{eq:newrep}):
    \begin{align}\label{eq:maindist}
q^{f,\theta_{\lambda}}_0=\sum_{i=1}^n\mu^{\theta_{\lambda}}_i(f)\hspace{0.02in}{p}_i=\underset{q\in \mathcal{Q}_{\theta}}{\text{\normalfont arg min}} \hspace{0.05in}\phi^{-1}_{\lambda}\Bigg(\int_S\phi_{\lambda}\big(u_0(f)\big)q\hspace{0.03in}d\nu\Bigg),
    \end{align}
    where $\mu_{i}(f)=\ell_i\frac{\mathds{1}_{\beta_i>0}}{\ell_0+\phi_{\lambda}(u_0(f))}\geq0$, for $i=1,\dots,n$, are such that $\int_Sq^{f,\theta_{\lambda}}_{0}d\nu=1$ and $q^{f,\theta_{\lambda}}_{0}\geq0$.
\end{lemma}
\par Each $i$'s weight $\mu^{\theta_{\lambda}}_i(f)$ in eq. (\ref{eq:maindist}) takes a simple form: It is proportional to the associated multiplier $\ell_i$. That is, the value of $i$'s probability judgment depends on the relative \textit{sensitivity} of the social welfare criterion to the constraint imposed by her reference model $p_i$.

\phantomsection\label{app:proofs}
\section*{Appendix C: Proofs for the Main Text}

\par Most of the proofs in this article will take the form of solving the following minimization
\begin{align}\label{eq:min}
    \underset{q\in D}{\text{ inf }}\hspace{0.04in}\mathcal{I}(q),
\end{align}
for some arbitrary functional $\mathcal{I}:\mathcal{V}\rightarrow\R$, where $\mathcal{V}$ is a reflexive Banach space (with norm $\lVert.\lVert$), and $D\subseteq \mathcal{V}$ is a closed and convex set. Then, $\mathcal{I}(.)$ is said to be \textit{proper} if it does not take
the value $-\infty$ and is not identically equal to $+\infty$. Moreover, $\mathcal{I}(.)$ is said to be \textit{coercive} over $D$ if $\text{lim}\hspace{0.02in}\mathcal{I}(u)=+\infty$, for $u\in D$ and $\lVert u\lVert\rightarrow\infty$.  The next result describes the sufficient conditions for the existence and uniqueness of a solution in (\ref{eq:min}).
\begin{lemma}\label{thm:exist}
    Suppose $q\mapsto\mathcal{I}(q)$ is convex, lower semicontinuous, and proper. Assume either $D$ is a bounded set or $\mathcal{I}(.)$ is coercive. Then, the minimization problem in (\ref{eq:min}) has at least one solution on $D$. In addition, the solution is unique if  $\mathcal{I}(.)$ is strictly convex on $D$.
\end{lemma}
\phantomsection\label{app:reflex}
This is a well-known result in the literature on convex optimization \citep[][Proposition 1.2, Chapter II.1]{calcvar76}. 
Throughout this appendix, we will apply Lemma \ref{thm:exist}  to establish existence (and sometimes uniqueness) of various solutions. This will then enable us to find closed-form expressions for these solutions by solving a differential equation called the \textit{Euler-Lagrange} equation \citep{var00}. 
\subsection*{Proofs of Observations \ref{thm:agg2} and \ref{thm:agg}}
Observations \ref{thm:agg2} and \ref{thm:agg} follow directly from \citet[][Theorems 1 and 2]{robust16}, respectively.


\subsection*{Proof of Proposition \ref{thm:welfare}}
--- \textit{Overview}: The intuition for the proof of Proposition \ref{thm:welfare} is \citet[][Corollary 5]{mmr06}: if the criteria of two variational preferences are equal (with equal utility functions), then their penalty functions are also equal. This is why uniqueness of the welfare-dominant set result is true in general, but since we restricted dominance to common-taste acts $F_{\star}\subseteq F$, \citet[][Corollary 5]{mmr06} cannot be applied, so we prove this directly using two-outcome common-taste acts, which we show are enough to identify the penalty function.
\begin{lemma}\label{thm:smallsets}
    If $Q\subseteq Q'$, then, for every $c$, $V^{c}_{0}(f|Q)\geq V^{c}_{0}(f|Q')$ $\forall f\in F$.
\end{lemma}
\begin{proof}[Proof of Lemma \ref{thm:smallsets}] This is proved as part of \citet[][Proposition 9]{hansenmiss22}.
\end{proof}

For any $A\subseteq S$ and $t\in\mathbb R$, define
\begin{align*}
\phi_Q(A,t):=\text{min}_{p\in\Delta}\{\hspace{0.02in}c_Q(p)+t\hspace{0.02in}p(A)\hspace{0.02in}\}.
\end{align*}
The next lemma shows that there always exist some two-outcome (nonconstant) acts in $F_\star$.
\begin{lemma}\label{lem:CT}
Suppose $c$-minimal agreement holds: there exist $x,y\in X$ with $u_i(x)>u_i(y)$ for all $i$. For any $\zeta\in(0,1]$, let $x_\zeta:=y+\zeta(x-y)$. Then, for every $A\subseteq S$, $x_\zeta Ay\in F_\star$.
\end{lemma}

\begin{proof}
Let $x,y\in X$ be as in c-minimal agreement and $Y:=\operatorname{co}(\{x,y\})$. 
$\forall z,z'\in Y$, there exist $\kappa,\kappa'\in[0,1]$ such that $z = y+\kappa(x-y)$ and
$z' = y+\kappa'(x-y)$. Since each $u_i$ is affine,
\begin{align*}
u_i(z)-u_i(z') = (\kappa-\kappa')\bigl(u_i(x)-u_i(y)\bigr).
\end{align*}
Because $u_i(x)>u_i(y)$ for all $i$, the sign of $u_i(z)-u_i(z')$ is the same for all
individuals, so for any $z,z'\in Y$ we have $z \succsim_i^\ast z'$ if and only if
$z \succsim_j^\ast z'$ for all $i,j$. By definition of common-taste acts, any two acts
$f,g$ with $f(S),g(S)\subseteq Y$ are common-taste acts. For any
$\zeta\in(0,1]$, $x_\zeta := y+\zeta(x-y)\in Y$, and for every $A\subseteq S$, $f := x_\zeta A y$ and the constant act $g := y$ satisfy $f(S),g(S)\subseteq Y$.
Thus, $f$ and $g$ are common-taste acts, so $x_\zeta A y \in F_\star$ for all $A\subseteq S$.
\end{proof}

\begin{lemma}\label{lem:L1}
Fix $x,y\in X$ with $u_0(x)>u_0(y)$ from c-minimal agreement and set
\begin{align}\label{eq:a}
a:=u_0(x)-u_0(y)>0.
\end{align}
Then, for every nonempty $Q\subseteq\Delta$ and every $A\subseteq S$, the two–outcome act $xAy$ satisfies
\begin{align}\label{eq:L1}
V^c_0(xAy| Q)=u_0(y)+\phi_Q(A,a).
\end{align}
\end{lemma}

\begin{proof}[Proof of Lemma \ref{lem:L1}]
Since $xAy$ pays $x$ on $A$ and $y$ on $A^c$, we have
\begin{align*}
    u_0(xAy)=u_0(y)+a\hspace{0.02in}\mathds{1}_A\quad\implies\quad \E_p[u_0(xAy)]=u_0(y)+a\hspace{0.02in}p(A)\quad \forall p\in\Delta.
\end{align*}
Plugging into \eqref{eq:rep2} and pulling out the constant $u_0(y)$ yields \eqref{eq:L1}.
\end{proof}

\begin{lemma}\label{lem:rd}
Fix $q\in\Delta$ and $A\subseteq S$. Define $g_q^A:[0,\infty)\to\mathbb R$ by
\begin{align}\label{eq:gq}
g_q^A(t):=\phi_{\{q\}}(A,t)=\text{\normalfont min}_{p\in\Delta}\{c(p,q)+tp(A)\}.
\end{align}
Then,
\begin{enumerate}
    \item $g_q^A(0)=0,\qquad 0 \leq g_q^A(t) \leq tq(A)\quad\text{for all }t\ge 0.$
    \item $\text{the minimizer $p_t\in\text{\normalfont arg min}_{p\in\Delta}\{c(p,q)+tp(A)\}$ exists for each $t\ge 0$}$.
    \item $\lim_{t\downarrow 0} \frac{g_q^A(t)}{t} = q(A).$
\end{enumerate}
\end{lemma}

\begin{proof}[Proof of Lemma \ref{lem:rd}]
\emph{Part 1.} Fix any $A\subseteq S$. Since $c(\cdot,q)\ge 0$, $g_q^A(t)\ge \text{min}_{p\in\Delta}\hspace{0.04in} tp(A)\geq0$ for all $t\geq0$. Evaluating \eqref{eq:gq} at $p=q$ gives $g_q^A(t)\leq c(q,q)+tq(A)=tq(A)$ for all $t\geq0$.

\emph{Part 2.} Since $\Delta$ is compact, the map $p\mapsto c(p,q)+tp(A)$ is lower semicontinuous, so the minimum is attained.

\emph{Part 3.} Let $p_t\in\text{arg min}_{p\in\Delta}\{c(p,q)+tp(A)\}$. From optimality,
\begin{align}\label{eq:opt_compact}
g_q^A(t)=c(p_t,q)+tp_t(A) \leq tq(A),
\end{align}
so $\frac{c(p_t,q)}{t}\leq q(A)-p_t(A)\leq 1$. In particular, $c(p_t,q)\to 0$ as $t\downarrow 0$. By lower semicontinuity and uniqueness of the minimizer of $c(\cdot,q)$ at $p=q$, any sequence $t_n\downarrow 0$ forces $p_{t_n}\to q$. Continuity of $p\mapsto p(A)$ on $\Delta$ then yields $p_t(A)\to q(A)$. Using $c(p_t,q)\ge 0$ and \eqref{eq:opt_compact},
\begin{align*}
   p_t(A) \leq  \frac{c(p_t,q)}{t}+p_t(A)=\frac{g_q^A(t)}{t} \leq q(A), 
\end{align*}
so letting $t\downarrow 0$ gives the right derivative of $g_q^A(t)$ at zero: $\lim_{t\downarrow 0} \frac{g_q^A(t)}{t} = q(A).$
\end{proof}

We prove Proposition \ref{thm:welfare} in the general case where  convexity of $q\mapsto c(.,q)$ is replaced with Axiom \ref{ax:hybrid}. By \citet[][Proposition 7]{hansenmiss22}, this convexity is equivalent to \citet[][Axiom A.9]{hansenmiss22}, which is stronger than Axiom \ref{ax:hybrid} (due to $F_{\star}$).
 \begin{proposition}[Impossibility I, extended]\label{thm:welfare1}
        Suppose $\succsim_{0,c,Q}$ satisfies Desideratum \ref{des:welfare}, Axiom \ref{ax:hybrid}, and the MBA profile $(\succsim_i)_{i=1}^n$ satisfies {\normalfont c}-minimal agreement with beliefs $(P_i)_{i=1}^n$. If $\succsim_{0,c,Q}$ satisfies revealed common-taste unambiguous Pareto dominance with respect to  $(\succsim_i)_{i=1}^n$, then $Q$ in (\ref{eq:modelselect3}) must be $Q=\{p_j\}$, where $p_j\in P_j$ is unique for some individual $j\in\{1,\dots,n\}$.
    \end{proposition}
\begin{proof}[Proof of Proposition \ref{thm:welfare1}]
Fix the notation $K:=\bigcup_{i=1}^n P_i$ and  $C:=\text{\normalfont co}(K)$ throughout this proof. We are given that $\succsim_{0,c,Q}$ satisfies revealed common-taste unambiguous Pareto dominance with respect to  $(\succsim_i)_{i=1}^n$, and $\succsim_{0,c,Q}$ satisfies Desideratum \ref{des:welfare}, i.e., $Q$ satisfies (\ref{eq:modelselect3}).
\par \textit{Step 1.}  Fix any $q^*\in Q$. By Lemma \ref{thm:smallsets}, we have $V^{c}_0(f| Q)\leq V^{c}_0(f|\{q^*\})$  $\forall f\in F$. Applying (\ref{eq:modelselect3}) with $Q'=\{q^*\}$ gives $V^{c}_0(f|\{q^*\})\leq V^{c}_0(f| Q)$ $\forall f\in F_\star$.
Hence, $V^{c}_0(f| Q)=V^{c}_0(f|\{q^*\})$ on $F_\star$, and for any $Q'\subseteq C$ and $f\in F_\star$, (\ref{eq:modelselect3}) with $Q$ implies
$V^{c}_0(f| Q')\leq V^{c}_0(f| Q)=V^{c}_0(f|\{q^*\})$, i.e., $\{q^*\}$ also satisfies (\ref{eq:modelselect3}). Since $q^\star$ was arbitrary, every $\{q\}$ with $q\in Q$ also satisfies (\ref{eq:modelselect3}).

\smallskip
\textit{Step 2.} Suppose $q_1,q_2\in Q$ both satisfy Desideratum \ref{des:welfare}. Then, for all $f\in F_\star$,
\begin{align}\label{eq:pointwise_eq}
V^c_0(f|\{q_1\})=V^c_0(f|\{q_2\}).
\end{align}
By c-minimal agreement, let $x,y\in X$ be as in Lemma \ref{lem:CT} and set $a:=u_0(x)-u_0(y)>0$ (see \eqref{eq:a}). For each $\zeta\in(0,1]$ and $A\subseteq S$, Lemma \ref{lem:CT} gives $x_\zeta Ay\in F_\star$, so by \eqref{eq:pointwise_eq} and Lemma \ref{lem:L1},
\begin{align}\label{eq:ray}
\phi_{\{q_1\}}(A,\zeta a)=\phi_{\{q_2\}}(A,\zeta a)\qquad\forall A\subseteq S, \forall \zeta\in(0,1].
\end{align}
Fix any $A$ and define $g_{q}^A(t):=\phi_{\{q\}}(A,t)$ for $t\ge 0$. By \eqref{eq:ray}, $g_{q_1}^A(t)=g_{q_2}^A(t)$ for all $t\in(0,a]$. By Lemma \ref{lem:rd}.3, their right derivatives at $0$ coincide and equal $q_1(A)$ and $q_2(A)$, respectively:
\[
q_1(A)=\lim_{t\downarrow 0}\frac{g_{q_1}^A(t)}{t}=\lim_{t\downarrow 0}\frac{g_{q_2}^A(t)}{t}=q_2(A).
\]
Since this holds for every $A\subseteq S$, we have $q_1=q_2$. Thus, any $Q$ satisfying (\ref{eq:modelselect3}) must be a singleton, i.e., $Q=\{q^*\}$ for a unique belief denoted $q^\ast\in C$.

\smallskip
\textit{Step 3.} Axiom \ref{ax:hybrid} implies that, for every $f\in F_\star$, the map $q\mapsto V^c_0(f|\{q\})$ is convex on $C$. If $q^\ast=\xi r+(1-\xi)w$ with $r\ne w$ and $\xi\in(0,1)$, then for every $f\in F_\star$,
\[
V^c_0(f|\{q^\ast\}) \leq \xi V^c_0(f|\{r\})+(1-\xi)V^c_0(f|\{w\}) \leq \text{max}\big\{V^c_0(f|\{r\}),V^c_0(f|\{w\})\big\}.
\]
But, by (\ref{eq:modelselect3}), with $Q'=\{r\}$ and $Q'=\{w\}$, we also have $V^c_0(f|\{r\})\leq V^c_0(f|\{q^\ast\})$ and $V^c_0(f|\{w\})\leq V^c_0(f|\{q^\ast\})$ for all $f\in F_\star$. Hence, equality holds throughout for every $f\in F_\star$, so both $\{r\}$ and $\{w\}$ satisfy (\ref{eq:modelselect3}). By Step 2’s uniqueness, $r=w=q^\ast$, a contradiction. Thus, $q^\ast$ is an extreme point of $C$, i.e., $q^\ast=p_j\in P_j$ for some individual $j\geq1$.
\end{proof}

\subsection*{Proof of Observation \ref{thm:neutral}}
\begin{proof}
   For any set of structured models $Q$, \citet[][Theorem 3.(2), Appendix A]{hansenmiss25} states that $V^{\lambda R}_0(.|Q)$ is a multiplier criterion ($Q$ is a singleton), if and only if it satisfies social ambiguity neutrality \citep[][A.12, Appendix A]{hansenmiss25}.
\end{proof}

\subsection*{Proof of  Proposition \ref{thm:max}}
We prove Proposition \ref{thm:max} when $S$ is finite, but the proof below extends to arbitrary $S$ by using \citet[][Proposition 14]{hansenmiss22}, when $\Delta$ is restricted to the subset of countably additive probability measures, and co(.) is replaced with $\overline{\text{co}}(.)$---the closed convex hull.
\begin{proof}
Set  $Q=\text{\normalfont co}\big(\bigcup_{i=1}^nP_i\big)$. Each $P_i$ is compact because it is a subset of $\Delta$. Thus, $\text{\normalfont co}\big(\bigcup_{i=1}^nP_i\big)$ is compact because it is the convex hull of a finite union of convex and compact sets. Since $\bigcup_{i=1}^nP_i$ is a compact subset of $\Delta$, the set function $\upsilon :2^S\rightarrow[0,1]$, defined by $\upsilon(E)=\text{min}_{q\in\bigcup_{i=1}^nP_i}q(E)$ for all events $E\in2^S$, is an exact capacity, which is continuous at $S$. This implies that  $\bigcup_{i=1}^nP_i\subseteq \text{core}(\upsilon)\subseteq\Delta$, and therefore $\text{\normalfont co}\big(\bigcup_{i=1}^nP_i\big)\subseteq \text{core}(\upsilon)\subseteq \Delta$. Putting all the above together and \citet[][Proposition 1.4.2]{dupuis97}, we get 
\begin{align*}
V^{\lambda R}_{0}\Bigg(f\Bigg|\bigcup_{i=1}^nP_i\Bigg)&=\underset{p\in\Delta}{\text{min}}\Big\{\E_{p}\big[u_0(f)\big]+\lambda\hspace{0.02in}\underset{q\in \bigcup_{i=1}^nP_i}{\text{min}}\hspace{0.02in} R(p\lVert q)\Big\}\\
&=\underset{q\in \bigcup_{i=1}^nP_i}{\text{min}}\hspace{0.03in}\phi^{-1}_{\lambda}\Big(\E_{q}\big[\phi_{\lambda}\big(u_0(f)\big)\big]\Big)\\
&=\phi^{-1}_{\lambda}\Big(\underset{q\in \bigcup_{i=1}^nP_i}{\text{min}}\hspace{0.03in}\E_{q}\big[\phi_{\lambda}\big(u_0(f)\big)\big]\Big)\\
&=\phi^{-1}_{\lambda}\Big(\underset{q\in \text{\normalfont co}(\bigcup_{i=1}^nP_i)}{\text{min}}\hspace{0.03in}\E_{q}\big[\phi_{\lambda}\big(u_0(f)\big)\big]\Big)\\
&=\underset{q\in \text{\normalfont co}(\bigcup_{i=1}^nP_i)}{\text{min}}\hspace{0.03in}\phi^{-1}_{\lambda}\Big(\E_{q}\big[\phi_{\lambda}\big(u_0(f)\big)\big]\Big)\\
&=\underset{p\in\Delta}{\text{min}}\Big\{\E_{p}\big[u_0(f)\big]+\lambda\hspace{0.02in}\underset{q\in \text{\normalfont co}(\bigcup_{i=1}^nP_i)}{\text{min}}\hspace{0.02in} R(p\lVert q)\Big\}=V^{\lambda R}_{0}\Bigg(f\Bigg|\text{\normalfont co}\Big(\bigcup_{i=1}^nP_i\Big)\Bigg),
\end{align*}
for every $f\in F$ and $\lambda\in(0,\infty]$, where $\phi_{\lambda}$ is the function defined in Section \ref{sec:analysis}.
\end{proof}

\subsection*{Proof of Proposition \ref{thm:hybrid}}
For any affine utility function $u:X\to\mathbb R$ and act $f$, write $u(f)\in\mathbb R^S$ for the statewise utility vector, and $\langle \varphi,p\rangle:=\sum_{s\in S}\varphi(s)p(s)$ for $\varphi\in\mathbb R^S$, $p\in\Delta$. Let $\mathbf{1}$ denote the vector of all ones, and $\text{ext}(A)$ denote the set of all the extreme points of $A$. Let $\lVert.\lVert_{\infty}$ denote the sup-norm.
\begin{definition}[Face and minimal face]\label{def:face}
\normalfont Let $D\subseteq\mathbb R^S$ be convex and nonempty.
A subset $\mathcal{F}\subseteq D$ is a \emph{face} of $D$ if $x=\gamma y+(1-\gamma)z\in \mathcal{F}$ with $y,z\in D$ and $\gamma\in(0,1)$ implies $y,z\in \mathcal{F}$.
For $x\in D$, the \emph{minimal face} $\mathcal{F}_x$ is the intersection of all faces of $D$ containing $x$.
\end{definition}
Proposition \ref{thm:hybrid} will exploit the existence of the supporting functionals in Lemma \ref{lem:support} below.
\begin{lemma}[Supporting functional at the minimal face]\label{lem:support}
Let $D\subseteq\mathbb R^S$ be compact and convex, and let $x\in D$.
Then, there exist $\psi\in\mathbb R^S$ and $m\in\mathbb R$ such that
\begin{align}\label{eq:support-ineq}
\langle \psi,y\rangle \ge m\quad\forall y\in D,
\qquad\text{and}\qquad
\{y\in D: \langle \psi,y\rangle=m\} \supseteq \mathcal{F}_x.
\end{align}
If, in addition, $D\subseteq\Delta$, then with $\varphi:=\psi-m\mathbf{1}$, we have
\begin{align}\label{eq:phiProps}
\langle \varphi,y\rangle \ge 0\quad\forall y\in D,
\qquad
\langle \varphi,y\rangle = 0\quad\forall y\in \mathcal{F}_x.
\end{align}
\end{lemma}
 Lemma \ref{lem:support} is a direct corollary of the supporting hyperplane theorem, but, for completeness, we prove it in Online Appendix \hyperref[app:proof]{F}.

\begin{lemma}\label{lem:faceK}
Let $W=\text{\normalfont co}(T)$ with $T$ compact.
If $\mathcal{F}$ is a nonempty face of $W$, then $\mathcal{F}\cap T\neq\varnothing$.
\end{lemma}
\begin{proof}[Proof of Lemma \ref{lem:faceK}]
    Every nonempty compact face $\mathcal{F}$ contains extreme points of $W$, and $W=\text{\normalfont co}(T)$ implies $\operatorname{ext}(W)\subseteq T$; hence $\mathcal{F}\cap T\supseteq \operatorname{ext}(\mathcal{F})\neq\varnothing$.
\end{proof}

\begin{lemma}[Carath\'eodory Theorem]\label{lem:caratheodory}
For any set $A$, if $z\in \text{\normalfont co}(A)\subseteq\mathbb R^S$, then $z$ is a convex combination of (at most $|S|+1$) finitely many points of $A$.
\end{lemma}

\begin{axiom}[\citet{hansenmiss22}, Axiom A.4]\label{ax:subjective}
   For all $f\in F$ and $x\in X$, $p\in \text{\normalfont co}(Q)$ if and only if $x\succ^{\scriptscriptstyle\wedge}_{0,c,Q} x^p_f \implies x\succ_{0,c,Q}f$.
\end{axiom}
 
Let $\text{ri}(.)$ denote the relative interior and $\text{rb}(.)$ denote the relative boundary. Recall the notation $K:=\bigcup_{i=1}^n P_i$ and $C:=\text{\normalfont co}(K).$ Note that $\text{ri}(C)$ contains all weighted averages of individuals' beliefs with \textit{strictly} positive weights, whereas its complement, $\text{rb}(C)$, contains all other weighted averages of individuals' beliefs (i.e., those with at least one zero weight).\footnote{For belief-aggregation purposes, $\text{rb}(C)$ is ``richer'' than $\text{ri}(C)$ in a combinatorial sense: $\text{rb}(C)$ contains up to $2^n - 2$ distinct ``sub-coalitions'' where at least one individual's beliefs are ignored by the social planner, whereas $\text{ri}(C)$ consists of a single homogeneous coalition where no individual's beliefs are ever ignored. Thus, extending Axiom \ref{ax:acc} to Bayesian planners with non-full-support priors would only strengthen the impossibility.}

\begin{lemma}\label{thm:rel}
   If $Q\subseteq C$ and Axiom \ref{ax:acc} holds, then $Q\subseteq \text{\normalfont rb}(C)$.
\end{lemma}
\begin{proof}[Proof of Lemma \ref{thm:rel}]
 By Definition \ref{def:bayes}, the set of all Bayesian planner's beliefs $p=\sum_{i=1}^n\pi_i\hspace{0.02in}p_i$ with $p_i\in P_i$ $\forall i$ and full-support prior $\pi=(x_i)_{i=1}^n$ is $\text{ri}(C)$. Fix $p\in \text{\normalfont ri}(C)$. Axiom \ref{ax:acc} says that there exists $(x,f_{\star})$  with $x\succ^{\scriptscriptstyle\wedge}_{0,c,Q} x^p_{f_{\star}}$ and $f_{\star}\succsim_{0,c,Q}x$. Applying the contrapositive of Axiom \ref{ax:subjective}, this means that $p\notin \text{co}(Q)$. Axiom \ref{ax:acc} says that this holds for every Bayesian planner's belief $p\in \text{\normalfont ri}(C)$, so $\text{co}(Q)\cap \text{ri}(C) =\varnothing$. Thus, $\text{co}(Q)\subseteq \text{rb}(C)$, and hence $Q\subseteq \text{rb}(C)$.
\end{proof}

Now, define $J^c_0(\varphi,q):=\text{min}_{p\in\Delta}\{\langle \varphi,p\rangle+c(p,q)\}$ for $q\in Q$, then
\begin{align}\label{eq:swap}
V^c_0(f|Q)=\text{min}_{q\in Q}J^c_0\big(u_0(f),q\big)\quad\forall f\in F
\end{align}
is the welfare criterion in eq. (\ref{eq:rep2}). We are now ready to prove Proposition \ref{thm:hybrid}.

\begin{proof}[Proof of Proposition \ref{thm:hybrid}]
This proof is constructive. Toward contradiction, assume there exists $q^\circ\in Q\setminus K$.
By \eqref{eq:socialrep2}, $q^\circ\in C$.
Let $\mathcal{F}_{q^\circ}$ be the minimal face of $C$ containing $q^\circ$. By Lemma \ref{thm:rel}, $q^\circ\in \text{rb}(C)$, so $\mathcal{F}_{q^\circ}$ is proper. 
By Lemma \ref{lem:faceK}, $\mathcal{F}_{q^\circ}\cap K\neq\varnothing$.
By Lemma \ref{lem:caratheodory} and Definition \ref{def:face},
\begin{align}\label{eq:face-rep}
q^\circ=\sum_{k=1}^m \lambda_kp_k,\quad \text{where}\quad
p_k\in K\cap \mathcal{F}_{q^\circ}; \lambda_k>0; \sum_{k=1}^m\lambda_k=1;  2\leq m\leq |S|+1,
\end{align}
and $m\ge 2$ since $q^\circ\notin K$. By Lemma \ref{lem:support} with $D=C$ and $x=q^\circ$, choose $\varphi\in\mathbb R^S$ such that
\begin{align}\label{eq:phi}
\langle \varphi,p\rangle\ge 0\quad\forall p\in C,\qquad
\langle \varphi,p_k\rangle=0\quad(k=1,\dots,m),
\end{align}
and $\varphi$ is nonconstant because $\mathcal{F}_{q^\circ}$ is a proper face. By c-minimal agreement, pick $a,b\in X$ with $\rho_i:=u_i(b)-u_i(a)>0$ $\forall i$ and set $\theta_i:=u_i(a)$ $\forall i$.
Define some nonconstant affine Bernoulli utility index $u^{\star}$ on the interval $[a,b]$ by $u^{\star}((1-r)a+rb):=r$ for all $r\in[0,1]$, i.e., we normalize it as $u^{\star}(a):=0$ and $u^{\star}(b):=1$ using the fact that $u^{\star}$ is assumed to be affine and nonconstant.
Now, since $u_i$ is affine and nonconstant $\forall i\geq1$, then on $[a,b]$, we can write
\begin{align*}
u_i\big((1-r)a+rb\big)=\rho_iu^{\star}((1-r)a+rb)+\theta_i=\rho_i r+\theta_i, \quad\forall i,r\in[0,1].
\end{align*}
By \eqref{eq:socialrep2}, $u_0=\sum_{i=1}^n\beta_iu_i+\gamma$, so on $[a,b]$ we have that, for all $r\in[0,1]$,
\begin{align*}
u_0\big((1-r)a+rb\big)=A r + B,\quad\text{where}\quad
A:=\sum_{i=1}^n\beta_i\rho_i>0,\quad B:=\sum_{i=1}^n\beta_i\theta_i+\gamma.
\end{align*}
Since \eqref{eq:phi} is preserved under positive rescalings, rescale $\varphi$ (by a positive factor, if necessary) so that $\|\varphi\|_\infty\leq A/2$.
For all $s\in S$, define $r_g(s):=\tfrac12$ and $r_f(s):=\tfrac12+\frac{\varphi(s)}{A}\in[0,1]$, and set
\begin{align*}
g(s):=(1-r_g(s))a+r_g(s)b,\qquad f(s):=(1-r_f(s))a+r_f(s)b.
\end{align*}
Notice that $g$ is a constant act, whereas $f$ is not, and both take values in $[a,b]$. Every $u_i$ is a positive affine transformation of $u^{\star}$ on $[a,b]\supseteq \text{co}(f(S)\cup g(S))$, so $f$ and $g$ are common-taste acts, i.e., $f,g\in F_{\star}$. Since $u_0(f(s))-u_0(g(s))=A(r_f(s)-r_g(s))=\varphi(s)$ for each $s$, then
\begin{align}\label{eq:u0diff}
u_0(f)-u_0(g)=\varphi.
\end{align}

Thus, for all $i$ and all $p\in P_i\subseteq C$, the inequality below holds by \eqref{eq:phi}
\begin{align*}
\E_p\big[u_i(f)-u_i(g)\big]=\langle u_i(f)-u_i(g),p\rangle
=\rho_i\langle u^{\star}(f)-u^{\star}(g),p\rangle
=\rho_i \Big\langle \frac{1}{A}\varphi,p\Big\rangle
=\frac{\rho_i}{A}\langle \varphi,p\rangle \ge 0,
\end{align*}
so $f\succsim^*_ig$ $\forall i$. Since $\succsim_{0,c,Q}$ satisfies revealed common-taste unambiguous Pareto dominance with respect to $(\succsim_i)_{i=1}^n$, $\succsim^*_0$ satisfies common-taste unambiguous Pareto dominance with respect to $(\succsim^*_i)_{i=1}^n$ (Definitions \ref{def:compareto}--\ref{def:revcompareto}). Thus, $f\succsim^*_0g$, which by Definition \ref{def:rev} implies that
\begin{align}\label{eq:inq}
    V^c_0(f|Q)\ge V^c_0(g|Q).
\end{align}
Now, $J^c_0(\varphi,p_k)=\text{min}_{p\in\Delta}\{\langle \varphi,p\rangle+c(p,p_k)\} \leq \langle \varphi,p_k\rangle+c(p_k,p_k) = 0$ $\forall k$, last equality holds by \eqref{eq:phi}.
We now have three things: (1) $f,g\in F_{\star}$, (2) the convex combination in \eqref{eq:face-rep} has at least $m\geq2$ distinct $p_k$'s, and  (3) $\varphi$ in \eqref{eq:u0diff} is nonconstant. Thus, we can apply the strict convexity in Desideratum \ref{des:hybrid} to $J^c_0$ as follows:
$J^c_0(\varphi,q^\circ)<\sum_{k=1}^m \lambda_kJ^c_0(\varphi,p_k) \leq 0,$
and hence
\begin{align}\label{eq:neg}
J^c_0(\varphi,q^\circ)<0.
\end{align}
\vspace{-0.1in}
Let $\zeta:=u_0(g(s))=A/2+B$ for all $s\in S$.
Using \eqref{eq:swap} and $u_0(f)=u_0(g)+\varphi$ from \eqref{eq:u0diff},
\begin{align*}
V^c_0(f|Q)
&= \text{min}_{q\in Q}\text{min}_{p\in\Delta}\{\langle u_0(g),p\rangle+\langle \varphi,p\rangle+c(p,q)\}
= \zeta + \text{min}_{q\in Q} J^c_0(\varphi,q),\\
V^c_0(g|Q)
&= \text{min}_{q\in Q}\text{min}_{p\in\Delta}\{\langle u_0(g),p\rangle+c(p,q)\}
= \zeta + \text{min}_{q\in Q}\text{min}_{p\in\Delta} c(p,q)=\zeta,
\end{align*}
since $\text{min}_{p\in\Delta} c(p,q)=c(q,q)=0$.
Using the fact that $q^\circ\in Q$ and \eqref{eq:neg}, respectively, yield $$V^c_0(f|Q)-V^c_0(g|Q)=\text{min}_{q\in Q}J^c_0(\varphi,q)\leq J^c_0(\varphi,q^\circ)<0,$$ which \textit{contradicts} $V^c_0(f|Q)\ge V^c_0(g|Q)$ in (\ref{eq:inq}).
Thus, we conclude that $Q\subseteq K$.
\end{proof}

\subsection*{Proof of Theorem  \ref{thm:pos}}
We start with a general result that will be useful to prove Theorem  \ref{thm:pos}. Let $V$ be a reflexive Banach space and $G : \mathcal{D} \to \mathbb{R}$ be of the Legendre type on an open convex domain $\mathcal{D} \subseteq V$, and its Hessian $\nabla^2 G(y)$ is positive definite for all $y \in \mathcal{D}$. Let $\mathcal{C}$ be a closed convex bounded subset of $\mathcal{D}$, and consider $x_1,\dots,x_n \in \mathcal{C}$.
For each $i$, let $r_i \ge 0$, and define the primal Bregman ball
$B^G_{r_i}(x_i)
    :=
    \bigl\{ y \in \mathcal{C} : D_G(x_i \lVert y) \leq r_i \bigr\},$
which is closed and convex. Define
\begin{align}
    \mathcal{S}
    :=
    \bigcap_{i=1}^n B^G_{r_i}(x_i)
    =
    \big\{ y \in \mathcal{C} : D_G(x_i \lVert y) \leq r_i \hspace{0.03in}\forall i \big\}.
\end{align}
Proposition \ref{thm:bregball} extends some insights in \citet[][Lemma 2 and Theorem 6]{small18} to cases where Bregman balls do not have the same radius in infinite-dimensional spaces.
\begin{proposition}
\label{thm:bregball}
Suppose $\mathcal{S}\subseteq\mathcal{C}$ is nonempty. Then, there exists a point $y^\star \in \mathcal{S}$ that can be written as a convex combination of $\{x_1,\dots,x_n\}$. That is, there exist weights $\zeta_1,\dots,\zeta_n \ge 0$ with $\sum_{i=1}^n \zeta_i = 1$ such that
$y^\star = \sum_{i=1}^n \zeta_i x_i\in\mathcal{S}$ or equivalently $\mathcal{S}\cap \text{\normalfont co}(\{x_1,\dots,x_n\})\neq\varnothing$.
\end{proposition}
The proof of Proposition \ref{thm:bregball} is a bit long, so it is in Online Appendix \hyperref[app:proof]{F}. To prove Theorem \ref{thm:pos} in the general case of nonfinite $S$ with $\Delta_d$---the set of density functions (Appendix \hyperref[app:compact]{A}), we assume this set is bounded. This holds when $S$ is finite, in which case $\Delta$ is the simplex.

\begin{proof}[Proof of Theorem \ref{thm:pos}]
    To match the variables in Proposition \ref{thm:bregball}, set $\mathcal{S}=\bigcap_{i=1,\beta_i>0}^nB^G_{\eta_i}(p_i)$, $x_i=p_i\in \Delta_d$ and $r_i=\eta_i$ for every $i$, and $\Delta_d=\mathcal{C}$. By Observation \ref{thm:agg}, $(u_0,P_0)$ are in (\ref{eq:socialrep}) and $\mathcal{S}\supsetneq P_0$ is nonempty. Thus, Proposition \ref{thm:bregball} yields $\bigcap_{i=1,\beta_i>0}^nB^G_{\eta_i}(p_i)\cap \text{co}(\{p_1,\dots,p_n\})\neq\varnothing$. 
\end{proof}

\subsection*{Proof of Corollary \ref{thm:pos1}}
\par The Bregman ball $B^G_{\eta_i}(p_i)$ in eq. (\ref{eq:breg}) is known as a \textit{primal} Bregman ball  to differentiate it from the so-called \textit{dual} Bregman ball $\hat{B}^G_{\eta_i}(p_i)$ defined as\footnote{\citet{topo18} provide graphical illustrations of primal and dual Bregman balls.}
$\hat{B}^G_{\eta_i}(p_i)=\big\{q\in\Delta:D_G(q\lVert p_i)\leq\eta_i\big\}.$
\begin{lemma}\label{thm:chernoff1}
    There exists a unique $r^*\geq0$ such that $\bigcap_{i=1,\beta_i>0}^nB^G_{r^*}(p_i)=\{q^{*}_0\}$, where $q^{*}_0$ is called the Chernoff point of the set $\{p_1,\dots,p_n\}$.
\end{lemma}
\begin{proof}[Proof of Lemma \ref{thm:chernoff1}]
    By \citet[][Section 4: Chernoff polytopes, p. 35:7]{small18}:
    \par (1) The smallest dual Bregman ball that contains the finite set $\{p_1,\dots,p_n\}$ exists, so let's denote it $\hat{B}^G_{r^*}(q^{*}_0)$ with radius $r^*\geq0$ and center $q^{*}_0$.
 \par (2) The constant $r^*\geq0$ is also the minimal radius such that $\bigcap_{i=1,\beta_i>0}^nB^G_{r^*}(p_i)=\{q^{*}_0\}$, and $q^{*}_0$ is the Chernoff point of the set $\{p_1,\dots,p_n\}$.
    \end{proof}

\begin{proof}[Proof of Corollary \ref{thm:pos1}]
 By Lemma \ref{thm:chernoff1}, there exists a unique constant $r^*\geq0$ such that the intersection $\bigcap_{i=1,\beta_i>0}^nB^G_{r^*}(p_i)=\{q^{*}_0\}$ in Theorem  \ref{thm:pos}, so $q^{*}_0\in\text{co}(\{p_1,\dots,p_n\})$, which is the unique Chernoff point of the set $\{p_1,\dots,p_n\}$. Thus, $P_0=\{q^{*}_0\}$, so the MBA preference $(u_0,P_0,\alpha_0)$ in Theorem  \ref{thm:pos} becomes SEU $(u_0,q^{*}_0)$, for all $\alpha_0$. 
\end{proof}

\subsection*{Proof of Observation \ref{thm:r}}
\begin{corollary}\label{thm:smallsets2}
    If $Q\subseteq Q'$, then, for every $c$, $V^{c}_{0}(Q)\geq V^{c}_{0}(Q')$.
\end{corollary}
\begin{proof}[Proof of Corollary \ref{thm:smallsets2}]
    Take sup over $f\in F$ on both sides of the inequality in Lemma \ref{thm:smallsets}.
\end{proof}
\begin{proof}[Proof of Observation \ref{thm:r}]
  Notice that the intersection of Bregman balls $\bigcap_{i=1,\beta_i>0}^nB^G_{r}(p_i)$ grows weakly as the common radius $r$ increases. Since the value function $V^{c}_{0}(.)$ is assumed to be well-defined and nonconstant, the intersection that maximizes $V^c_{0}(.)$ is a singleton (Corollary \ref{thm:smallsets2}). By Lemma \ref{thm:chernoff1}, there exists a unique $r^*\geq0$ at which the intersection is a singleton.
\end{proof}

\subsection*{Proof of Theorem \ref{thm:rep}} 
As stated in the sketch (Appendix \hyperref[sec:sketch]{B}), proving Theorem \ref{thm:rep} follows two main steps. Throughout, fix any act $f\in F$ and parameters $\theta_{\lambda}=(\beta,\eta,\lambda)\in \R^{2n+1}_+$. 
The goal is to show existence and uniqueness of a belief $q_0^{f,\theta_{\lambda}}\in \mathcal{Q}_{\theta}$---the solution to the inner minimization in eq. (\ref{eq:rep}). Since $\succsim_{0,\lambda R}$ satisfies \textit{neutral unambiguous Pareto dominance} with respect to $(\succsim_i)_{i=1}^n$, then Observation \ref{thm:agg} yields that $Q\subseteq\bigcap^n_{i=1,\beta_i>0}\varGamma_{\eta_i}(p_i)=\mathcal{Q}_{\theta}$. Then, it is assumed that $Q=\mathcal{Q}_{\theta}$.
\par {\large\textbf{-- First Step:}} The sketch in Appendix \hyperref[sec:first]{B.I} showed that we can solve the inner minimization over implausible models
by applying \citet[][Proposition 1.4.2]{dupuis97}. We provide an alternative proof that uses elementary results from calculus of variations.
\begin{lemma}\label{thm:inner}
    For any plausible model $q\in \mathcal{Q}_{\theta}$, we have
    \begin{align*}
    \underset{p\in\Delta_d}{\text{\normalfont min}}\Bigg\{\int_Su_0(f)p\hspace{0.03in}d\nu+\lambda\hspace{0.02in} R(p\lVert q)\Bigg\}=\phi^{-1}_{\lambda}\Bigg(\int_S\phi_{\lambda}\big(u_0(f)\big)\hspace{0.03in}q\hspace{0.03in}d\nu\Bigg),
\end{align*}
and the minimum is attained uniquely at the implausible model $p^f_0(.|q)=\frac{\phi_\lambda(u_0(f))}{\E_q[\phi_\lambda(u_0(f))]}q\in\Delta_d$.
\end{lemma}
\begin{proof}[Proof of Lemma \ref{thm:inner}]
   Fix any plausible model $q\in \mathcal{Q}_{\theta}$. It suffices to prove this result on the subset $\Delta_d^+\subseteq\Delta_d$---the set containing models that are absolutely continuous with respect to $q$. This way $R(p\lVert q)$ is finite for all $p\in\Delta_d^+$, where $\Delta_d^+$ is convex and compact subset of the reflexive Banach space $\Delta_d$. The minimization can then be rewritten as
\begin{align}\label{eq:obj}
    \underset{p\in\Delta_d^+}{\text{min}} \int_S U_f(p|q)\hspace{0.03in}d\nu,
\end{align}
where $U_f(p|q):=u_0(f)p+\lambda p\hspace{0.01in}\text{log}\frac{p}{q}$ for all $f\in F$. By Lemma \ref{thm:exist}, there exists a unique solution $p^f_0$ to this minimization in $\Delta_d^+$ because $\int_SU_f(p|q)d\nu$ in eq. (\ref{eq:obj}) is proper, which holds since $R(p\lVert q)$ is a nonnegative bounded, strictly convex and continuous function in $p$, and $u_0(f)$ is also bounded. Since $U_f(p|q)$ is continuous in $p$, for every $f\in F$, we can therefore apply standard results from calculus of variations \citep[][]{var00}, which state that a minimizer of eq. (\ref{eq:obj}) is also a solution to the following Euler-Lagrange equation
\begin{align*}
   \nabla_p U\big(p^f_0\big|q\big)+\ell G=0,
\end{align*}
where $\nabla_p$ denotes the derivative operator with respect to $p$, and the constant $\ell$ denotes the Lagrange multiplier corresponding to the constraint  $G:=\int_Sp^f_0\hspace{0.02in}d\nu=1$. Taking the derivative and solving the equation above for $p^f_0\in \Delta_d^+$, we get the unique solution given by
\begin{align*}
    p^f_0(s|q)=q(s)\text{ exp}\Big(-u_0\big(f(s)\big)/\lambda-\ell-1\Big)=\frac{-\phi_{\lambda}\big(u_0(f(s))\big)}{\E_{q}\big[-\phi_{\lambda}(u_0(f))\big]}q(s)\in \Delta_d^+ \quad \forall s\in S,
\end{align*}
 where $p^f_0\geq0$ and the Lagrange multiplier $\ell=\text{log}\hspace{0.02in}\E_{q}\big[-\phi_{\lambda}\big(u_0(f)\big)\big]-1$ is pinned down by the normalizing constraint $G$. Plugging $p^f_0$ in eq. (\ref{eq:obj}) results in the desired expression. 
\end{proof}

{\large\textbf{-- Second Step:}} This is the main step of Theorem \ref{thm:rep} (Appendix \hyperref[sec:second]{B.II}). We recall that the goal is to identify  $q^{f,\theta_{\lambda}}_{0}\in \mathcal{Q}_{\theta}$ and $\mathcal{Q}_{\theta}=\bigcap^n_{i=1,\beta_i>0}\varGamma_{\eta_i}(p_i)$. First, we write the main minimization problem in eq. (\ref{eq:newrep}) more explicitly as follows
\begin{align}\label{eq:finalrep}
       V^{\lambda R}_{0}(f|\mathcal{Q}_{\theta})=\underset{q\in \mathcal{Q}_{\theta}}{\text{min}}\hspace{0.03in}\underbrace{-\lambda\text{ log}\Bigg(\int_S\text{exp}\big(-u_0(f)/\lambda\big)\hspace{0.03in}q\hspace{0.03in}d\nu\Bigg)}_{\mathcal{H}(q)}.
\end{align}
 
\begin{lemma}\label{thm:exist2}
  For $f\in F$, there exists a unique solution in eq. (\ref{eq:finalrep}).
\end{lemma}
\begin{proof}[Proof of Lemma \ref{thm:exist2}]
\par It suffices to verify that $\mathcal{H}(q)$ in eq. (\ref{eq:finalrep}) and $\mathcal{Q}_{\theta}$ satisfy the conditions of $\mathcal{I}(q)$ and $D$ in Lemma \ref{thm:exist}, respectively. First, the natural log function is strictly concave so $-\text{log}$ is strictly convex and continuous in $q$. The exponential function inside the integral is nonnegative and bounded for all acts $f\in F$, so $\mathcal{H}(q)$ is proper. Second, by (\ref{eq:socialrep}), $\mathcal{Q}_{\theta}=\bigcap^n_{i=1,\beta_i>0}\varGamma_{\eta_i}(p_i)\neq\varnothing$, where each $\varGamma_{\eta_i}(p_i)$ in eq. (\ref{eq:const}) is convex and has radius $\eta_i\geq0$, so $\mathcal{Q}_{\theta}$ is convex and bounded. Thus, we conclude from Lemma \ref{thm:exist} that $\mathcal{H}(q)$ in eq. (\ref{eq:finalrep}) attains its minimum value at
a unique belief in $\mathcal{Q}_{\theta}$.
    \end{proof}
\begin{proof}[Proof of Lemma \ref{thm:sol}]
    Since the natural log function is a monotonic transformation and $\lambda>0$,
    \begin{align*}
        q^{f,\theta_{\lambda}}_{0}=\underset{q\in \mathcal{Q}_{\theta}}{\text{arg min}}-\lambda\text{ log}\Bigg(\int_S\text{exp}\big(-u_0(f)/\lambda\big)\hspace{0.03in}q\hspace{0.03in}d\nu\Bigg)=\underset{q\in \mathcal{Q}_{\theta}}{\text{arg min}}-\int_S\text{exp}\big(-u_0(f)/\lambda\big)\hspace{0.03in}q\hspace{0.03in}d\nu.
    \end{align*}
   By (\ref{eq:socialrep}), $\mathcal{Q}_{\theta}=\bigcap^n_{i=1,\beta_i>0}\varGamma_{\eta_i}(p_i)\neq\varnothing$, so any minimizer must satisfy each individual $i$'s constraint in eq. (\ref{eq:const}) provided $i$ has nonzero weight in the social utility, and be a valid density. 
    Since Lemma \ref{thm:exist2} guarantees the existence of a unique solution---denoted $q^{f,\theta_{\lambda}}_{0}$---in $\mathcal{Q}_{\theta}$, we can assume, without loss of generality, that each constraint  is binding, i.e.
$R(p_i\lVert q)=\eta_i$ in $\varGamma_{\eta_i}(p_i)$, because reducing any non-binding constant $\eta_i$ would result in the same optimum. Specifically, we can write more compactly the minimization of interest as
    \begin{align*}
   q^{f,\theta_{\lambda}}_{0}=\underset{q}{\text{arg min}} -\int_S\text{exp}\big(-u_0(f)/\lambda\big)\hspace{0.03in}q\hspace{0.03in}d\nu\quad \text{s.t.}\quad\begin{cases}
R(p_i\lVert q)= \eta_i, \beta_i>0, \forall i;\\
        \int_{S} q\hspace{0.03in}d\nu=1,  q\geq0.
\end{cases}      
    \end{align*}
 By the Lagrange theorem, we can now write the Lagrangian in eq. (\ref{eq:lag1}) explicitly as
    \begin{align*}
         \mathscr{L}({q})=-\int_S\text{exp}\big(-u_0(f)/\lambda\big)\hspace{0.03in}q\hspace{0.03in}d\nu+\sum_{i=1}^n\ell_i\mathds{1}_{\beta_i>0}\Big(\int_S {p}_i\text{ log}\frac{p_i}{q}\hspace{0.03in}d\nu-\eta_i\Big)+\ell_0\Big(\int_Sq\hspace{0.03in}d\nu-1\Big),
    \end{align*}
    where each $\ell_i$ is the Lagrange multiplier of individual $i$'s constraint in eq. (\ref{eq:const}), the indicator $\mathds{1}_{\beta_i>0}$ verifies whether $i$ has nonzero weight in the social utility in (\ref{eq:socialrep}) for each $i=1,\dots,n$, and $\ell_0$ denotes the Lagrangian of the constraint $\int_Sqd\nu=1$. Our goal now is to minimize this Lagrangian with respect to ${q}$, so all constants can be omitted and instead minimize 
      \begin{align}\label{eq:lag2}
         \mathscr{L}_*({q})=\int_S\Bigg(\underbrace{-\text{exp}\Big(-u_0\big(f(s)\big)\big/\lambda\Big)q(s)+\sum_{i=1}^n\ell_i\mathds{1}_{\beta_i>0} {p}_i(s)\text{ log}\frac{p_i(s)}{q(s)}+\ell_0q(s)}_{\psi(q|s)}\Bigg)d\nu(s).
    \end{align}
 Thus, we can apply techniques from calculus of variations to $\mathscr{L}_*({q})$. Specifically, since $\psi(q|s)$ in eq. (\ref{eq:lag2}) is continuous in $q$, a necessary condition for the existence of a solution $q^{f,\theta_{\lambda}}_{0}$ to eq. (\ref{eq:finalrep}) is that $q^{f,\theta_{\lambda}}_{0}$ must be a stationary point of the functional $\mathscr{L}_*({q})$ \citep{var00} and hence has to be a solution to the following Euler-Lagrange equation
\begin{align}\label{eq:EL}
    \nabla_{{q}}\psi-\frac{\partial}{\partial {s}}\big(\nabla_{{q}'}\psi\big)=0,
\end{align}
where ${q}':=\frac{\partial q}{\partial s}$. Since $\psi(q|s)$ in eq. (\ref{eq:lag2}) is not a function of ${q}'$, the second term in eq. (\ref{eq:EL}) vanishes (because $\nabla_{q'}\psi=0$). Thus, the Euler-Lagrange equation in eq. (\ref{eq:EL}) reduces to $\nabla_q\psi=0$, which can be solved uniquely as follows
\begin{align}
   \nabla_{q}\psi= -\text{exp}\big(-u_0(f)/\lambda\big)-\sum_{i=1}^n\ell_i\mathds{1}_{\beta_i>0}\frac{p_i}{q^{f,\theta_{\lambda}}_{0}}+\ell_0&=0\nonumber\\
      q^{f,\theta_{\lambda}}_{0}\ell_0-q^{f,\theta_{\lambda}}_{0}\text{exp}\big(-u_0(f)/\lambda\big)&=\sum_{i=1}^n\ell_i\mathds{1}_{\beta_i>0}p_i\nonumber\\
   q^{f,\theta_{\lambda}}_{0}&=\sum_{i=1}^n\frac{\ell_i\mathds{1}_{\beta_i>0}}{\ell_0+\phi_{\lambda}(u_0(f))}p_i,\label{eq:final3}
\end{align}
where the existence of $q^{f,\theta_{\lambda}}_{0}$ implies that the fraction in eq. (\ref{eq:final3}) is always well-defined.
 Further, since $q^{f,\theta_{\lambda}}_{0}$
 is an element of $\mathcal{Q}_{\theta}\subseteq \Delta_d$, it must be a valid density, so it follows that $$1=\int_Sq^{f,\theta_{\lambda}}_{0}d\nu=\sum^n_{i=1}\ell_i\mathds{1}_{\beta_i>0}\int_S\frac{1}{\ell_0+\phi_{\lambda}(u_0(f))}p_id\nu,$$ which holds whenever  each Lagrange multiplier $\ell_i$ satisfies, for $i=1,\dots,n$,
 $$\ell_i=\Big(\sum^n_{j=1}\mathds{1}_{\beta_j>0}\Big)^{-1}\Bigg(\int_S\frac{1}{\ell_0+\phi_{\lambda}(u_0(f))}p_id\nu\Bigg)^{-1},$$ so let $\mu^{\theta_{\lambda}}_i(f):=\frac{\ell_i\mathds{1}_{\beta_i>0}}{\ell_0+\phi_{\lambda}(u_0(f))}$ such that $q^{f,\theta_{\lambda}}_{0}=\sum_{i=1}^n\mu^{\theta_{\lambda}}_i(f)p_i$. 
If any $\mu^{\theta_{\lambda}}_i(f)$'s are negative in the optimal solution $\hat{q}=\sum_{i=1}^n\mu^{\theta_{\lambda}}_i(f)p_i$ in eq. (\ref{eq:final3}), we can always truncate them to zero and adjust the remaining nonnegative $\mu_j(f)$'s (for $j\neq i)$ such that $q^{f,\theta_{\lambda}}_{0}$ integrates to 1. This would result in a function that satisfies all the constraints and that is uniformly larger than $\hat{q}$ and therefore would reduce the objective function in eq. (\ref{eq:finalrep}), implying that $\hat{q}$ could not have been optimal, which is a contradiction. Thus, $\mu^{\theta_{\lambda}}_i(f)\geq0$ for all $i=1,\dots,n$.
\end{proof}

\subsection*{Proofs of Corollaries \ref{thm:constant}--\ref{thm:bound}}
Corollary \ref{thm:constant} is a special case of Theorem \ref{thm:pos}. Corollary \ref{thm:bound} is proved in Online Appendix \hyperref[app:proof]{F}.

\subsection*{Proof of Proposition \ref{thm:accurate}}

\begin{proof}
    We aim to solve the following optimization problem
    \begin{align}\label{eq:entrop}
        \underset{q\in \mathcal{Q}_{\theta}}{\text{\normalfont min}}\hspace{0.02in}R(p^*\lVert q).
    \end{align}
Observe that this minimization is reminiscent of Lemma \ref{thm:sol}, but the only difference is that the objective functions are swapped, i.e., $R(p^*\lVert q)$ versus $\E_{q}\big[\phi_{\lambda}(u_0(f))\big]$, respectively. Therefore, we need to check that eq. (\ref{eq:entrop}) satisfies all the conditions in Lemma \ref{thm:exist}. This follows immediately, since the relative entropy $R(p^*\lVert q)$ is a nonnegative, bounded, and strictly convex function in $q$, and hence it is continuous in $q$ \citep[e.g.,][Lemma 2.1]{calcvar76}. 
Thus, there exists a unique solution $q^*_{\sigma}\in \mathcal{Q}_{\theta}$ to eq. (\ref{eq:entrop}) by Lemma \ref{thm:exist}. Following exactly the outline of the second step of the proof of Lemma \ref{thm:sol}, the functional to be minimized here is
   \begin{align*}
         \mathscr{L}^*({q})=\int_S\Bigg(\underbrace{p^*(s)\text{log}\frac{p^*(s)}{q(s)}+\sum_{i=1}^n\ell_i\mathds{1}_{\beta_i>0} {p}_i(s)\text{ log}\frac{p_i(s)}{q(s)}+\ell_0q(s)}_{\psi^*(q|s)}\Bigg)d\nu(s).
    \end{align*}
Here, the Euler-Lagrange equation becomes $\nabla_q\psi^*(q|.)=0$, where the function $\psi^*(q|.)=p^*\text{log}\frac{p^*}{q}+\sum_{i=1}^n\ell_i\mathds{1}_{\beta_i>0}\text{log}\frac{p_i}{q}+\ell_0q$ is defined above. Solving this first-order condition yields
\begin{align}
   \nabla_q\psi^*= -\frac{p^*}{q^*_{\sigma}}-\sum_{i=1}^n\ell_i\mathds{1}_{\beta_i>0}\frac{p_i}{q^*_{\sigma}}+\ell_0=0\implies\
   q^*_{\sigma}=\frac{1}{\ell_0}p^*+\sum_{i=1}^n\frac{\ell_i\mathds{1}_{\beta_i>0}}{\ell_0}p_i,\label{eq:mod2}
\end{align}
so define $\varphi_0:=1/\ell_0$ and $\varphi_i:=\frac{\ell_i\mathds{1}_{\beta_i>0}}{\ell_0}$, for $i=1,\dots,n$. Since $q^*_{\sigma}\in \mathcal{Q}_{\theta}$, $1=\int_Sq^*_{\sigma}d\nu=\sum_{i=0}^n\varphi_i$, so the $\varphi_i$'s must sum to 1. To prove that these weights $\{\varphi_i\}_{i=0}^n$ must be nonnegative, we proceed as follows. Suppose to the contrary that some of the $\varphi_i$'s are negative in the optimal solution $\bar{q}=\varphi_0p^*+\sum_{i=1}^n\varphi_ip_i$ in eq. (\ref{eq:mod2}) with $\sum_{i=0}^n\varphi_i=1$. Then, we can always make $\bar{q}$ uniformly larger by restricting any negative weights to zero and renormalizing the remaining weights such that they sum to 1. This process would result in a density that satisfies
the constraints while reducing the objective function (eq. (\ref{eq:entrop})) and hence, the original solution $\bar{q}$ could not have been optimal, which is a contradiction. Thus, we conclude that $\varphi_i\geq0$ for all $i=0,\dots,n$. Then, let $\sigma:=\varphi_0$ and $1-\sigma=\sum_{i=1}^n\varphi_i$. Thus, we can define $\mu^{\sigma,\theta}_i:=\frac{\ell_i\mathds{1}_{\beta_i>0}}{(1-\sigma)\ell_0}\geq0$, for each $i=1,\dots,n$, which satisfy $\sum_{i=1}^n\mu^{\sigma,\theta}_i=1$, and now we can write $q^*_{\sigma}$ in eq. (\ref{eq:mod2}) as $q^*_{\sigma}=\sigma p^*(s)+(1-\sigma)q^{\theta}_\sigma,$
where $q^{\theta}_{\sigma}=\sum_{i=1}^n\mu^{\sigma,\theta}_ip_i$. When $\sigma=0$, $q^{\theta}_{\sigma}=q^{\theta}_0$. 
\end{proof}

\subsection*{Proof of Proposition \ref{thm:mindist}}
\begin{proof}
    Since this is a  minimization, it suffices to prove the result on the subset $\Delta_d^+\subseteq\Delta_d$---the set containing the models with respect to which all the $p_i$'s are absolutely continuous. This way $R(p_i\lVert q)\geq0$ is finite for all $q\in\Delta_d^+$ and all $i=1,\dots,n$. Just like in Theorem \ref{thm:rep} and Proposition \ref{thm:accurate}, we can now apply Lemma \ref{thm:exist} to conclude that there exists a unique solution $q_*\in\Delta_d^+$, since $\Delta_d^+$ is a convex and compact subset of the reflexive Banach space $\Delta_d$, and $R(p_i\lVert q)$ is bounded, strictly convex, and continuous in $q$, for all $i=1,\dots,n$. Then, the functional to be minimized in this case becomes
       \begin{align*}
         \bar{\mathscr{L}}({q})=\int_S\Bigg(\underbrace{\sum_{i=1}^n\mu^{\theta}_i {p}_i(s)\text{ log}\frac{p_i(s)}{q(s)}+\bar{\ell}q(s)}_{\bar{\psi}(q|s)}\Bigg)d\nu(s),
    \end{align*}
    so the Euler-Lagrange equation is $\nabla_q\bar{\psi}(q|.)=0$, where $\bar{\psi}(q|.)=\sum_{i=1}^n\mu^{\theta}_i {p}_i\text{ log}\frac{p_i}{q}+\bar{\ell}q$ is defined above, and $\bar{\ell}$ denotes the Lagrange multiplier for the constraint $\int_Sqd\nu=1$ since $q\in\Delta_d^+$. The solution to this first-order condition is
    \begin{align*}
        \nabla_q\bar{\psi}=-\sum_{i=1}^n\mu^{\theta}_i\frac{p_i}{q_*}+\bar{\ell}=0 \implies
        q_*=\frac{1}{\bar{\ell}}\sum_{i=1}^n\mu^{\theta}_ip_i.
    \end{align*}
    Since $q_*\in\Delta_d^+$, $\bar{\ell}=1$, so $q_*=\sum_{i=1}^n\mu^{\theta}_ip_i=q^{\theta}_0\in\mathcal{Q}_{\theta}$ in eq. (\ref{eq:model}) is the unique minimizer.
\end{proof}

\subsection*{Proof of Corollary \ref{thm:mindist2}}

\begin{proof}
    We prove this result as follows:
    \begin{align*}
        \sum_{i=1}^n\mu^{\theta}_iR(p_i\lVert q)&=\sum_{i=1}^n\mu^{\theta}_i\Bigg(\int_Sp_i\hspace{0.03in}\text{log}\hspace{0.03in}p_i\hspace{0.03in}d\nu-\int_Sp_i\hspace{0.03in}\text{log}\hspace{0.03in}q\hspace{0.03in}d\nu\Bigg)=-\sum_{i=1}^n\mu^{\theta}_iH(p_i)-\int_S\Bigg(\sum_{i=1}^n\mu^{\theta}_ip_i\Bigg)\text{log}\hspace{0.03in}q\hspace{0.03in}d\nu\\
        &=-\sum_{i=1}^n\mu^{\theta}_iH(p_i)-\int_Sq^{\theta}_0\hspace{0.03in}\text{log}\hspace{0.03in}q\hspace{0.03in}d\nu+\Bigg(\int_Sq^{\theta}_0\hspace{0.03in}\text{log}\hspace{0.03in}q^{\theta}_0\hspace{0.03in}d\nu-\int_Sq^{\theta}_0\hspace{0.03in}\text{log}\hspace{0.03in}q^{\theta}_0\hspace{0.03in}d\nu\Bigg)\\
        &=-\sum_{i=1}^n\mu^{\theta}_iH(p_i)+H(q^{\theta}_0)+R(q^{\theta}_0\lVert q),
    \end{align*}
  where $q^{\theta}_0=\sum_{i=1}^n\mu^{\theta}_ip_i$ from Proposition \ref{thm:mindist}.  Thus, $\sum_{i=1}^n\mu^{\theta}_iR(p_i\lVert q^{\theta}_0)=-\sum_{i=1}^n\mu^{\theta}_iH(p_i)+H(q^{\theta}_0)=0$  if and only if $H(q^{\theta}_0)=\sum_{i=1}^n\mu^{\theta}_iH(p_i)$, since $R(q^{\theta}_0\lVert q)=0$ if and only if $q=q^{\theta}_0$.
\end{proof}

\subsection*{Proof of Proposition \ref{thm:mono}}
\begin{proof}
 Individual $i$'s equality constraint in eq. (\ref{eq:const}) is
    $\eta_i=\int_{{S}} {p}_i\text{ log}\frac{p_i}{q}\hspace{0.03in}d\nu,$
 then we substitute the function $\bar{q}=\sum_{k=1}^n\mu_kp_k$ in place of ${q}$ above, where $\bar{q}$ is only known to satisfy the individual equality constraint in eq. (\ref{eq:const}) for all $i$ with $\sum_{k=1}^n\mu_k=1$, and obtain $\eta_i=\int_{{S}}{p}_i\text{ log}\frac{p_i}{\sum_{k=1}^n\mu_{k}p_k}\hspace{0.03in}d\nu.$
Let $\varphi_{ij}(s)=p_i(s)-p_j(s)$, for all $i\neq j$, so $\int_S\varphi_{ij}(s)d\nu(s)=0$. We can then write $\bar{q}(s)=p_j(s)+\sum_{k\neq j}\mu_k\varphi_{kj}$, and $p_i(s)=\bar{q}(s)+\varphi_{ij}(s)-\sum_{k\neq j}\mu_k\varphi_{kj}(s)\geq0$ $\forall s\in S$, which is nonnegative since each reference model $p_i\in \Delta_d$ is a valid density, so \begin{align}\label{eq:ineq}
    \bar{q}(s)\geq -\Big(\varphi_{ij}(s)-\sum_{k\neq j}\mu_k\varphi_{kj}(s)\Big).
\end{align} Since $|\eta_i-\bar{\eta}_i|<\bar{\delta}_i$, for some $\bar{\delta}_i<\infty$, we can take a derivative of $\eta_i$ with respect to $\mu_i$ to get
\begin{align*}
    \frac{\partial\eta_i}{\partial\mu_i}&=\frac{\partial}{\partial\mu_{i}}\Bigg[\int_{{S}}{p}_i\text{ log}\frac{p_i}{\sum_{k=1}^n\mu_{k}p_k}\hspace{0.03in}d\nu\Bigg]=-\int_{{S}}{p}_i\frac{\varphi_{ij}}{\bar{q}}\hspace{0.03in}d\nu=-\int_{{S}}\Big(\bar{q}+\varphi_{ij}-\sum_{k\neq j}\varphi_{kj}\Big)\frac{\varphi_{ij}}{\bar{q}}\hspace{0.03in}d\nu\\
    &=-\int_{{S}}\bar{q}\frac{\varphi_{ij}}{\bar{q}}\hspace{0.03in}d\nu-\int_{{S}}\Big(\varphi_{ij}-\sum_{k\neq j}\mu_k\varphi_{kj}\Big)\frac{\varphi_{ij}}{\bar{q}}\hspace{0.03in}d\nu\leq-\int_{{S}}\varphi_{ij}\hspace{0.03in}d\nu+\int_{{S}}\bar{q}\frac{\varphi_{ij}}{\bar{q}}\hspace{0.03in}d\nu=0,
\end{align*}
for any $j\neq i$, where the inequality holds by (\ref{eq:ineq}).  Thus, $\eta_i$ is nonincreasing in $\mu_i$, so its inverse is also nonincreasing at all points where its derivative exists, and hence
\begin{align}\label{eq:mono}
    \frac{\partial\mu_i}{\partial \eta_i}=\Big(\frac{\partial\eta_i}{\partial\mu_i}\Big)^{-1}\leq 0,
\end{align}
for each $i=1,\dots,n$. Since these weights $\{\mu_i\}_{i=1}^n$ must sum to 1, i.e., $\sum_{i=1}^n\mu_i=1$, we can take the derivative on both sides of this equation with respect to $\eta_i$ to get
$$ 0=\frac{\partial}{\partial \eta_i}\sum_{i=1}^n\mu_i=\frac{\partial}{\partial \eta_i}\Big(\mu_i+\sum_{j\neq i}\mu_{j}\Big)=\frac{\partial\mu_i}{\partial \eta_i}+\frac{\partial}{\partial \eta_i}\sum_{j\neq i}\mu_{j},$$
and thus $\frac{\partial}{\partial \eta_i}\sum_{j\neq i}\mu_{j}\geq0$ (by eq. (\ref{eq:mono})), i.e., the sum of remaining weights $\mu_j$'s must be increasing in $\eta_i$, for all $j\neq i$. Now, recall that $\bar{q}=\sum_{k=1}^n\mu_kp_k$ is assumed to satisfy all $n$ equality constraint in eq. (\ref{eq:const}). At one extreme, if $\eta_i=0$ (for some $i\in\{1,\dots,n\}$), then $R(p_i\lVert \bar{q})=0$, which holds if and only if $p_i=\bar{q}=\sum_{k=1}^n\mu_kp_k$, and hence $\mu_i=1$ and $\mu_k=0$ for all $k\neq i$. At the other extreme, if $\eta_i=\infty$, then 
$R(p_i\lVert \bar{q})=\infty$, which happens whenever $p_i$ is not absolutely continuous with respect to $\bar{q}=\sum_{k=1}^n\mu_kp_k$, so it must be that $\mu_i=0$ and $\sum_{k\neq i}\mu_k=1$. Since eq. (\ref{eq:mono}) shows that $\mu_i$ is monotonic in $\eta_i$ (holding fixed $\eta_j$ for each $j\neq i$), we therefore conclude that $\mu_i\in[0,1]$, for all $i=1,\dots,n$, and hence $\bar{q}=\sum_{i=1}^n\mu_ip_i\in \mathcal{Q}_{\theta}$.
\end{proof}

\subsection*{Proof of Proposition \ref{thm:optim}}

\begin{proof}
    The optimization in eq. (\ref{eq:opt}) can be written explicitly as follows
    \begin{align*}
        {F}_0=\underset{f\in F}{\text{ arg sup }}- \int_S\text{exp}\big(-u_0(f)/\lambda\big)\hspace{0.03in}q^{\theta}_0\hspace{0.03in}d\nu=\underset{f\in F}{\text{arg inf }}\underbrace{\int_S\text{exp}\big(-u_0(f)/\lambda\big)\hspace{0.03in}q^{\theta}_0\hspace{0.03in}d\nu}_{\mathcal{M}(f)},
    \end{align*}
    where $\lambda>0$ and $q^{\theta}_0$ is in eq. (\ref{eq:model}), and now the right-hand-side resembles the general minimization in eq. (\ref{eq:obj}). Since $u_i$ is strictly concave in $f$ for all $i=1,\dots,n$, then so is $u_0$ because it is a linear combination of the $u_i$'s with the nonnegative weights in (\ref{eq:socialrep}). Further, the function $\text{exp}(-x)$ is strictly convex and strictly decreasing in $x$, so $\text{exp}\big(-u_0(f)/\lambda\big)$ is strictly convex in $f$ (by strict concavity of $u_0$). Thus, $\mathcal{M}(f)$ is strictly convex and continuous in $f$, and is also a nonnegative and bounded function on $F$. We can therefore apply Lemma \ref{thm:exist} to establish that $\mathcal{M}(f)$ attains its minimum uniquely at some act $f_0$ in $F$. Thus, ${F}_0=\{f_0\}$ in eq. (\ref{eq:opt}), and therefore $f_0$ is admissible by \citet[][Proposition 8.(ii)]{hansenmiss22}.
    \par As in the proof of Lemma \ref{thm:sol}, we can now apply results from calculus of variations \citep{var00}. 
    Specifically, $f_0$ must be a solution to the Euler-Lagrange equation $\nabla_f\psi_*=0$, where $\psi_*(f|.)=\text{ exp}\big(-u_0(f)/\lambda\big)\hspace{0.03in}q^{\theta}_0$. Solving this equation yields
    \begin{align*}
       \nabla_f\psi_*=\frac{-u_0'(f_0)}{\lambda}\text{ exp}\big(-u_0(f_0)/\lambda\big)\hspace{0.03in}q^{\theta}_0=0 \implies \sum_{i=1}^n\beta_iu'_i(f_0)=0,       
    \end{align*}
    where we use the fact that $u_i$ is differentiable for any $i$ with $\beta_i>0$, with derivative $u'_i:=\nabla u_i$, and for all such $i$,  $p_i(s)>0$ for all $s\in S$, which implies that $q^{\theta}_0(s)>0$ $\forall s\in S$ in eq. (\ref{eq:model}). 
\end{proof}

\begin{singlespace}
\bibliography{ref}
\bibliographystyle{apalike}
\end{singlespace}

\clearpage

\phantomsection\label{app:online}
\begin{center}
{\LARGE\bf Online Appendix: \par ``Robust Aggregation of Preferences''}
\end{center}
The online appendix is organized as follows. Online Appendix \hyperref[sec:axiom]{A} provides a microfoundation of our welfare criterion. Online Appendix \hyperref[sec:app2]{B} presents more applications of our welfare criterion in Ellsberg experiments and discusses estimation of the social preference. Online Appendix \hyperref[sec:ext]{C} explores several extensions of our framework. 
Online Appendix \hyperref[app:mba]{D} describes the class of MBA preferences. Online Appendix \hyperref[app:dom]{E} illustrates the welfare-dominant belief in Desideratum \ref{des:welfare} (Section \ref{sec:impo1}). Lastly, Online Appendix \hyperref[app:proof]{F} collects the omitted proofs. 

\phantomsection\label{sec:axiom}
\section*{Online Appendix A: Microfoundation}
This appendix describes simple behavioral axioms that characterize our welfare criterion with respect to the individuals’ preferences. Unlike Section \ref{sec:model}, we will treat the profile of individuals' preferences as the only primitives and derive from them the welfare criterion in Observation \ref{thm:phi}. As noted before, however,  Pareto-type conditions with MBA preferences lead to impossibility results. To avoid these complications, our ensuing microfoundation abstracts from conflict of interests by focusing on the case where all individuals share the same utility function but have different beliefs \citep[e.g.,][]{gil11,agg21,miss24}. \par On one hand, let each individual $i=1,\dots,n$ have a $\phi$-divergence preference
\begin{align*}
V^{\lambda D_{\phi}}_{i}(f|p_i)=\underset{p\in\Delta}{\text{min}}\hspace{0.02in}\Big\{\E_p\big[u(f)\big]+\lambda D_{\phi}(p\lVert p_i)\Big\}\quad \forall f\in F,
\end{align*}
which were introduced as a special case of the weighted divergence in \citet[][eq. (16)]{mmr06}. Here, each individual $i$ has a unique structured model $p_i$, which is called a ``reference'' or ``baseline'' model.
However, each $i$ is concerned that her $p_i$ is misspecified,\footnote{\citet{miss23} find experimental evidence that people are willing to pay on average 8.4\% of their expected payoff to avoid being faced with ambiguity and an extra 5.3\% to avoid facing misspecification.} and this concern is quantified by a common parameter $\lambda \in(0,\infty]$. \par On the other hand, the social planner  has an arbitrary variational preference $\succsim_0$:
 \begin{align}\label{eq:var}
    V_{0}(f)=\underset{p\in\Delta}{\text{min}}\hspace{0.02in}\Big\{\E_p\big[u(f)\big]+c_0(p)\Big\}\quad \forall f\in F,
\end{align}
 where $c_0(.)$ is any convex, lower
semicontinuous, and grounded (achieves value zero) function, and $u(X)$ is unbounded in $\R$. Notice that without additional assumptions on eq. (\ref{eq:var}), the class of variational preferences is not enough to capture misspecification. This is because the set $\{p\in\Delta:c_0(p)=0\}$ consists of structured models, but they are not explicitly characterized within the general representation in eq. (\ref{eq:var}).
\par The following axiom will help us link the social preference to the individuals' preferences.
 \begin{axiom}[Unambiguous Pareto]\label{def:AUP}\normalfont
     For every $f \in F$ and $x \in X$, if $f \succsim_i x$ for
all $i =1,\dots,n$, then $f \succsim_0 x$. 
 \end{axiom}
This axiom is ``simple'' in the sense that it only requires the social planner and the
individuals to compare an arbitrary act $f$ with a constant act $x$, so it is not very ``cognitively demanding.'' It prescribes that if all individuals favor an ambiguous act over an unambiguous one,
then so should the social planner. It is weaker than the standard Pareto principle, which requires dominance with respect to all acts.  \citet[][]{miss24} also uses this axiom to aggregate the beliefs of individuals who have identical tastes but different sets of beliefs, and their preferences are represented by \citeauthor{hansenmiss22}'s (\citeyear[][eq. (1)]{hansenmiss22}) general criterion. \par The next result shows that a social planner with representation $V_0(f)$ in eq. (\ref{eq:var}) satisfies the axiom in Axiom \ref{def:AUP} whenever $V_0(f)$ coincides with our social criterion in eq. (\ref{eq:rep}).

\begin{proposition}\label{thm:axiom}
$\succsim_0$ satisfies  Unambiguous Pareto if and only if $V_0(.)=V^{\lambda D_{\phi}}_{0}(.|Q_0)$, where 
\begin{align*}
     V^{\lambda D_{\phi}}_{0}(f|Q_0)=\underset{p\in\Delta}{\text{\normalfont min}}\Big\{\E_p\big[u(f)\big]+\lambda\hspace{0.02in}\underset{q\in Q_0}{\text{\normalfont min}}\hspace{0.02in} D_{\phi}(p\lVert q)\Big\}\quad\forall f\in F, \quad \text{and}\quad Q_0\subseteq\text{\normalfont co}\big(\{p_1,\dots,p_n\}\big).
\end{align*}
\end{proposition}
 Proposition \ref{thm:axiom} is a special case of \citet[][Proposition 3.1]{miss24} where individuals have $\phi$-divergence preferences. It shows that respecting Unambiguous Pareto when individuals have $\phi$-divergence preferences is equivalent to the social welfare criterion in Observation \ref{thm:phi} with an aggregation of structured models. Moreover, the social planner inherits her concerns for misspecification (captured by $\lambda$) from individuals. Notice also that the restriction of social beliefs $Q_0\subseteq\text{co}\big(\bigcup_{i=1}^n\{p_i\}\big)$ in Proposition \ref{thm:axiom} is reminiscent of Observation \ref{thm:agg2}.
 \par A sharper characterization arises when we further impose a \textit{cautious} axiom.
   \begin{axiom}[Ambiguity Avoidance]\label{def:caution}\normalfont
   For every $f \in F$ and $x \in X$, if there exists an $i$ such that $x \succ_i f$, then $x \succ_0 f$. 
 \end{axiom}
\par  Ambiguity Avoidance implies a  high degree of caution when dealing with social uncertainty, hence its popularity in the decision-theory literature  \citep{gil10,util16,hansenmiss22}. It states that if at least one individual strictly prefers a constant act $x$ to an
uncertain act $f$, then so should the social planner. This behavior highlights that the social planner highly values each individual's probability assessment.
\begin{corollary}\label{thm:avoid}
    In Proposition \ref{thm:axiom}, if $\succsim_0$ also satisfies Ambiguity Avoidance and $D_{\phi}=R$, then
\begin{align*}
     V^{\lambda R}_{0}(f|Q_0)=\underset{1\leq i\leq n}{\text{\normalfont min}}\hspace{0.03in}\phi^{-1}_{\lambda}\Big(\E_{{p}_i}\big[\phi_{\lambda}(u(f))\big]\Big)\quad\forall f\in F,\quad \text{and}\quad Q_0=\text{\normalfont co}\big(\{p_1,\dots,p_n\}\big).
 \end{align*} 
\end{corollary}
 \par Due to Ambiguity Avoidance, the social planner worries more about ambiguity than all individuals combined because $\{p_1,\dots,p_n\}\subsetneq Q_0$.  This result continues to hold even when Unambiguous Pareto is replaced with the standard Pareto principle \citep{miss24}. 
 \begin{corollary}\label{thm:meu}
     In  Corollary \ref{thm:avoid}, if $\lambda =\infty$, then $ V^{\lambda R}_{0}(f|Q_0)=\underset{1\leq i\leq n}{\text{\normalfont min}}\hspace{0.03in}\E_{{p}_i}[u(f)]$, for all $f\in F$.
 \end{corollary} 
Section \ref{sec:discussion} shows that $\lambda=\infty$  in Corollary \ref{thm:avoid} corresponds to an MEU social planner. Thus, Corollary \ref{thm:meu} resembles closely \citet[][Theorem 2]{util16}---a social planner who aggregates SEU individuals has an MEU representation and evaluates each act according to the minimum expected social utility over the set of individuals' reference models. 
\par Let's illustrate another sense in which the structured models $Q$ in eq. (\ref{eq:rep2}) are important. Given any $x\succ_0y$, a \textit{bet} on any $A\subseteq S$ is the act $xAy$ that takes value $x$ if $s\in A$ and otherwise $y$. Notice that this definition extends to settings where $S$ is not finite (Appendix \hyperref[app:compact]{A}).
 \begin{definition}[Social confidence]\normalfont
A preference $\succsim$ with beliefs $Q$ satisfies \textit{social confidence} if, given any $x\succsim y$, $q(A)\geq q(B)$ for all $q\in Q$ implies $xAy\succsim xBy,$ for all $A,B\subseteq S$. \end{definition}
This is reminiscent of  Pareto dominance under ambiguity and captures \citeauthor{hill19}'s (\citeyear{hill19}) notion of ``credal statements,'' i.e., a decision maker with preference $\succsim$ has a higher degree of confidence in event $A$ than $B$ whenever there is \textit{unanimity} among all the beliefs in $Q$. 
\begin{obs}
   The preference $\succsim_{0,D_{\phi},Q}$  in Observation \ref{thm:phi} satisfies social confidence. 
 \end{obs}
 This is \citet[][Proposition 3]{hansenmiss22}. It states that our social planner believes the models in $Q$, although incorrect, are  \textit{useful} in the sense of George E. P. Box because she is willing to choose bets on events that they unanimously rank as more likely. 
\par  To summarize, we have shown how to link our social welfare criterion (eq. (\ref{eq:rep})) directly to the individuals' preferences using simple behavioral axioms, which highlights its necessity in social planning when all individuals are concerned about misspecification.

\phantomsection\label{sec:app2}
\section*{Online Appendix B: Additional Applications}
This appendix considers three additional applications of our framework. 
Online Appendix \hyperref[sec:ells]{B.I} revisits a classic two-color \citeauthor{ellsb61}'s (\citeyear{ellsb61}) urn experiment to observe our social planner's behavior in the canonical environment of ambiguous decision-making.  Online Appendix \hyperref[sec:measure]{B.II} uses this experiment to discuss the empirical measurement of all the parameters in our criterion using a ``revealed-preference'' method, which may be useful in applied settings where numerical values of parameters are needed. Online Appendix \hyperref[sec:asset]{B.III} considers a manager of a financial institution who wishes to price an asset by choosing a stochastic discount factor. Before doing so, she seeks the advice of several investors (or stakeholders) who have diverse beliefs about the relevant market forces. We show that our criterion is an extension of the popular \citeauthor{hansen97}'s (\citeyear{hansen91,hansen97}) distance, and its special cases coincide with prominent aggregations in econometrics and finance  \citep[][]{asset21}.

\phantomsection\label{sec:ells}
\subsection*{B.I: Ellsberg Paradox in Social Choice}

\par This application illustrates the betting behavior of our social planner in a classic Ellsberg's two-color urns \citep{ellsb61}. Within each urn, the standard multiplier preference is known to coincide with SEU, but \citet{strz11} shows that the former is a good model of what happens across the urns. This section takes his insights a step further by showing that the entropic preference in eq. (\ref{eq:rep}) is a good model to aggregate what happens across individuals. Here, let  $f_s:=f(s)$ be a finite-valued function, $f:S\rightarrow X$, where $X:=\Delta(Z)$ is the set of all simple probability distributions on the set of monetary payoffs $Z\subseteq\R$, and elements of $X$ are called \textit{lotteries}  \citep{ansc63}.
 
\par Consider two urns containing colored balls. Urn I contains 100 red and black balls in unknown proportion, while Urn II is known to contain 50 red and 50 black balls. The social planner has to bet on the color of the ball drawn from each urn.\footnote{This choice environment should be interpreted as a canonical (or abstract) representation of any situation involving ambiguous collective decision-making such as the AI-regulation problem in the Introduction.} The outcome of her bet will constitute the outcome for $n=2$ individuals, i.e., she is betting on their behalf. Although this environment may seem abstract, it is often used in social choice as a canonical representation of situations involving collective ambiguity \citep[e.g.,][]{ells22}. 

\begin{remark}[Predictions]\label{rem:ells}\normalfont \citet{ellsb61} made three key observations in this situation:
    \begin{enumerate}
    \item[(1)] Most people are indifferent between betting on red from Urn I and on black from Urn I. This suggests that they view these two contingencies as interchangeable in the absence of evidence against symmetry.
    \item[(2)] Most people are indifferent between betting on red from Urn II and on black from Urn II; this preference can be justified by their knowledge of the composition of Urn II.
    \item[(3)] Most people strictly prefer betting on red from Urn II to betting on red from Urn I, thereby displaying ambiguity aversion. 
\end{enumerate}
\end{remark}
 Our framework is consistent with these insights when each MBA coefficient $\alpha_i(f)$ (eq. (\ref{eq:mba})) captures ambiguity aversion, e.g., $\alpha_i(f)=1$, for all $f$, which is MEU. Thus, \citet{util16} argue that it is desirable that a social planner displays ambiguity aversion since the individuals will bear the outcomes of her bets. We now show that our social planner represented by eq. (\ref{eq:repclose}) is consistent with this since she is averse to ambiguity. \par Let $S=\{R,B\}$, where $R$ and $B$ denote a red and black ball being drawn from Urn I, respectively, and let $\delta_z$ denote the lottery  paying off $z\in Z\subseteq\R$ with probability 1. Then, betting \$100 on red from Urn I corresponds to an act $f_R=(\delta_{100}, \delta_0)$, whereas betting \$100 on black from Urn I corresponds to an act $f_B=(\delta_0, \delta_{100})$. In contrast, betting \$100 on red from Urn II corresponds to a lottery $\pi_R= \frac{1}{2} \delta_{100}+\frac{1}{2} \delta_{0}$, while betting \$100 on black from Urn II corresponds to a lottery $\pi_B= \frac{1}{2} \delta_{0}+\frac{1}{2} \delta_{100}$, and hence $\pi_R=\pi_B$. 
\par Let $V^{\lambda R}_{0}(.):=V^{\lambda R}_{0}(.|\mathcal{Q}_{\theta})$ be the criterion in eq. (\ref{eq:repclose}) and set $\mu_i:=\mu^{\theta}_i$.  By Remark \ref{rem:ells}.(2), individuals' beliefs about the composition of Urn II will agree, so the social criterion satisfies $V^{\lambda R}_{0}(\pi_R)=V^{\lambda R}_{0}(\pi_B)=\phi_{\lambda}\big(\frac{1}{2}u_0(100)+\frac{1}{2}u_0(0)\big)$, for all $\mu_i\in[0,1]$. However, there need not be such individual-level agreements in Urn I, so the social criteria for $f_R$ and $f_B$ are
\begin{align*}
V^{\lambda R}_{0}(f_R)&=\mu_1\big[p_{1}\phi_{\lambda}(u_0(100))+(1-p_{1})\phi_{\lambda}(u_0(0))\big]+\mu_2\big[q_{2}\phi_{\lambda}(u_0(100))+(1-q_{2})\phi_{\lambda}(u_0(0))\big],\\
    V^{\lambda R}_{0}(f_B)&=\mu_1\big[(1-p_{1})\phi_{\lambda}(u_0(100))+p_{1}\phi_{\lambda}(u_0(0))\big]+\mu_2\big[(1-q_{2})\phi_{\lambda}(u_0(100))+q_{2}\phi_{\lambda}(u_0(0))\big],
\end{align*}
where $p_i$ denotes $i$'s reference model of the probability that the ball drawn from Urn I is red.
\par On one hand, suppose the two individuals' reference models completely agree (or coincide) in Urn I, i.e., $p_1=p_2\in[0,1]$. In this case, the social criterion in eq. (\ref{eq:repclose}) reduces to the standard multiplier criterion as in \citet{strz11}. By indifference in Urn I (Remark \ref{rem:ells}.(1)), $V^{\lambda R}_{0}(f_R)=V^{\lambda R}_{0}(f_B)$, which implies $p_1=p_2=1/2$ for all $\mu_i\in[0,1]$, and hence $V^{\lambda R}_{0}(f_R)=V^{\lambda R}_{0}(f_B)=\frac{1}{2}\phi_{\lambda}(u_0(100))+\frac{1}{2}\phi_{\lambda}(u_0(0))$. Then, $\pi_R\sim_0\pi_B\succ_0 f_B\sim_0 f_R$ follows by Jensen's inequality, for all $\lambda<\infty$, $\beta_i\geq0$, and $\gamma\in\R$ in (\ref{eq:socialrep}). Thus, the social planner, who is betting on behalf of the two individuals, prefers risky bets over probabilistically equivalent uncertain bets, which is consistent with Ellsberg's prediction in Remark \ref{rem:ells}.(3).
\par On the other hand, suppose the two individuals' reference models completely disagree in Urn I, i.e., $p_1=1-p_2\in[0,1]$, so our criterion differs from the standard multiplier criterion. From Remark \ref{rem:ells}.(1), each individual would be indifferent between betting on red or black in Urn I, and hence the same holds for the social planner by unambiguous Pareto dominance, i.e., $V^{\lambda R}_{0}(f_R)=V^{\lambda R}_{0}(f_B)$, which implies $\mu_i=1/2$, so $V^{\lambda R}_{0}(f_R)=V^{\lambda R}_{0}(f_B)=\frac{1}{2}\phi_{\lambda}(u_0(100))+\frac{1}{2}\phi_{\lambda}(u_0(0))$. It therefore follows again by Jensen's inequality that $\pi_R\sim_0\pi_B\succ_0 f_B\sim_0 f_R$, for all $\lambda<\infty$, $\beta_i>0$, and $\gamma\in\R$. In this case, we have also deduced that $\beta_i>0$ in the social utility $u_0$ ((\ref{eq:socialrep})) because $\mu_i=1/2>0$, for $i=1,2$. The fact that $\mu_i=1/2$ is intuitive because it indicates that the social planner's optimal utilitarian weighting rule to deal with reference models that completely disagree is simply the 50:50 rule.
\par Thus, whether or not individuals' beliefs agree in Urn I, the bets of a social planner represented by eq. (\ref{eq:repclose}) remain cautious and hence are robust to individuals' disagreements. This cautious behavior is not a coincidence because it can be formalized as a behavioral axiom that links our social criterion to the individuals’ preferences (Online Appendix \hyperref[sec:axiom]{A}).

\phantomsection\label{sec:measure}
\subsection*{B.II: Empirical Measurement of Parameters}
The Ellsberg experiment above provides a simple choice environment to infer the parameters in our criterion. We now build on \citet{strz11} to show that the intensity of the preference for betting on Urn I versus Urn II---the premium the social planner is willing to pay to switch between these two bets---is related to the parameter $\lambda\in(0,\infty]$ in eq. (\ref{eq:repclose}). 
\par Suppose each individual $i$, for $i=1,2$, has a constant relative risk aversion utility function $u_i(x)=(\omega_i+x)^{1-\varphi_i}$, with initial wealth denoted $\omega_i$. Let $c_i$ denote individual $i$'s \textit{certainty equivalent} of $\pi_R$ and $\pi_B$, i.e.,  the amount of money that, when received for sure, would make individual $i$ indifferent to $\pi_R$ and $\pi_B$. Formally, $c_i$ solves $$(\omega_i+c_i)^{1-\varphi_i}=\frac{1}{2}(\omega_i+100)^{1-\varphi_i}+\frac{1}{2}\omega^{1-\varphi_i}_i,$$ and let $\widehat{\varphi}_i:=\varphi_i(c_i)$ denote the solution to this equation, for $i=1,2$, so the individual curvature parameter $\varphi_i$ can be computed using the observed value of $i$'s certainty equivalent $c_i$. To infer the remaining parameters in the social utility $u_0$ in (\ref{eq:socialrep}), assume for simplicity that $\gamma=0$ and $\beta_1+\beta_2=1$  so that $u_0(x)=\sum_{i=1}^2\beta_i u_i(x)$. Following Online Appendix \hyperref[sec:ells]{B.I}, the social planner is represented by the criterion in eq. (\ref{eq:repclose}) with $\lambda<\infty$. Since $V^{\lambda R}_{0}(\pi_R)=V^{\lambda R}_{0}(\pi_B)=\phi_{\lambda}\big(\frac{1}{2}u_0(100)+\frac{1}{2}u_0(0)\big)$ in Urn II, let $c_0$ denote the social planner's certainty equivalent of $\pi_R$ and $\pi_B$, which solves
$$\sum_{i=1}^2\beta_i(\omega_i+c_0)^{1-\widehat{\varphi}_i}=\frac{1}{2}\sum_{i=1}^2\beta_i(\omega_i+100)^{1-\widehat{\varphi}_i}+\frac{1}{2}\sum_{i=1}^2\beta_i\omega^{1-\widehat{\varphi}_i}_i,$$
so let $\widehat{\beta}_i:=\beta_i(c_0,c_1,c_2)$ denote the solution to this equation, where $\widehat{\beta}_1+\widehat{\beta}_2=1$. That is, the values of the individuals' and social planner's certainty equivalents of $\pi_R$ and $\pi_B$ facilitate the computation of the social utility weights $\beta$ in (\ref{eq:socialrep}).
\par To infer $\lambda$, consider the case in Online Appendix \hyperref[sec:ells]{B.I} where individuals' beliefs completely disagree in Urn I, i.e., $p_1=1-p_2\in[0,1]$. We deduced that $\mu_i=1/2$ in this case, so $q_0=\frac{1}{2}(p_1+p_2)=\frac{1}{2}$ and hence $V^{\lambda R}_{0}(f_R)=V^{\lambda R}_{0}(f_B)=\frac{1}{2}\phi_{\lambda}(u_0(100))+\frac{1}{2}\phi_{\lambda}(u_0(0))$. Then, let $\tau$ denote the social planner's certainty equivalent of $f_R$ and $f_B$, i.e.,  the amount of money that, when received for sure, would make her indifferent to $f_R$ and $f_B$. Formally, $\tau$ solves
\begin{align*}
    \phi_{\lambda}\Bigg(\sum_{i=1}^2\widehat{\beta}_i(\omega_i+\tau)^{1-\widehat{\varphi}_i}\Bigg)=\frac{1}{2}\phi_{\lambda}\Bigg(\sum_{i=1}^2\widehat{\beta}_i(\omega_i+100)^{1-\widehat{\varphi}_i}\Bigg)+\frac{1}{2}\phi_{\lambda}\Bigg(\sum_{i=1}^2\widehat{\beta}_i\omega^{1-\widehat{\varphi}_i}_i\Bigg),
\end{align*}
and let $\widehat{\lambda}:=\lambda(\tau,c_0,c_1,c_2)$ denote the solution to this equation. Thus, the observed value $\tau$ of the social planner's certainty equivalent of $f_R$ and $f_B$ along with all the values $\{c_i\}_{i=0}^2$ of the certainty equivalents of $\pi_R$ and $\pi_B$  make it possible to compute $\lambda$. These insights continue to hold even when the individuals' beliefs agree in Urn I, i.e., $p_1=p_2$ (Online Appendix \hyperref[sec:ells]{B.I}).
\begin{remark}[Dimensionality]\normalfont
    This application indicates the following estimation challenge at the societal level. The larger the society, the higher-dimensional the social weights $\beta$ in $u_0$ ((\ref{eq:socialrep})) becomes, so more individuals' certainty equivalents need to be elicited to infer all the parameters in our social criterion (eq. (\ref{eq:repclose})). As shown in this application, dimensionality can be reduced when the social planner sets $\gamma=0$ and $\sum_{i=1}^n\beta_i=1$ in $u_0$. 
\end{remark}
In summary, the above analysis outlines a ``revealed preference'' method that can be used in practice to estimate or infer---from observed choice data---our social planner's utility function and all other (behavioral) parameters in our multiplier welfare criterion.

\phantomsection\label{sec:asset}
\subsection*{B.III Asset Pricing}
 We build on \citet{asset21} to illustrate how to apply our framework for asset pricing. Consider a financial institution composed of $n$ investors (or stakeholders) and a manager. The latter wishes to price an asset, so she consults the investors before making a decision. The manager's goal is to choose a stochastic discount factor (SDF) $m\in F$ that prices the asset correctly, i.e.
$\E_p[v_m]=a,$
where $v_m=Wm$, $a$ denotes a nonzero payoff, $W$ denotes the returns on the asset, and we follow \citet{asset21} by omitting time indices to ease notation. If $W$ is gross returns, then $a=1$. \par Any candidate SDF $f(s)$ is a function of an unknown state $s\in S$.  
The investors are allowed to have diverse preferences over the SDFs. However, they all wish to price the asset correctly such that the pricing error, denoted $\xi_p(f)$, of the asset is zero, i.e.,
$\xi_p(f):=\E_p[v_f]-a=0.$ When this equality does not hold, the SDF $f$ is said to be misspecified, which could cause major financial losses, so investors may not have conflict of interests. However, the investors have conflicting beliefs about the state, so the manager has to aggregate these beliefs. Meanwhile, the behavioral finance literature has raised concerns for misspecification because investors are prone to several psychological biases. 
\par To fix ideas, a natural starting point is the case when $n=1$. Let $p_1$ denote investor 1’s reference model of the empirical model. Then, \citet[][Section 3.1]{asset21} shows that the information-theoretic version of \citeauthor{hansen97}'s (\citeyear{hansen91,hansen97}) distance is to find a model $p$ with minimal entropy divergence from $p_1$, i.e.,
\begin{align}\label{eq:entropy}
    \underset{p\in\Delta}{\text{ min }}R(p\lVert p_1)\quad \text{s.t.}\quad \xi_p(f)=0,
\end{align}
\citep[][eq. (22)]{asset21}, i.e., the minimizer is a model consistent with the asset pricing restrictions.  Next, we extend this idea to the case where there are $n>1$ investors who have conflicting beliefs about the empirical distribution. 
\par Now, let $n>1$ and each investor  $i$'s set of beliefs be $P_i=\varGamma_{\eta_i}(p_i)$, for $i=1,\dots,n$, which have a nonempty intersection $\mathcal{Q}_{\theta}$, i.e., the set of investors' plausible beliefs is nonempty. The manager, $i=0$, on the other hand, is represented by our criterion in eq. (\ref{eq:rep}), so the optimization in eq. (\ref{eq:entropy}) can be extended using our insights to
\begin{align*}
    \underset{q\in \mathcal{Q}_{\theta}}{\text{ min }} \underset{p\in\Delta}{\text{ min }}R(p\lVert q)\quad \text{s.t.}\quad \xi_p(f)=0,
\end{align*}
where the double minimization is now reminiscent of the squared \citeauthor{hansen97}'s (\citeyear[][eq. (52)]{hansen97}) distance, and hence the above can be viewed as its information-theoretic extension when $n>1$ \citep[][eq. (20)]{asset21}. To apply Corollary \ref{thm:constant}, let's assume $\mathcal{Q}_{\theta}=\{q^{\theta}_0\}$, for $q^{\theta}_0=\sum_{i=1}^n\mu^{\theta}_ip_i$ in eq. (\ref{eq:model}), which coincides with the aggregation presented in \citet[][eq. (15)]{asset21}. Then, we need to solve  
\begin{align*}
\underset{p\in\Delta}{\text{ min }}R\big(p\lVert q^{\theta}_0\big)\quad \text{s.t.}\quad \xi_p(f)=0.
\end{align*}
For all $f\in F$ and $\theta_{\ell}\in\R^{2n+1}_+$, the unique solution to the minimization above is
$$p^{f,\theta_{\ell}}_0(s)=\frac{e^{-v_f/\ell}}{\E_{q^{\theta}_0}\big[e^{-v_f/\ell}\big]}\hspace{0.02in}q^{\theta}_0(s) \quad \forall s\in S$$
\citep[see,][Proposition 1.4.2]{dupuis97}, where the constant $\ell$ is uniquely identified by the constraint $\xi_{p^{f,\theta_{\ell}}_0}(f)=0$. Thus, as in \citet[][Section 3.1]{asset21}, each SDF $f$ defines a unique probability distribution $p^{f,\theta_{\ell}}_0$.  \citeauthor[][]{asset21}'s (\citeyear[][eq. (7)]{asset21}) aggregation procedure is the minimization in Proposition \ref{thm:mindist}, where the relative entropy is replaced with the generalized entropy divergence discussed in Online Appendix \hyperref[sec:extballs]{C.I}.
\phantomsection\label{sec:ext}
\section*{Online Appendix C: Some General Extensions}

Online Appendix \hyperref[sec:extballs]{C.I} considers an extension of the analysis in Section \ref{sec:tension}--\ref{sec:prop} when individuals' sets of beliefs are more general than entropy balls. Alternatively, Online Appendix \hyperref[sec:extpar]{C.II} explores settings where the social planner is allowed to have her own subjective beliefs. 

\phantomsection\label{sec:extballs}
\subsection*{C.I: General Sets of Beliefs}

This online appendix extends our main results to settings where the individuals' sets of beliefs are not necessarily entropy balls. The resulting aggregation of beliefs will no longer be utilitarian, but instead, it will resemble some aggregations that are popular in econometrics \citep{asset21} and in the inequality literature \citep{maa86}. 

\par We start by introducing a broad family of divergences called ``$\rho$-divergences'' \citep{zhang06}, which encompasses the relative entropy and other popular divergences.
\begin{definition}\label{def:alpha}\normalfont
\citet[][eq. (4)]{zhang06} defines the family of $\rho$-divergences as
\begin{align*}
    D_{\rho}(p\lVert q)=\frac{1}{\rho(1-\rho)}\E_{p}\Bigg[1-\Big(\frac{q}{p}\Big)^{\rho}\Bigg],
\end{align*}
for any constant $\rho\in(0,1)$, and any $p,q\in\Delta_d$.    
\end{definition}
\citet[][]{zhang06} shows that the $\rho$-divergence is closely related to the R\'enyi divergence---a popular divergence in statistics. For some examples, $\rho\rightarrow0$ corresponds to the relative entropy, i.e., $R(p\lVert q)=\underset{\rho\rightarrow0}{\text{lim}}\hspace{0.02in}D_{\rho}(p\lVert q)$, and the $\rho=1/2$ corresponds to the square of the Hellinger divergence, which is also popular in statistics. If we made a change of variables with respect to $\rho$, as $\rho=-\kappa$ for $\kappa\in\R$, the $\rho$-divergence
 would coincide with the so-called generalized entropy divergence considered in \citet[][eq. (8)]{asset21} and  \citet[][eq. (1a)]{maa86} in the study of multi-dimensional inequality. Under this change of variables, $\rho=-1$ would correspond to the relative entropy. Hence, we focus on the $\rho$-divergence, for $\rho\in(0,1)$, in what follows.
 \par Each $i$'s entropy ball $\varGamma_{\eta_i}(p_i)$ in eq. (\ref{eq:const}) can now be generalized to the following ball
\begin{align}\label{eq:const2}
    \mathcal{D}^{\rho}_{\tau_i}(p_i)=\Big\{{q}\in \Delta_d: D_{\rho}({p}_i\lVert{q})\leq \tau_i \Big\},
\end{align}
which is a closed and convex set, where $\tau_i\geq0$ is the analogue of the radius $\eta_i$, for $i=1,\dots,n$. For parsimony, we fix $\rho$ for all $i=0,\dots,n$, otherwise this framework would feather $n$ extra parameters. The preference of an MEU individual $i$ with set of beliefs defined by $\mathcal{D}^{\rho}_{\tau_i}(p_i)$ is analogous to the so-called ``divergence preference.'' For notation, the intersection of individuals' balls from Observation \ref{thm:agg} becomes $\mathcal{Q}^{\rho}_{\theta}:=\bigcap_{i=1,\beta_i>0}^n\mathcal{D}^{\rho}_{\tau_i}(p_i)$, where, in this notation, we write $\theta=(\beta,\tau)\in\R^{2n}_+$.
\par We recall that the main step in the proof of Theorem \ref{thm:rep} was the inner minimization (or projection) over $\mathcal{Q}_{\theta}$ since the outer minimization over $\Delta_d$ in eq. (\ref{eq:rep}) can be handled by applying techniques from \citet[][Proposition 1.4.2]{dupuis97}. This indicates that the main result to generalize is Proposition \ref{thm:accurate}, whose corresponding minimization becomes
\begin{align}\label{eq:min2}
    \underset{{q}\in\mathcal{Q}^{\rho}_{\theta}}{\text{min}}\hspace{0.05in}D_{\rho}(p^*\lVert {q}),
\end{align}
where the truth $p^*\in\Delta_d$ is absolutely continuous with respect to all the models in $\mathcal{Q}^{\rho}_{\theta}$. The next result is the analogue of Proposition \ref{thm:accurate} in this more general setting.
\begin{proposition}\label{thm:min2}
    There exists a unique solution $q_\rho\in\mathcal{Q}^{\rho}_{\theta}$ to the minimization in eq. (\ref{eq:min2}):
    \begin{align*}
        q_\rho\propto\Bigg(\sum_{i=1}^{n+1}\sigma_i p_{i}^{1-\rho}\Bigg)^{\frac{1}{1-\rho}},
    \end{align*}
    where $p_{n+1}:=p^*$ and the $\sigma_i$'s are some constants such that $\int_S q_\rho d\nu=1$, for any $\rho\in(0,1)$.\end{proposition}

The optimal belief $q_\rho$ is non-utilitarian and is less tractable compared to the utilitarian belief $q_0$ in eq. (\ref{eq:model}). When $\sigma_{n+1}=0$ (the weight associated with $p^*$), $q_\rho$ coincides with the aggregations in \citet[][eq. (9)]{asset21} and \citet[][eq. (5)]{maa86} after applying a change of variable with respect to $\rho$. When $\rho\rightarrow0$, it follows that $\mathcal{Q}^{\rho}_{\theta}\rightarrow\mathcal{Q}_{\theta}=\bigcap_{i=1,\beta_i>0}^n\varGamma_{\eta_i}(p_i)$ and hence $q_{\rho}\rightarrow q^*_{\sigma}$, where $q^*_{\sigma}$ is the solution from Proposition \ref{thm:accurate}, so all our main insights can be recovered from this general framework whenever $\rho\rightarrow0$.

We conclude with an analogue of our comparative statics results in Proposition \ref{thm:mono}.

\begin{proposition}\label{thm:mono2}
   For each $i$, $\sigma_{i}$ in Proposition \ref{thm:min2} decreases monotonically in $\tau_{i}$ for all $\rho$.
\end{proposition}
This result indicates that although the weights $\sigma_i$ in Proposition \ref{thm:min2} are nearly arbitrary, each one decreases whenever the corresponding radius $\tau_i$ increases. This therefore shows that our social planner always favors more confident (or knowledgeable) individuals regardless of the form of their sets of beliefs. Another extension of our framework is to allow each radius $\tau_i(f)$ to depend on act $f\in F$, which would capture \citeauthor{hill13}'s (\citeyear{hill13}) insights suggesting that acts involve various stakes, so beliefs should be considered depending on confidence levels. This extension requires a much more detailed analysis, so it is left for future research.

\phantomsection\label{sec:extpar}
\subsection*{C.II: Subjective Social Belief}
We now explore a setting where the social planner has her own subjective belief $p_0:=p_0(s|\vartheta)$ that is parameterized by a vector of parameters $\vartheta\in \Pi_0\subseteq\R^k$.  
She does not fully trust $p_0$, however, so she consults $n$ individuals and wishes to find a belief $q$ that minimizes $R(p_0\lVert q)$ over $\Pi_0$ subject to the individuals' constraint preferences in the spirit of Proposition \ref{thm:accurate}. Specifically, her goal is  to solve the following minimization problem
\begin{align*}
    \underset{\vartheta\in \Pi_0}{\text{min}}\hspace{0.03in}R\big(p_0(.|\vartheta)\big\lVert q\big) \quad \text{s.t.} \quad q\in\mathcal{Q}_{\theta},
\end{align*}
where we recall that $\mathcal{Q}_{\theta}=\bigcap_{i=1,\beta_i>0}^n\varGamma_{\eta_i}(p_i)$ is the intersection of the individuals' entropy balls. The corresponding Lagrangian, which is analogous to that of Proposition \ref{thm:accurate}, becomes
\begin{align*}
    \mathscr{L}_{\vartheta}(q)=R(p_0\lVert q)+\sum_{i=1}^n\mathds{1}_{\beta_i>0}\ell_i\big(R(p_0\lVert q)-\eta_i\big)+\ell_0\Big(\int_Sq\hspace{0.03in}d\nu-1\Big),
\end{align*}
where the $\ell_i$'s denote the Lagrange multipliers. After simplifying this Lagrangian, we get that the minimization above is equivalent to the maximization of the following function
\begin{align*}
    \mathscr{V}_{\vartheta}(q)&=\int_Sp_0(s)\hspace{0.03in}\text{log}\hspace{0.03in}q(s)\hspace{0.03in}d\nu(s)+\sum_{i=1}^n\ell_i\mathds{1}_{\beta_i>0} \int_Sp_i(s)\hspace{0.03in}\text{log}\hspace{0.03in}q(s)\hspace{0.03in}d\nu(s)-\ell_0\int_Sq(s)d\hspace{0.03in}\nu(s)\\
    &=\sum_{i=0}^n\tau_i \int_Sp_i(s)\hspace{0.03in}\text{log}\hspace{0.03in}q(s)\hspace{0.03in}d\nu(s)-\ell_0\int_Sq(s)\hspace{0.03in}d\nu(s),
\end{align*}
where $\tau_0=1$ and $\tau_i=\mathds{1}_{\beta_i>0}\ell_i$, for $i=1,\dots,n$. From Proposition \ref{thm:accurate}, we recall that that $\ell_0$ must
have the same sign as every one of the multipliers $\ell_i$'s as well as $1$ implying that all these
multipliers are nonnegative. The optimal solution therefore maximizes the function
\begin{align*}
    \sum_{i=0}^n\tau_i \int_Sp_i(s)\hspace{0.03in}\text{log}\hspace{0.03in}q(s)\hspace{0.03in}d\nu(s),
\end{align*}
since the $\tau_i$'s are all nonnegative, and recall that $p_0(s):=p_0(s|\vartheta)$. Suppose now that $q(s):=p_0(s|\vartheta^*)$, where $\vartheta^*$ denotes the unknown parameter of interest to the social planner. Then, in order to solve this optimization problem using existing methods, we need the following standard assumptions to hold. For notation, let $\mathscr{Q}$ denote the space spanned by $\{p_1,\dots,p_n\}$ with respect to the inner product $\langle \overline{q},\underline{q}\rangle=\int_S\overline{q}(s)\underline{q}(s)\hspace{0.03in}d\nu(s)$.
\begin{enumerate}
    \item Given all the individuals' constraint preferences specified in $\mathcal{Q}_{\theta}$, the map $\vartheta\mapsto R\big(p_0(.|\vartheta)\big\lVert q\big)$ has a unique minimum, denoted $\vartheta_0\in\Pi_0$.
    \item The derivative of $\text{\normalfont log}\hspace{0.03in}p_0(.|\vartheta)$ with respect to $\vartheta$ exists a.e. [$\nu$] and can be taken inside the integral sign in $R(p_i\lVert q)$ for $i=1,\dots,n$.
    \item For $j=1,\dots,k$, $\frac{\partial \text{log}\hspace{0.03in}p_0(.|\vartheta^*)}{\partial \vartheta^*_{j}}$ does not lie in the hyperplane of functions that are orthogonal to any non-null element of $\mathscr{Q}$.
\end{enumerate}

\begin{proposition}
Under the assumptions above, the
unique optimal parameter $\vartheta_0$ satisfies 
\begin{align*}
        \vartheta_0=\underset{\vartheta\in\Pi_0}{\text{\normalfont arg max}}\hspace{0.03in} \sum_{i=0}^n\tau_i \int_S\text{\normalfont log}\hspace{0.03in}p_0(s|\vartheta)\hspace{0.03in}d\vartheta_i(s).
    \end{align*}
\end{proposition}

This result follows by the standard Lagrange argument \citep[e.g.,][Section 2]{beavis90}. For the rest of this online appendix, we discuss how estimation can be performed. \par Suppose each individual (including the social planner) observes independent and
identically distributed samples $s_{i1},\dots,s_{im_i}$ drawn from the density $p_i$, for $i=0,\dots,n$. That is, each $s_{ij}$, for  $j =
1,\dots, m_i$, is assumed to have a density function $p_i$, for $i = 0,\dots, n$.  Each observation may be a vector, but they all have the same dimension. Further, we also assume
the samples observed by different individuals are independent of each other. For notation, let $\mb{s}_i =
(s_{i1},\dots,s_{im_i})$ and $\mb{s} = (\mb{s}_1,\dots,\mb{s}_n)$. Then, following \citeauthor{akaike77}'s (\citeyear{akaike77}) approach, the social planner can estimate $\vartheta^*$ by seeking the parameter value $\vartheta$ that maximizes 
\begin{align*}
       \hspace{0.03in} \sum_{i=0}^n\tau_i \int_S\text{\normalfont log}\hspace{0.03in}p_0(s|\vartheta)\hspace{0.03in}d\widehat{\vartheta}_i(s),
    \end{align*}
where $\widehat{\vartheta}_i$ denotes $i$'s empirical distribution function for $i=0,\dots,n$. Then,  for a realization of the random sample $\mb{s}$, the
weighted likelihood (WL), denoted $\mathscr{W}$, can be written as
\begin{align*}
    \mathscr{W}(\vartheta|\mb{s})=\prod_{i=0}^n\prod_{j=0}^{m_i}p_0(s_{ij}|\vartheta)^{\tau_i/m_i},
\end{align*}
so the estimate of the parameter can be obtained by solving the following maximization
$$\hat{\vartheta}_0=\underset{\vartheta\in\Pi_0}{\text{arg max}}\hspace{0.05in} \mathscr{W}(\vartheta|\mb{s}).$$
To find this weighted likelihood estimator (WLE), let
\begin{align*}
    \text{log}\hspace{0.03in}\mathscr{W}(\vartheta|\mb{s})=\sum_{i=0}^n\sum_{j=0}^{m_i}\frac{\tau_i}{m_i}p_0(s_{ij}|\vartheta),
\end{align*}
and hence the WL equation  $\frac{\partial}{\partial \vartheta}\text{log}\hspace{0.03in}\mathscr{W}(\vartheta|\mb{s})=0$ can be solved. The social planner's optimal belief becomes the density $p_0(.|\hat{\vartheta}_0)$. For illustration, consider the extreme case where the social planner ignores all the individuals' constraint preferences, i.e., $\tau_i = 0$ for $i =1,\dots, n$. Her goal simplifies to minimizing $R(p_0\lVert q)$, so the WL function can be simplified to $\prod_{j=1}^{m_0}p_0(s_{0j}|\vartheta)^{1/m_0}$. In this special case, we get
\begin{align*}
    \text{log}\hspace{0.03in}\mathscr{W}(\vartheta|\mb{s})=\frac{1}{m_0}\sum_{j=0}^{m_0}p_0(s_{0j}|\vartheta),
\end{align*}
and therefore $\hat{\vartheta}_0$ would coincide with the classical MLE. The next example illustrates how $\hat{\vartheta}_0$ can be derived in closed-form, and a special case is a popular estimator from statistics.
\begin{ex}[James-Stein estimator]\normalfont
    Suppose $s_i\sim p_i=\mathcal{N}(\vartheta_i,1)$, for all $i=0,\dots,n$, i.e., each individual (including the social planner) draws an independent signal from a normal distribution with mean $\vartheta_i\in\R$ and unit variance, for $n\geq3$. The WL function becomes
        $$\text{log}\hspace{0.03in}\mathscr{W}(\vartheta|\mb{s})=-\frac{n}{2}\text{log}\hspace{0.03in}2\pi-\frac{1}{2}\sum_{i=0}^n\varphi_i(s_i-\vartheta)^2,$$
    so the WLE is $\hat{\vartheta}_0=\sum_{i=0}^n\varphi_is_i$, where the $\varphi_i$'s are weights. For instance, whenever these weights satisfy $\varphi_{0}=1-\frac{n-1}{n}B^{JS}$ and $\varphi_i=B^{JS}/n$, for $i=1,\dots,n$, where $B^{JS}=(n-3)/\sum_{i=1}^n(s_i-\overline{s})^2$ and  $\overline{s}=\frac{1}{n}\sum_{i=0}^ns_i$. Then, our WLE $\hat{\vartheta}_0$ coincides with the so-called \textit{James-Stein} estimator 
   $\vartheta^{JS}_0=\overline{s}+(1-B^{JS})(s_0-\overline{s})$, which is very popular in practice because it dominates the sample mean when $n\geq3$ \citep{stein61}. The social planner's optimal belief therefore becomes $p_0\big(.|\vartheta^{JS}_0\big)=\mathcal{N}\big(\vartheta^{JS}_0,1\big).$
    
\end{ex}

\phantomsection\label{sec:dynam}

\phantomsection\label{app:mba}
\section*{Online Appendix D: MBA Preferences}

This appendix aims to briefly describe the axioms of the MBA preferences defined in Section \ref{sec:ind}. All the details that follow can be found in \citet[][Section 2]{cerre11}.  To this end, let $S$ be the set of states of nature, which is endowed with an algebra $\Sigma$. Further, let $B_0(\Sigma, \Gamma)$ denote the set of simple $\Sigma$-measurable functions on $S$ with values in the interval $\Gamma\subseteq\R$, where $B_0(\Sigma, \Gamma)$ is endowed with the topology induced by the sup-norm. Also, let $ba_1(\Sigma)$ done the set of all finitely additive probability distributions on $\Sigma$. Endow $ba_1(\Sigma)$ with the (relative) weak* topology that is the topology induced by $B_0(\Sigma, \R)$. We may notice that $ba_1(\Sigma)$ is compact under this topology. Now,  a functional $I:B_0(\Sigma, \Gamma)\rightarrow\R$ is said to be
\begin{itemize}
    \item \textit{monotonic} if $I(a)\geq I(b)$ whenever $a\geq b$;
    \item \textit{continuous} if it is sup-norm continuous;
    \item \textit{normalized} if $I(\gamma\mathds{1}_S)=\gamma$, for all $\gamma\in\Gamma$.
\end{itemize}
Then, a preference relation $\succsim$ on $F$, is said to be a ``Monotonic, Bernoulian, and Archimedean'' (MBA) preference if it satisfies the following axioms:
\begin{itemize}
    \item[---]\textit{Axiom 1} (Weak order): the binary relation $\succsim$ is non-trivial, complete, and transitive;
    \item[---]\textit{Axiom 2} (Monotonicity): if $f,g\in F$ and $f(s)\geq g(s)$ for all $s\in S$, then $f\succsim g$;
    \item[---]\textit{Axiom 3} (Risk Independence): if $x,y,z\in X$ and $\gamma\in(0,1]$, then $x\succ y$ implies $\gamma x+ (1-\gamma)z\succ \gamma y+ (1-\gamma)z$;
    \item[---]\textit{Axiom 4} (Archimedean): if $f,g,h\in F$ and $f\succ g\succ h$, then there exists $a,b\in(0,1)$ such that $a f+(1-a)h\succ g\succ b f+(1-b)h$.
\end{itemize}

The first two axioms characterize the class of so-called \textit{rational preferences}, whereas the last two are tailored for the \citet{ansc63} framework. The Archimedean axiom is a mild continuity condition. These four axioms imply the existence of (1) a Bernoulli utility index on $X$, that is, a utility function $u: X \rightarrow \R$, which is affine and represents the restriction of $\succsim$ on $X$; (2) a certainty equivalent $x_f$ for all acts $f\in F$. Most importantly, \citet[][Propositions 1, 2, and 5]{cerre11} provide the axiomatization of MBA preferences in Definition \ref{def:rev} under these four axioms, where the functional $I$ is uniquely determined by the choice of the utility function $u$, where $I(u(x_f))=I\big(u(x_f)\mathds{1}_S\big)=u(x_f).$

\phantomsection\label{app:mbamiss}
\subsection*{D.I: MBA and Misspecification}

In Section \ref{sec:social}, we remarked that since the general decision criterion under misspecification introduced in \citet[][eq. (1)]{hansenmiss22}: 
\begin{align}\label{eq:rep3}
V^{c}(f| Q)=\underset{p\in\Delta}{\text{min}}\Big\{\E_{p}\big[u(f)\big]+\underset{q\in  Q}{\text{min}}\hspace{0.03in} c(p,q)\Big\}\quad \forall f\in F,
\end{align}
is a special case of a variational preference, it admits an MBA representation \citep[][]{cerr15}. We formalize this here. For any closed set $D\subsetneq ba_1(\Sigma)$ and every $f\in F$, define $\underline{D}(u(f)):=\text{min}_{q\in D}\E_{q}[u(f)]$ and $\overline{D}(u(f)):=\text{max}_{q\in D}\E_{q}[u(f)]$.  \citet[][Proposition 5]{cerre11} shows that any MBA preference relation $\succsim$ can be represented by
\begin{align}\label{eq:mba2}
    I(u(f))=\alpha(f)\underline{ P}(u(f))+(1-\alpha(f))\overline{P}(u(f)) \quad \forall f\in F,
\end{align}
where $ P\subsetneq ba_1(\Sigma)$ is a nonempty, unique, closed and convex set. Here, $\alpha:B_0(\Sigma,u(X))\rightarrow[0,1]$ is defined as $\alpha(f)=\frac{\overline{P}(u(f))-I(u(f))}{\overline{P}(u(f))-\underline{ P}(u(f))}$, for all $f\in F$ \citep[][eq. (2)]{cerre11}. \par Our goal is to express $V^{c}(f| Q)$ in eq. (\ref{eq:rep3}) in the form of $I(u(f))$ in eq. (\ref{eq:mba2}). For this to be possible, we need $V^{c}(f| Q)=I(u(f))$, so there must exist a unique set $ P\subsetneq ba_1(\Sigma)$ such that $\underline{ P}(u(f))\leq V^{c}(f| Q)\leq \overline{P}(u(f))$  \citep[][ Corollary 3]{cerre11}. For all $f\in F$, we know $V^{c}(f| Q)\leq \text{min}_{q\in  Q}\E_{q}[u(f)]$ \citep[][eq. (23)]{hansenmiss22}, so it must be that $ Q\subsetneq  P$ to ensure $\alpha(f)\in[0,1]$, for all $f\in F$. We have therefore established that, for the decision criterion $V^{c}(f| Q)$ in eq. (\ref{eq:rep3}), there exists a unique, closed and convex set $ P\subsetneq ba_1(\Sigma)$ where $ Q\subsetneq  P$ and a coefficient $\alpha^{c}(f)=\frac{\overline{P}(u(f))-V^{c}(f| Q)}{\overline{P}(u(f))-\underline{ P}(u(f))}\in[0,1]$ such that 
\begin{align*}
    V^{c}(f| Q)=\alpha^{c}(f)\underline{ P}(u(f))+(1-\alpha^{c}(f))\overline{P}(u(f))\quad \forall f\in F.
\end{align*}
Thus, $(u, P,\alpha^c)$ is the MBA representation of $V^{c}(.| Q)$ in eq. (\ref{eq:rep3}) where $ Q\subsetneq  P\subsetneq ba_1(\Sigma)$. The exact form of the MBA set of beliefs $P$ is given in \citet[][eq. (3)]{cerr15}, whereas the set of structured models $Q$ is defined to satisfy $\{p\in\Delta:\text{min}_{q\in Q}c(p,q)=0\}$.
\begin{remark}\normalfont
  Notice that the strict inclusions $ Q\subsetneq  P\subsetneq ba_1(\Sigma)$ capture the idea of misspecification concerns of a decision maker. Specifically, it indicates that the decision maker represented by $V^{c}(.| Q)$ in eq. (\ref{eq:rep3}) pays special attention to structured models $ Q$. However, she is concerned that the models in $ Q$ may be misspecified, so she also entertains the unstructured models contained in $ P\supsetneq  Q$, which may have some epistemic content. Then, the strict inclusion $ P\subsetneq ba_1(\Sigma)$ reflects the fact that she cannot entertain all unstructured models because she generally views them as statistical artifacts that lack substantive motivation.    
\end{remark}

\phantomsection\label{app:dom}
\section*{Online Appendix E: Welfare-Dominant Belief}
This online appendix leverages the entropic welfare criterion in eq. (\ref{eq:rep}) to illustrate the welfare-dominant set $Q$ in Desideratum \ref{des:welfare}. Recall the notation $C:=\text{co}\big(\bigcup_{i=1}^nP_i\big)$.

\par\noindent--- \textit{Overview}: Under the entropic criterion, evaluating $f$ at a singleton $\{q\}$ reduces to a certainty equivalent of an exponential transform of $u_0(f)$ under $q$.
On the common-taste domain where higher states (in a common-state order) yield higher $u_0$, the exponential transform is statewise decreasing. Any belief that shifts probability mass toward better states lowers the exponential moment and raises the certainty equivalent.
If there is a single first-order stochastic dominance (FOSD)-maximal belief in $C$, that belief maximizes value for every act in $F_\star$; taking $Q$ to be a singleton consisting of this belief will satisfy Desideratum \ref{des:welfare}. We provide an interpretation of this set in the context of AI at the end of this appendix.
\par Recall that, for $\lambda>0$ and $Q=\{q\}$, we have the following closed-form expression
\[
V^{\lambda R}_0(f|\{q\})
=\underset{p\in \Delta}{\text{min}}\big\{\E_p[u_0(f)]+\lambda R(p\Vert q)\big\}
=-\lambda\log \E_q\left[e^{-u_0(f)/\lambda}\right].
\]
\par In the result below, we assume there is a \textit{common-state} order on $S$, i.e., a total order 
$\geq_S$ on the finite state space 
$S$, which can be interpreted as ranking states from worse to better. Then, for any beliefs $p,p'\in \Delta$, we say: 
$p$ \textit{FOSD} $p'$ \textit{with respect to} $\geq_S$ \textit{if} $\sum_{s\geq_St}p(s)\geq \sum_{s\geq_St}p'(s)$ \textit{holds for every threshold state} $t\in S$, \textit{with strict inequality for some} $t$.
\begin{obs}\label{thm:entropic}
Fix a common-state order $\geq_S$. Assume: (i) $P_i=\{p_i\}$ are totally ordered by FOSD with respect to $\geq_S$, with a greatest element $p^{\star}$, and (ii) for all $f\in F_\star$, $u_0(f(\cdot))$ is nondecreasing in $\geq_S$. Then, the set $Q:=\{p^{\star}\}$ satisfies \textnormal{Desideratum \ref{des:welfare}}.
\end{obs}

\begin{proof}
Fix $f\in F_\star$ and set $g_f(s):=e^{-u_0(f(s))/\lambda}$, which is nonincreasing in the common-state order.
By FOSD, $\E_{p^{\star}}[g_f]\leq \E_q[g_f]$ for all $q\in C$; hence
$V^{\lambda R}_0(f|\{q\})=-\lambda\log\E_q[g_f]\leq -\lambda\log\E_{p^{\star}}[g_f]=V^{\lambda R}_0(f|\{p^{\star}\})$.
Then, notice also that the following inequality
\begin{align*}
    V^{\lambda R}_0(f| Q')=\text{min}_{q\in Q'}V^{\lambda R}_0(f|\{q\})\leq V^{\lambda R}_0(f|\{p^{\star}\})=V^{\lambda R}_0(f| Q)
\end{align*}
holds for all $Q'\subseteq C$. Since $f\in F_{\star}$ was arbitrary, the above holds uniformly across $F_{\star}$.
\end{proof}
We end by providing a real-life interpretation for the welfare-dominant belief $p^{\star}$.
\par\noindent--- \textit{Interpretation}: Consider a Federal Open Market Committee (FOMC) meeting where members of the Federal Reserve are evaluating a new policy to manage the macro consequences of Generative AI. Each regional Fed research team brings a structural model of the AI transition. They disagree about how likely the “AI boom” is versus scenarios where AI disrupts labor markets and compresses demand. However, they all agree on one thing: conditional on any given state of the world, a particular \textit{pro-cyclical} policy---say, allowing credit to expand more aggressively in AI-intensive sectors in good times---is strictly better for social welfare than the status-quo policy. In other words, on the common-taste domain $F_{\star}$, everyone ranks, state by state, the pro-cyclical policy above the status quo.
\par The FOMC chair is concerned about misspecification and uses our entropic criterion. Starting from a baseline model $q$, she lets Nature distort the probabilities toward worse macro states, subject to the relative-entropy penalty, and she evaluates each policy using the resulting worst-case certainty equivalent. If she anchored this procedure on a very pessimistic belief, she would inevitably \textit{double count}  caution in the following sense: the baseline model already puts a lot of weight on bad AI states, and then the robustness operator tilts further toward those same bad states. This can therefore make the welfare criterion reject even policies that all advisors view as strictly better in every state (i.e., consensual policies). To prevent this, a chair who satisfies Desideratum \ref{des:welfare} would choose the most FOSD-optimistic belief in the convex hull of advisors' beliefs, $p^{\star}$, i.e., the belief that, within the common-state order (from ``very bad AI transition'' to ``very good AI transition''), puts relatively more probability on AI outcomes that all advisors agree are better in every state. 
\par In summary, $p^{\star}$ is not driven by ``blind optimism.'' Instead, it is the unique structured model that makes robustness tests as favorable as possible to policies that all advisors agree are better in every state, while the entropic worst-case still builds in caution about tail risks.

\phantomsection\label{app:proof}
\section*{Online Appendix F: Omitted Proofs}

\subsection*{Proof of Lemma \ref{lem:support}}
\begin{proof}[Proof of Lemma \ref{lem:support}]
Consider the normal cone
$N_D(x):=\big\{\psi\in\mathbb R^S: \langle \psi,y-x\rangle\ge 0 \forall y\in D\big\}$,
which is nonempty for compact convex $D$.
Pick any $\psi\in N_D(x)$ and set $m:=\langle \psi,x\rangle$.
Then, for all $y\in D$, $\langle \psi,y\rangle\ge \langle \psi,x\rangle=m$, hence the affine hyperplane $H:=\{y:\langle \psi,y\rangle=m\}$ supports $D$ and contains $x$.
Let $L:=D\cap H=\{y\in D:\langle \psi,y\rangle=m\}$.
We claim $L$ is a face containing $x$.
To see this, suppose $z\in L$ and $z=\gamma y+(1-\gamma)z'$ with $y,z'\in D$ and $\gamma\in(0,1)$.
Then,
$m=\langle \psi,z\rangle=\gamma\langle \psi,y\rangle+(1-\gamma)\langle \psi,z'\rangle\ge \gamma m+(1-\gamma)m=m$,
so $\langle \psi,y\rangle=\langle \psi,z'\rangle=m$ and $y,z'\in L$.
Thus, $L$ is a face containing $x$.
By minimality, $\mathcal{F}_x\subseteq L$, which proves \eqref{eq:support-ineq}.
If $D\subseteq\Delta$, then $\langle \mathbf{1},y\rangle=1$ for all $y\in D$; with $\varphi:=\psi-m\mathbf{1}$ we get
$\langle \varphi,y\rangle=\langle \psi,y\rangle-m\ge 0$ on $D$ and equality on $L\supseteq \mathcal{F}_x$, hence \eqref{eq:phiProps}. \end{proof}

\subsection*{Proof of Proposition \ref{thm:bregball}}
Recall some definitions. Let be $V$ a reflexive Banach space and $G : \mathcal{D} \to \mathbb{R}$ be of the Legendre type on an open convex domain $\mathcal{D} \subseteq V$, and its Hessian $\nabla^2 G(y)$ is positive definite for all $y \in \mathcal{D}$. Let $\mathcal{C}\subseteq\mathcal{D}$ be a closed convex set.  For $x,y \in \mathcal{C}$, their primal Bregman divergence is
\begin{align*}
    D_G(x \lVert y)
    :=
    G(x) - G(y) - \langle \nabla G(y), x - y \rangle.
\end{align*}
Fix $n \ge 1$, and let $x_1,\dots,x_n \in \mathcal{C}$.
For each $i$, let $r_i \ge 0$, and define the primal Bregman ball
\begin{align*}
    B^G_{r_i}(x_i)
    :=
    \bigl\{ y \in \mathcal{C} : D_G(x_i \lVert y) \leq r_i \bigr\},
\end{align*}
which is closed and convex. Define
\begin{align*}
    \mathcal{S}
    :=
    \bigcap_{i=1}^n B^G_{r_i}(x_i)
    =
    \Bigl\{ y \in \mathcal{C} : D_G(x_i \lVert y) \leq r_i \quad \text{for all } i=1,\dots,n \Bigr\},
\end{align*}
Define the function $\Phi : \mathcal{C} \to \mathbb{R}$ by
\begin{align}
    \Phi(y)
    :=
    \underset{1 \leq i \leq n}{\text{max}} \Bigl( D_G(x_i \lVert y) - r_i \Bigr).
    \label{eq:defPhi}
\end{align}
\begin{lemma}
\label{lem:sublevel}
For any $y \in \mathcal{C}$, we have
\begin{align*}
    y \in \mathcal{S}
    \quad\Longleftrightarrow\quad
    \Phi(y) \leq 0.
\end{align*}
\end{lemma}

\begin{proof}[Proof of Lemma \ref{lem:sublevel}]
($\Rightarrow$)
If $y \in \mathcal{S}$, then $D_G(x_i \lVert y) \leq r_i$ $\forall i$. Hence, $D_G(x_i \lVert y) - r_i \leq 0$ $\forall i,$
so $\Phi(y)
    = \text{max}_{i} \big(D_G(x_i \lVert y) - r_i\big)
    \leq 0.$

($\Leftarrow$)
If $\Phi(y) \leq 0$, then for each $i$,
\[
    D_G(x_i \lVert y) - r_i
    \le
    \text{max}_j \Bigl(D_G(x_j \lVert y) - r_j \Bigr)
    = \Phi(y)
    \leq 0,
\]
so $D_G(x_i \lVert y) \leq r_i$ for all $i$, hence $y \in \mathcal{S}$.
\end{proof}

\begin{proof}[Proof of Proposition \ref{thm:bregball}]

Since $G$ is of the Legendre type, $y \mapsto D_G(. \lVert y)$ is convex, and $\Phi$ in eq. (\ref{eq:defPhi}) is lower semicontinuous and proper, and because the pointwise maximum of finitely many convex functions is convex, $\Phi$ is also convex. Now, consider the minimization problem
\begin{align*}
    \inf_{y \in \mathcal{C}} \Phi(y).
\end{align*}

\begin{lemma}
\label{lem:attains}
There exists $y^\star \in \mathcal{C}$ such that
\[
    \Phi(y^\star)
    = \inf_{y \in \mathcal{C}} \Phi(y).
\]
Moreover, $\Phi(y^\star) \leq 0$, so $y^\star \in \mathcal{S}$.
\end{lemma}

\begin{proof}[Proof of Lemma \ref{lem:attains}]
Since $\mathcal{S}$ is nonempty, pick $\bar{y} \in \mathcal{S}$. By Lemma \ref{lem:sublevel}, $\Phi(\bar{y}) \leq 0$. Thus, $\inf_{y \in \mathcal{C}} \Phi(y)
    \le
    \Phi(\bar{y})
    \leq 0.$
Observe that $\Phi$ is convex, lower semicontinuous, proper, and $\mathcal{C}$ is a closed convex and bounded subset of a reflexive Banach space. Thus, we can apply Lemma \ref{thm:exist} to conclude that $\Phi$ attains its minimum at some $y^\star \in \mathcal{C}$. By the two previous inequalities, $\Phi(y^\star)
    = \inf_{y \in \mathcal{C}} \Phi(y)
    \leq 0,$
hence, by Lemma \ref{lem:sublevel}, $y^\star \in \mathcal{S}$.
\end{proof}

Now, let's analyze the \emph{active set} at $y^\star$ in Lemma \ref{lem:attains}: 
\begin{align*}
    A(y^\star)
    :=
    \Bigl\{
        i \in \{1,\dots,n\}
        :
        \Phi(y^\star)
        =
        D_G(x_i \lVert y^\star) - r_i
    \Bigr\}.
\end{align*}
 $A(y^\star)$ contains indices $i$ where the maximum in eq. \eqref{eq:defPhi} is achieved at $y^\star$. Since $G$ has a Hessian, it is twice Fr\'echet differentiable, so $m_i(y):= D_G(x_i \lVert y) - r_i$ is  Fr\'echet differentiable for all $i$. Since $\Phi$ in eq. \eqref{eq:defPhi} is the pointwise maximum of the finitely many convex, Fr\'echet differentiable functions $m_i(.)$, its subdifferential at $y^\star$ is 
\begin{align}
    \partial \Phi(y^\star)
    =
    \text{co}
    \Big(\Bigl\{
        \nabla m_i(y^\star)
        :
        i \in A(y^\star)
    \Bigr\}\Big)
    =
    \text{co}
    \Big(\Bigl\{
        \nabla_y D_G(x_i \lVert y^\star)
        :
        i \in A(y^\star)
    \Bigr\}\Big),
    \label{eq:subdiff}
\end{align}
which holds by applying \citet[][Exercise 4.1.44.(b)]{convex10} and dropping their ``weak*-closure'' because $\partial m_i(y^\star)=\{\nabla m_i(y^\star)\}$ is a singleton for all $i$, so $\{\nabla m_i(y^\star):i \in A(y^\star)\}$ is a finite set of gradients. Since $y^\star$ minimizes the convex function $\Phi$, the first-order optimality condition in \citet[][Exercise 4.1.49]{convex10} implies that
\begin{align}
    0 \in \partial \Phi(y^\star)+N_{\mathcal{C}}(y^\star),
    \label{eq:optimality}
\end{align}
where $N_{\mathcal{C}}(y^\star)$ denotes the normal cone at $y^\star$.
By \eqref{eq:subdiff}, there exists
$\zeta\in \Delta(A(y^\star))$ such that
\begin{align}
   g:=\sum_{i \in A(y^\star)} \zeta_i \nabla_y D_G(x_i \lVert y^\star)\in \partial \Phi(y^\star).
    \label{eq:zerograd}
\end{align} 
Now, we use the specific structure of the Bregman divergence. For all $x,y$, the first Fr\'echet derivative of $D_G(x \lVert y)
    = G(x) - G(y) - \langle \nabla G(y), x-y\rangle$ can be shown to be
\begin{align}
    \nabla_y D_G(x \lVert y)
    =
    \nabla^2 G(y)\bigl(y - x\bigr),
    \label{eq:gradBreg}
\end{align}
where $\nabla^2 G(y)$ is the Hessian of $G$ at $y$, which is assumed to be positive definite and hence invertible. Substituting \eqref{eq:gradBreg} into \eqref{eq:zerograd} yields
\begin{align}\label{eq:gg}
    g= \sum_{i \in A(y^\star)} \zeta_i \nabla^2 G(y^\star)\bigl(y^\star - x_i\bigr)
    = \nabla^2 G(y^\star)
       \Bigl(
           y^\star-
           \sum_{i \in A(y^\star)} \zeta_i x_i
       \Bigr),
\end{align}
using that $\sum_{i \in A(y^\star)} \zeta_i=1$. Choose any $\hat{g}\in N_{\mathcal{C}}(y^\star)$. Then, using \eqref{eq:optimality} with $g$ and $\hat{g}$, we get 
\begin{align*}
    - \nabla^2 G(y^\star)
       \Bigl(
           y^\star-
           \sum_{i \in A(y^\star)} \zeta_i x_i
       \Bigr)=\hat{g}.
\end{align*}
Let $\bar{x}:=\sum_{i\in A(y^\star)}\zeta_i x_i$, so $\bar{x}\in \mathcal{C}$ because $x_i\in \mathcal{C}$ for all $i\in A(y^\star)$ and $\mathcal{C}$ is convex. Rearranging the equation above and using the fact that $\nabla^2 G(y^\star)$ is invertible, we get 
\begin{align}\label{eq:diff}
    \bar{x}-y^\star=[\nabla^2 G(y^\star)]^{-1}\hat{g}.
\end{align}
By definition of normal cone, $\langle \hat{g},z-y^\star \rangle\leq0$ for all $z\in\mathcal{C}$ \citep[][Exercise 4.1.33]{convex10}. Choosing $z=\bar{x}$, we must have $\langle \hat{g},\bar{x}-y^\star \rangle\leq0$. Substituting the expression for $\bar{x}-y^\star$ (from eq. (\ref{eq:diff})) yields
\begin{align}\label{eq:ineqhat}
    \big\langle \hat{g},[\nabla^2 G(y^\star)]^{-1}\hat{g} \big\rangle\leq0.
\end{align}
Since $\nabla^2 G(y^\star)$ is positive definite, its inverse is also positive definite and therefore the quadratic form $\langle \hat{g},[\nabla^2 G(y^\star)]^{-1}\hat{g}\rangle$ is strictly positive for all $\hat{g}\neq0$. The inequality in (\ref{eq:ineqhat}) implies that $\hat{g}=0$, so plugging this in eq. (\ref{eq:diff}) yields $$y^\star=\bar{x}=\sum_{i\in A(y^\star)}\zeta_i x_i.$$
 By Lemma \ref{lem:attains}, $y^\star \in \mathcal{S}$, so $y^\star\in \mathcal{S}\cap \text{co}(\{x_1,\dots,x_n\})$.
\end{proof}

\subsection*{Proof of Corollary \ref{thm:bound}}
To prove Corollary \ref{thm:bound}, we need the definition of a \textit{functional} derivative.
\begin{definition}[Functional Derivative]\normalfont\label{def:deriv}
Let $W:\Delta_d\rightarrow\R$ be a functional. Given a function $h\in \Delta_d$, the functional derivative of $W$
 at $h$, denoted $\frac{\partial W}{\partial h}$, is defined as the function satisfying
\begin{align*}
    \int_S \xi(s)\hspace{0.03in}\frac{\partial W}{\partial h}(s)\hspace{0.03in}d\nu(s)=\underset{\varepsilon\rightarrow0}{\text{lim}}\frac{W(h+\varepsilon \xi)-W(h)}{\varepsilon}=\frac{d}{d\varepsilon}W(h+\varepsilon \xi)\Big|_{\varepsilon=0},
\end{align*}
where $\varepsilon$
 is a scalar and $\xi$
 is an arbitrary function.\hfill 
\end{definition} 
Given this definition, we can define the following quantity that will be useful 
\begin{align}\label{eq:inner}
    \Lambda\big(pq\lVert hq\big):=\int_S(q-h)\frac{\partial R(p\lVert q)}{\partial q}\hspace{0.03in}d\nu=-\int_S(1-h/q)p\hspace{0.03in}d\nu,
\end{align}
for all $p,q,h\in\Delta_d$, where $\frac{\partial R(p\lVert q)}{\partial q}=-p/q$ (Definition \ref{def:deriv}). The next two lemmas are useful.
\begin{lemma}\label{thm:convex}
 Let $p\neq q\in\Delta_d$ and $w_{\varphi} := p + \varphi(q - p)$, for a constant $\varphi\in [0, 1]$. Then,
$R(p\lVert w_{\varphi})$ is strictly convex with respect to $\varphi$. 
\end{lemma}
\begin{proof}[Proof of Lemma \ref{thm:convex}]
    If $\varphi_1\neq\varphi_2\in[0,1]$, then $w_{\varphi_1}\neq w_{\varphi_2}$ follows since $p\neq q$. Let $\alpha\in(0,1)$ be a constant and
    $R\big(p\lVert
    \alpha w_{\varphi_1}+(1-\alpha)w_{\varphi_2}\big)<\alpha R(p\lVert
     w_{\varphi_1})+(1-\alpha)R(p\lVert
 w_{\varphi_2}),$
    which holds since the relative entropy is strictly convex in both arguments. Then, the result follows from $ \alpha w_{\varphi_1}+(1-\alpha)w_{\varphi_2}=p+(q-p)\big(\alpha \varphi_1+(1-\alpha)\varphi_2\big)=w_{\alpha \varphi_1+(1-\alpha)\varphi_2}$ by definition of $w_{\varphi}$.
\end{proof}
\begin{lemma}\label{thm:pyth}
    For $p,q,h\in \Delta_d$, $R(p\lVert h)\geq R(p\lVert q)-\Lambda(pq\lVert hq)$, with equality if and only if $q=h$.
\end{lemma}
\begin{proof}[Proof of Lemma \ref{thm:pyth}]
    Let $h_{\varphi}:=q+\varphi(h-q)$ and $\mathcal{R}(\varphi):=R(p\lVert h_{\varphi})$, for a constant $\varphi\in[0,1]$, and suppose $q\neq h$. Then, $\mathcal{R}$ is strictly convex in $\varphi$ by Lemma \ref{thm:convex}. Let $\zeta\big(f(s)\big)=h(s)-q(s)$ in the definition of functional derivative (Definition \ref{def:deriv}), then
    \begin{align*}
        \mathcal{R}'(s)=\frac{d}{d\gamma}R(p\lVert h_{\varphi+\gamma})\Big|_{\gamma=0}=\int_{S}(h-q)\frac{\partial R(p\lVert h_{\varphi})}{\partial h_{\nu}}\hspace{0.03in}d\nu,
    \end{align*}
    where $h_{\varphi+\gamma}=h_{\varphi}+\gamma(h-q)$. For any $\varphi>0$, we have $\mathcal{R}(\varphi)>\mathcal{R}(0)+\mathcal{R}'(0)(\varphi-0)$, since $\mathcal{R}$ is strictly convex. When $\varphi=1$, $\mathcal{R}(1)=R(p\lVert h)$, and when $\varphi=0$, $\mathcal{R}(0)=R(p\lVert q)$, so 
    \begin{align*}
        \mathcal{R}'(0)=\int_{S}(h-q)\frac{\partial R(p\lVert q)}{\partial q}\hspace{0.03in}d\nu=-\Lambda(pq\lVert hq),
    \end{align*}
  where $\Lambda(pq\lVert hq)$ is defined in eq. (\ref{eq:inner}), and hence it follows that $R(p\lVert h)>R(p\lVert q)-\Lambda(pq\lVert hq)$. When $q=h$, the result follows trivially with equality in which case $\Lambda(pq\lVert hq)=0$.
\end{proof}

For example, setting $p=h$ in Lemma \ref{thm:pyth} yields $\Lambda(pq\lVert pq)>R(p\lVert q)$ whenever $p\neq q$. 
\begin{proof}[Proof of Corollary \ref{thm:bound}]
    When $\sigma=0$, $q^{\theta}_0=\text{arg min}_{q\in \mathcal{Q}_{\theta}}R(p^*\lVert q)$ by Proposition \ref{thm:accurate}. We can now use the same steps as in Lemma \ref{thm:pyth}. For a constant $\varphi\in[0,1]$, let $w_{\varphi}:=q^{\theta}_0+\varphi (q-q^{\theta}_0)$ and $\mathcal{R}(\varphi):=R(p^*\lVert w_{\varphi})$. Then, $\mathcal{R}(0)$ is the minimum for all $\varphi\in[0,1]$, so $\mathcal{R}'(0)\geq0$, and hence
    \begin{align*}
        0\leq \mathcal{R}'(0)=\int \big(q-q^{\theta}_0\big)\frac{\partial R(p^*\lVert q^{\theta}_0)}{\partial q^{\theta}_0}\hspace{0.03in}d\nu=-\Lambda\big(p^*q^{\theta}_0\lVert qq^{\theta}_0\big),
    \end{align*}
    then defining $\kappa^*_q:=-\Lambda(p^*q^{\theta}_0\lVert qq^{\theta}_0\big)\geq0$ yields the desired result.
\end{proof}

\subsection*{Proofs of Proposition \ref{thm:axiom} and Corollaries \ref{thm:avoid}--\ref{thm:meu}}
Recall \citet[][Proposition 3.1]{miss24}, where every individual $i$ has a decision criterion
$V_i(f)=\underset{p\in\Delta}{\text{min}}\big\{\E_p[u(f)]+\underset{q\in Q_i}{\text{min}}\hspace{0.04in}d_i(p, q)\big\},$
where the $Q_i$'s are compact and convex sets, $d_i(.,.)$ is a jointly lower semicontinuous and convex function, and it
satisfies $d_i(p,q)=0$ if and only if $p=q$. Notice that all these conditions are satisfied by the $\phi$-divergence $D_{\phi}(.\lVert.)$.
\begin{proof}
---Proposition \ref{thm:axiom}: We can apply \citet[][Proposition 3.1.]{miss24} to obtain that satisfying Axiom \ref{def:AUP} is equivalent to $c_0(p)=\text{ min}_{q\in Q_0}\lambda D_{\phi}(p\lVert q)$, where $Q_0\subseteq\text{co}(\{p_1,\dots,p_n\})$. 
\par\noindent---Corollary \ref{thm:avoid}: This result follows from \citet[][Theorem 1]{miss24} because Axioms \ref{def:AUP}--\ref{def:caution} imply that $Q_0=\text{co}(\{p_1,\dots,p_n\})$. Thus, by Proposition \ref{thm:max}, we have that, for all $f\in F$, $$V^{\lambda R}_{0}\big(f\big|\text{co}(\{p_1,\dots,p_n\})\big)=V^{\lambda R}_{0}\big(f\big|\{p_1,\dots,p_n\}\big),$$
so $V^{\lambda R}_{0}(f|Q_0)=\text{min}_{i\leq n}\hspace{0.02in}\phi^{-1}_{\lambda}\big(\E_{p_i}[\phi_{\lambda}(u(f))]\big)$ by \citet[][Proposition 1.4.2]{dupuis97}. 

\par\noindent---Corollary \ref{thm:meu}: This result follows directly from \citet[][Corollary 1]{miss24}.
\end{proof}

\subsection*{Proof of Proposition \ref{thm:min2}}

Following the arguments in the proof of Proposition \ref{thm:accurate}, the set of structured models $\mathcal{Q}^{\rho}_{\theta}$ is convex, closed, and bounded for all $\rho\in(0,1)$. Moreover, the objective function $D_{\rho}(p^*\lVert {q})$ is bounded, non-negative, strictly convex, and continuous in $q$. We can therefore apply Lemma \ref{thm:exist} to establish that $D_{\rho}(p^*\lVert {q})$ attains a unique minimum on $\mathcal{Q}^{\rho}_{\theta}$, for all $\rho\in(0,1)$. To ease notation, in what follows, let $p_{n+1}:=p^*$. 
\par  By Lagrange's theorem, our optimization problem can be written as the minimization or the following functional  
\begin{align*}
    \mathscr{L}_{\rho}({q})=\frac{1}{\rho(1-\rho)}\int_{{S}} p_{n+1}({s})\Bigg[1-\Big(&\frac{{q}({s})}{p_{n+1}({s})}\Big)^{\rho}\Bigg]\hspace{0.03in}d\nu({s})+\ell_0\Bigg[\int_{{S}} {q}({s})\hspace{0.03in}d\nu({s})-1\Bigg]\\&+\sum_{i=1}^n\ell_i\Bigg\{\frac{1}{\rho(1-\rho)}\int_{{S}} p_i({s})\Bigg[1-\Big(\frac{{q}({s})}{p_i({s})}\Big)^{\rho}\Bigg]\hspace{0.03in}d\nu({s})-\tau_{i}\Bigg\},
\end{align*}
where  $\ell_i$ denote the Lagrange multipliers of $i$'s constraint (\ref{eq:const2}) and $\ell_0$ denotes the Lagrange multiplier of constraint the normalizing constraint $\int_Sq\hspace{0.02in}d\nu=1$. Omitting constants, this Lagrangian can be simplified to the functional
\begin{align*}  \mathscr{L}^*_{\rho}({q})=\int_S\underbrace{\Bigg\{-\frac{p_{n+1}({s})}{\rho(1-\rho)}\Big(\frac{{q}({s})}{p_{n+1}({s})}\Big)^{\rho}+\ell_0 {q}({s})-\sum_{i=1}^n\frac{\ell_ip_i({s})}{\rho(1-\rho)} \Big(\frac{{q}({s})}{p_i({s})}\Big)^{\rho}\Bigg\}}_{:=\hat{\psi}_{\rho}({q}|s)}\hspace{0.04in}d\nu({s}).
\end{align*}
The Euler-Lagrange equation becomes $\nabla_q\hat{\psi}_{\rho}(q|.)=0$, so solving this equation yields
\begin{align*}
  -\frac{p_{n+1}^{1-\rho}{q}_\rho^{\rho-1}}{1-\rho} +\ell_0 -\sum_{i=1}^n\ell_i \frac{p_{n+1}^{1-\rho}{q}_\rho^{\rho-1}}{1-\rho}&=0\nonumber\\
 (1-\rho)\ell_0{q}_{\rho}^{1-\rho} &=p_{n+1}^{1-\rho}+\sum_{i=1}^n\ell_ip_i^{1-\rho}\nonumber\\{q}_\rho&=\Bigg(\sum_{i=1}^{n+1}\sigma_{i}p_i^{1-\rho}\Bigg)^{\frac{1}{1-\rho}},
\end{align*}
for all $\rho\in(0,1)$, where the weights $\{\sigma_{i}\}_{i=1}^{n+1}$ are given by 
\begin{align*}
    \sigma_{i}=\frac{\ell_i}{(1-\rho)\ell_0}, \text{ for all }i\neq n+1, \quad \text{and}\quad \sigma_{n+1}&=\frac{1}{(1-\rho)\ell_0}.
\end{align*}
Since ${q}_\rho\in\mathcal{Q}^{\rho}_{\theta}$, it must be a valid density for all $\rho\in(0,1)$, i.e., the $\sigma_i$'s must be chosen such that $\int_S{q}_\rho\hspace{0.02in} d\nu=1$, and hence
\begin{align*}
    q_\rho\propto \Bigg(\sum_{i=1}^{n+1}\sigma_{i}p_i^{1-\rho}\Bigg)^{\frac{1}{1-\rho}},
\end{align*}
 i.e., $q_{\rho}$ is equal to the right hand side up to a fixed multiplicative factor, for all $\rho\in(0,1)$.
  
\subsection*{Proof of Proposition \ref{thm:mono2}}

    As noted in the proof of Proposition \ref{thm:mono}, we continue to work with equality constraints. The $i$-th (equality) constraint in eq. (\ref{eq:const2}) is
    $$\tau_{i}=\frac{1}{\rho(1-\rho)}\int_{{S}} p_i\Bigg[1-\Big(\frac{{q}}{p_i}\Big)^{\rho}\Bigg]\hspace{0.03in}d\nu=\frac{1}{\rho(1-\rho)}\int_{{S}}\Big[p_i-p_i^{1-\rho}{q}^{\rho}\Big]\hspace{0.03in}d\nu,$$
for $i=1,\dots,n$. Replacing ${q}_\rho$ for ${q}$ above, for some $\sigma_i$'s that satisfy all the equality constraints (by Proposition \ref{thm:min2}) yields 
\begin{align*}
    \tau_{i}=\frac{1}{\rho(1-\rho)}\int_{{S}}\Bigg[p_i-p_i^{1-\rho}\Bigg(\sum_{k=1}^n\sigma_{k}p_k^{1-\rho}\Bigg)^{\frac{\rho}{1-\rho}}\Bigg]\hspace{0.03in}d\nu,
\end{align*}
for all $\rho\in(0,1)$, and now taking the derivative with respect to $\sigma_{i}$ for $i=1,\dots,n$,
\begin{align*}
    \frac{\partial\tau_{i}}{\partial\sigma_{i}}&=\frac{-1}{\rho(1-\rho)}\frac{\partial}{\partial\sigma_{i}}\Bigg[\int_{{S}}p_i^{1-\rho}\Big(\sigma_{i}p_i^{1-\rho}+\sum_{k\neq i}\sigma_{k}p_k^{1-\rho}\Big)^{\frac{\rho}{1-\rho}}\hspace{0.03in}d\nu\Bigg]\\
    &=\frac{-1}{(1-\rho)^2}\int_{{S}}p_i^{2(1-\rho)}\Big(\sigma_{i}p_i^{1-\rho}+\sum_{k\neq i}\sigma_{k}p_k^{1-\rho}\Big)^{\frac{2\rho-1}{1-\rho}}\hspace{0.03in}d\nu\\
    &=\frac{-1}{(1-\rho)^2}\int_{{S}}p_i^{2(1-\rho)}{q}_\rho^{2\rho-1}\hspace{0.03in}d\nu\\
    &\leq0,
\end{align*}
where the last inequality holds because $p_i$ and ${q}_\rho$ are valid densities, for all $i=1,\dots,n$ and all $\rho\in(0,1)$. Since this transformation is monotonic, then
\begin{align*}
    \frac{\partial\sigma_{i}}{\partial \tau_{i}}=\frac{1}{\frac{\partial\tau_{i}}{\partial\sigma_{i}}}\leq0,
\end{align*}
for all $i=1,\dots,n$ and all $\rho\in(0,1)$, which concludes the proof.

\end{document}